\documentclass{aa}
\usepackage{xcolor}
\usepackage[breaklinks=true, colorlinks, citecolor={blue}, urlcolor={blue}]{hyperref}
\usepackage{graphicx}
\graphicspath{ {./images/} }
\usepackage{amssymb}
\usepackage{booktabs}
\usepackage{afterpage}
\usepackage[varg]{txfonts}
\usepackage{supertabular}
\usepackage[normalem]{ulem}
\usepackage{csquotes}
\usepackage{mathtools}
\usepackage{float}
\usepackage{natbib}
\usepackage{footmisc}
\usepackage{soul}

\usepackage[switch]{lineno}
\usepackage{multirow}
\usepackage{multicol}
\usepackage{amsfonts}
\usepackage{eucal}
\usepackage{tabularx}
\usepackage{siunitx}
\usepackage{geometry}
\usepackage{pdflscape}
\usepackage{subcaption} 
\usepackage{comment}
\usepackage{pifont}
\newcommand{\cmark}{\ding{51}}%
\newcommand{\xmark}{\ding{55}}%

\usepackage{appendix}
\usepackage{natbib,twoopt}
\bibpunct{(}{)}{;}{a}{}{,} 
\makeatletter
\newcommandtwoopt{\citeads}[3][][]{\href{http://adsabs.harvard.edu/abs/#3}%
{\def\hyper@linkstart##1##2{}%
\let\hyper@linkend\@empty\citealp[#1][#2]{#3}}}
\newcommandtwoopt{\citepads}[3][][]{\href{http://adsabs.harvard.edu/abs/#3}%
{\def\hyper@linkstart##1##2{}%
\let\hyper@linkend\@empty\citep[#1][#2]{#3}}}
\newcommandtwoopt{\citetads}[3][][]{\href{http://adsabs.harvard.edu/abs/#3}%
{\def\hyper@linkstart##1##2{}%
\let\hyper@linkend\@empty\citet[#1][#2]{#3}}}
\newcommandtwoopt{\citeyearads}[3][][]%
{\href{http://adsabs.harvard.edu/abs/#3}
{\def\hyper@linkstart##1##2{}%
\let\hyper@linkend\@empty\citeyear[#1][#2]{#3}}}

\makeatother

\bibpunct{(}{)}{;}{a}{}{,}

\bibpunct{(}{)}{;}{a}{}{,} 

\newcommand{\order}[1]{} 

\def\gsim{ \lower .75ex \hbox{$\sim$} \llap{\raise .27ex \hbox{$>$}} }

\def\lsim{ \lower .75ex\hbox{$\sim$} \llap{\raise .27ex \hbox{$<$}} }

\begin{document}
\title{Gamma-ray burst prompt emission spectra at high energies}
\titlerunning{Prompt emission at high energies}

\author{Samanta Macera\inst{\ref{inst1},\ref{inst2}}\thanks{\email{\href{mailto:samanta.macera@gssi.it}{samanta.macera@gssi.it}}} 
    \and Biswajit Banerjee\inst{\ref{inst1},\ref{inst2}, \ref{inst3}}
    \and Alessio Mei\inst{\ref{inst1},\ref{inst2}}
    \and Pawan Tiwari\inst{\ref{inst1},\ref{inst2}}
    \and Gor Oganesyan\inst{\ref{inst1},\ref{inst2},\ref{inst3}} 
    \and Marica Branchesi\inst{\ref{inst1},\ref{inst2}, \ref{inst3}}}
    
\institute{Gran Sasso Science Institute, Viale F. Crispi 7, I-67100, L'Aquila (AQ), Italy
\label{inst1}
\and
INFN - Laboratori Nazionali del Gran Sasso, I-67100, L’Aquila (AQ), Italy\label{inst2}
\and 
INAF - Osservatorio Astronomico d’Abruzzo, Via M. Maggini snc, I-64100 Teramo, Italy
\label{inst3}
}
\authorrunning{Macera et. al 2024}

\date{}

\abstract{
Despite more than fifty years of gamma-ray burst (GRB) observations, several questions regarding the origin of the prompt emission, particularly at high energies, remain unresolved. We present a comprehensive analysis of 35 GRBs observed by \textit{Fermi}/GBM and \textit{Fermi}/LAT over the past 15 years, focusing on the nature of high-energy (HE, E$>$100 MeV) emission during the prompt emission phase. Our study combines temporal and spectral analyses to investigate the synchrotron origin of the observed emission spanning the energy range from 10 keV to 100 GeV and explore the possible contribution of additional spectral components. Temporal modeling of \textit{Fermi}/LAT light curves for 12 GRBs in our sample reveals deviations from standard afterglow scenarios during the early phases, suggesting a significant contamination from prompt emission. We find that most GRB spectra align with synchrotron emission extending to GeV energies, with the slope $p$ of the non-thermal electron distribution clustering around $p\sim2.7$, consistently with theoretical predictions. For three GRBs, an additional power law component is required to explain the high-energy emission, but the nature and temporal evolution of this component remain unclear due to the limited quality of \textit{Fermi}/LAT data. When the power law component is needed, the synchrotron spectrum shows a sharp MeV suppression. It could be explained by the pair loading effects in the early afterglow. These findings emphasize the importance of multi-wavelength observations in unveiling the mechanisms driving early HE prompt emission in GRBs. We briefly discuss the implications of our findings for future very-high-energy (VHE, E$>$100 GeV) gamma-ray observatories, such as the Cherenkov Telescope Array, and address the detection prospects of additional non-thermal components in GRB spectra.}

\maketitle
\section{Introduction}\label{sec:intro}

Gamma-ray burst (GRB) prompt emission consists of brief flashes of MeV radiation overall lasting fractions of seconds. Due to its short duration, typically spanning seconds to minutes, it is predominantly observed within the 10 keV to 10 MeV energy range, determined by the sensitivity of the triggering instruments. 
This emission is thought to be produced by the non-thermal particles accelerated in the dissipative zones within the relativistic jets (e.g. \citealt{Rees:1994nw, Sari:1997kn, Paczynski1994ApJ...427..708P, Daigne:2000xg}). Due to the lack of knowledge on the GRB jet composition from first principles, the dissipative processes responsible for the GRB production cannot be determined a priori \citep{Meszaros:1996sv}. Two main prompt emission sites are typically proposed: optically thick regions below the photosphere \citep{Meszaros:1999gb} and optically thin regions above it. 
Either mildly relativistic shocks or magnetic reconnection  \citep{Thompson1994MNRAS.270..480T, Spruit:2000zm} processes could be responsible for the heating of charged particles that power the observed radiation. 

In the standard GRB model, dissipation of the relativistic unsteady jet above the photosphere produces non-thermal electrons via internal shocks. In this scenario, the prompt emission is produced by the synchrotron radiation of shock-accelerated electrons \citep{Rees:1994nw,tavani1996shock}. However, the apparent inconsistency of the low-energy spectra of GRBs with the expected fast-cooling synchrotron spectral shape \citep{Crider:1996gq,Preece1998,Ghisellini2000,Ghirlanda:2002mt, Kaneko:2006ru, nava2011spectral, Gruber:2014iza} has opened up more complex scenarios. Several solutions within the synchrotron radiation model have been proposed \citep[i.e.]{lloyd2000synchrotron, Peer:2006pbr, Burgess:2018dhc}, including effects of incomplete cooling of electrons \citep{Kumar:2008rv, Daigne:2010fb, Beniamini:2013ue, Daigne:2024zso}. 
Alternatively, some authors suggest that the GRB spectra are formed in optically thick media, where dissipative photospheres provide hot electrons that enable multiple Compton scatterings of low-energy photons \citep{Ghisellini:1999tq, Peer:2003yot, Rees:2004gt}. 

For a long time, the prompt emission spectra of long GRBs were best characterized by the Band function \citep{Band:1993eg}. This model consists of two smoothly connected power law segments, with a characteristic energy falling in the range $\sim$ 100 keV$-$1 MeV. Typically, the photon indices below and above the peak energy in the $\nu F_{\nu}$ spectrum are $-1$ and $-2.3$, respectively.
This empirical model can effectively describe most GRB prompt spectra. However, despite extensive observations and modeling, a physical interpretation of the Band function remains elusive. As an empirical model, the parameters of the Band function are not constrained by physical assumptions. Furthermore, due to its simplicity, the Band function, with its two power law segments, cannot capture many significant features of the GRB spectrum, such as possible energy breaks.

In some GRBs observations, the spectrum was significantly deviating from a Band function, particularly with excess of low energy photons ($\leq 30 {\rm keV}$).
These deviations have led to suggestions of a thermal (blackbody) component superimposed on the Band function \citep{Ghirlanda:2002zx, Ryde:2004ir, Guiriec:2015ppa}.

Recently, breaks in the GRB prompt emission spectra have been found \citep{Ravasio:2019kiw}, also with the help of the inclusion of X-ray and optical data \citep{Oganesyan:2017ork, Oganesyan:2019fpa}.
This finding necessitated the addition of a third low-energy power law segment. Consequently, the spectra are now best described by three smoothly connected power law segments with indices $\alpha_{1}$, $\alpha_{2}$ and $\beta$, where the break energy $E_{b}$ ranges from a few keV to hundreds of keV. The values of $\alpha_{1}$ and $\alpha_{2}$ have been found to be consistent with marginally fast cooling synchrotron radiation \citep{Ravasio:2019kiw}.

The main challenge in identifying GRB prompt emission mechanism lies mainly in the lack of early multi-wavelength (MWL) observations. Observations at lower energies (optical and X-rays) and in the MeV gamma-ray range suggest that some GRB spectra are consistent with moderately fast-cooling synchrotron radiation models \citep{Oganesyan:2017ork, Oganesyan:2017eds, Ravasio:2019kiw}. Systematic modeling of GRB spectra using physically motivated synchrotron models is essential to advance our understanding of these enigmatic emissions.  

In general, the high-energy part of the spectrum is crucial for distinguishing between different GRB emission mechanisms, as different models provide different prediction in this energy range \citep{Rees:1994nw, Rees:2004gt, Peer:2008udu}. In particular, in the synchrotron scenario the high-energy spectrum is essential for constraining the particle energy distribution. In the optically thin scenarios (both in internal shocks or reconnection) the highest energy photons are strictly related to the size of the emitting region and  VHE photons can in principle be produced by the inverse Compton scattering \citep[see][for description of possible scenarios]{Banerjee:2022gkv}. However, it is important to consider that this part of the spectrum can also be influenced by other processes, such as pair annihilation, which may alter its observed characteristics. Furthermore, it remains uncertain whether the synchrotron model alone can fully account for the overall shape of the GRB spectrum, including its width and the transition to higher energies \citep{vanEerten:2015paa, Burgess:2014dsa}. In the optically thick models, the highest-energy photons are limited by the opacity of the last scattering surface. Therefore, it is essential to characterize the high and VHE portions of the prompt emission. 
Little is known about the prompt emission spectra at the HE and VHE gamma-rays.
Only recently, for GRB 221009A, early emission from GRBs in the VHE domain (0.3 to above 5 TeV) has been discovered, thanks to the observation by LHAASO \citep{LHAASO:2023kyg, LHAASO:2023lkv}. The detection of the VHE emission overlaps with the prompt emission detected by \textit{Fermi}/GBM. One of the possible explanations for multi-TeV emission is the synchrotron self-Compton radiation of the afterglow \citep[SSC; ][]{Derishev:2023xyx, Banerjee:2024hxp}, since no evidence of TeV light curve variability associated to the prompt emission has been found \citep{LHAASO:2023kyg}.

The Large Area Telescope \citep[LAT;][]{Atwood2009ApJ...697.1071A} onboard the Fermi Observatory enables the wide field coverage of the sky in the 30 MeV - 300 GeV energy range. The large effective area of LAT allows for the detection of short-lasting ($\sim$ 10 s) HE and VHE sources at a level of $\sim$ 10 mCrab. Provided that each point of the sky is observed by LAT in every 3 hours, many GRBs have been covered in their prompt emission phase for the last 15 years. The HE emission of GRBs has been extensively studied by many authors \citep[e.g.]{hascoet2011fermi, Vianello:2017yfd, Chand2020, Mei:2022ncd, Ravasio:2023vgh}. In most cases, the long-lasting (up to $10^{4}$ s) GeV emission discovered by LAT was attributed to synchrotron radiation from the electrons accelerated in the external shock (afterglow radiation) \citep[e.g.]{ackermann2013first, Ajello:2019avs}. Some authors found early GeV radiation to be consistent with the high-energy extension of the prompt emission spectrum \citep[][and references therein]{Nava:2014zua, Nava:2018qkq, Miceli:2022efx}. Usually, the prompt GeV spectrum is soft and is interpreted as suppressed by the photon annihilation. In these studies the prompt emission spectra are modeled by the empirical two-power law functions, and the afterglow emission component is treated separately. 
The fundamental problem in identifying the origin of the early GeV radiation is the limitation of the LAT sensitivity. It is difficult to trace the temporal evolution of the GeV emission and therefore to distinguish the onset of the afterglow from a possible prompt emission contribution. 

In this paper, we consider the high-energy emission temporally coincident with the prompt emission of GRBs detected for the last 15 years. We make use of the publicly available data from the Fermi Gamma-ray Burst Monitor \citep[GBM; 8 keV - 40 MeV;][]{Meegan2009ApJ...702..791M} and Fermi-LAT ($\sim$30 MeV - 300 GeV). We investigated the nature of the early GeV emission in the context of a synchrotron radiation model. We also search for an additional HE/VHE spectral component in the Fermi-LAT data in the prompt emission phase. The paper is organized as follows. In Section 2, we describe the joint keV-GeV sample selection criteria. In Section 3, we provide details of the acquisition and treatment of the data. The temporal analysis of the GeV emission is provided in Section 4. The description and the implementation of the tested models for the joint keV-GeV spectra is reported in Section 5. The outcome of the multi-wavelength prompt emission modeling is presented in Section 6 and discussed in Section 7. We draw the conclusions of our work in Section 8.

\section{Sample selection}\label{sec:samples}

The GRBs included in our study are primarily detected by \textit{Fermi}/GBM.  We select GRBs from the GBM trigger catalog\footnote{https://heasarc.gsfc.nasa.gov/W3Browse/fermi/fermigbrst.html}, ensuring that the probability of the trigger being a GRB exceeds 95\% (probability that the GBM flight system classified the trigger as a GRB is more than 95\%). We further refine our selection to include GRBs with a localization accuracy below 0.5$^{\circ}$, aligning with the point spread function of LAT at around 1\,GeV. From this refined sample, we choose those that demonstrate detection with LAT within their prompt phase of emission (within T$_{90}$) and detected with a test statistic TS$>$25, indicating a detection significance greater than 5$\sigma$. A significant fraction of these GRBs were not within the FoV of LAT during the prompt emission phase (within T$_{\rm 90}$) hence not included in the sample. This selection described above resulted in 35 GRBs, observed from July 2008 to October 2022. Among these, 19 GRBs are localized by \textit{Swift}, with 16 being localized by \textit{Swift}/XRT. Furthermore, \textit{Fermi}/LAT localized 15 GRBs, and GRB~110721A was localized through IPN triangulation.

Since the goal of this work is to characterize the high-energy emission during the prompt emission phase, the temporal binning of the combined keV-GeV dataset is driven by the \textit{Fermi}/LAT data because of its limited sensitivity. We define time intervals according to their LAT likelihood ratio, i.e. test statistics (TS). We select time-bins where ${\rm TS}  \geq 10$, which correspond to a LAT detection significance $>3 \sigma$, as detailed in Section ~\ref{Sec_Data}. 12 out of the 35 GRBs in the initial sample exhibit a joint keV-GeV emission which enables us  to identify at least two time bins where both GBM and LAT data are available, according to our LAT selection criterion previously described. We refer to these GRBs as the \textit{Sample-1} and we report them in Table \ref{tab:Sample1}. 
Given the identification of multiple time-bins for each GRB in Sample-1, we carried out a temporal and (time-resolved) spectral analysis for a total of 67 spectra. The time-resolved GRBs, the respective T$_{90}$, and redshift (when available) are reported in Tab.~\ref{tab:Sample1}. 

For the remaining 23 GRBs, it was possible to identify only a single bin with joint GBM-LAT detection. We refer to these GRBs as Sample-2, which is reported in the Tab.~\ref{tab:Sample2}. In case of the GRBs in Sample-2, the presence of a single bin prevents a temporal analysis.

\section{Data extraction and analysis}\label{Sec_Data}

\subsection{Fermi-GBM data}
For each GRB in our sample, we retrieved the \textit{Fermi}/GBM (8 keV - 40 MeV) data from the Fermi GBM Burst catalog\footnote{\url{https://heasarc.gsfc.nasa.gov/W3Browse/Fermi/fermigbrst.html}} and performed a standard reduction using the Fermi science tool \textsc{GTBURST}\footnote{\url{https://fermi.gsfc.nasa.gov/ssc/data/analysis/scitools/gtburst.html}}. We considered data from the two sodium iodide (NaI, 8-900 keV) and one bismuth germanate (BGO, 0.3-40 MeV) detectors with the lowest viewing angles. The background analysis is performed by choosing custom background intervals that lead to a convergent polynomial fit. Source intervals are chosen according to the LAT time-bins, in order to perform a simultaneous \textit{Fermi}/GBM and \textit{Fermi}/LAT spectral analysis. The data reduction outputs are the source spectra, background spectra and the weighted response files that are later used for the spectral analys is through the \textsc{XSPEC} software\footnote{\url{https://heasarc.gsfc.nasa.gov/xanadu/xspec/}}  provided by the High Energy Astrophysics Science Archive Research Center (HEASARC). We ignore the energy channels outside 8-900 keV for the NaI detectors, together with the 30-40 keV interval to avoid the iodine K-edge line at 33.17 keV. For the BGO detectors, we ignored the channels outside the 300 keV - 40 MeV interval. For the spectral analysis of this dataset, we applied a Poisson-Gaussian statistics (\texttt{pgstat}).

\begin{table*}[ht!]
\caption{Properties of GRBs in Sample-1, including their redshift (z), duration ($T_{90}^{\rm GBM}$), availability of LLE data, number of analyzed bins, and the results from timing analysis: \textit{Fermi}/LAT peak flux ($F_{\rm peak}^{\rm LAT}$), peak time ($t_{\rm peak}^{\rm LAT}$), and temporal index of the LAT light curves.}\label{tab:Sample1}
\centering
\begin{tabular} {c r r c r c r c } \hline
\multirow{2}{*}{GRB name}& \multirow{2}{*}{z}&  \multirow{2}{*}{T$^{\rm GBM}_{90}$ [s]} & \multirow{2}{*}{LLE} & \multirow{2}{*}{N$_{\rm bin}$} & F$^{\rm LAT}_{\rm peak}$          & \multirow{2}{*}{t$^{\rm LAT}_{\rm peak}$ [s]}  & \multirow{2}{*}{$\gamma$} \\
                         &                           &   &                 &              & [erg cm$^{-2}$ s$^{-1}$] &   &  \\ \hline
\vspace{0.1cm}
GRB080916C &    4.35&62.9&\cmark&10& $1.48^{+0.27}_{-0.27} \times 10^{-6}$ & $6.61^{+0.76}_{-0.76}$ & $1.46^{+0.08}_{-0.07}$ \\
\vspace{0.1cm}
GRB090902B &    1.82& 21.9&\cmark&11& $4.57^{+0.63}_{-0.74} \times 10^{-6}$ & $9.33^{+1.07}_{-1.07}$ & $1.73^{+0.10}_{-0.09}$ \\
\vspace{0.1cm}
GRB090510 &     0.9 &0.96&\cmark&2& $1.66^{+0.46}_{-0.50} \times 10^{-5}$ & $0.81^{+0.09}_{-0.11}$ & $1.79^{+0.16}_{-0.14}$ \\
\vspace{0.1cm}
GRB090926A &   2.11&13.7&\cmark&10& $2.82^{+0.78}_{-0.65} \times 10^{-6}$ & $8.13^{+1.12}_{-0.94}$ & $1.56^{+0.23}_{-0.14}$ \\
\vspace{0.1cm}
GRB110731A &   2.83&7.5&\cmark&2& $1.48^{+0.61}_{-0.92} \times 10^{-6}$ & $6.17^{+2.70}_{-2.27}$ & $2.32^{+1.09}_{-0.70}$ \\
\vspace{0.1cm}
GRB130427A &   0.34&138.2&\cmark&3& $1.41^{+0.20}_{-0.20} \times 10^{-6}$ & $14.79^{+1.36}_{-1.36}$ & $1.37^{+0.05}_{-0.04}$ \\
\vspace{0.1cm}
GRB131108A &   2.40&18.2&\cmark&5& $1.45^{+0.33}_{-0.40} \times 10^{-6}$ & $2.95^{+0.82}_{-0.68}$ & $1.56^{+0.17}_{-0.12}$ \\
\vspace{0.1cm}
GRB160509A &  1.17 &369.6&\cmark&3& $9.33^{+2.15}_{-3.01} \times 10^{-7}$ & $23.44^{+3.24}_{-3.78}$ & $3.37^{+0.84}_{-0.75}$ \\
\vspace{0.1cm}
GRB160625B &   1.41&453.4&\cmark&6& $1.05^{+0.36}_{-0.46} \times 10^{-7}$ & $316.23^{+36.41}_{-50.97}$ & $2.62^{+1.20}_{-0.76}$ \\
\vspace{0.1cm}
GRB170214A &   2.52&122.9&\cmark&3& $1.62^{+0.45}_{-0.49} \times 10^{-7}$ & $87.10^{+12.03}_{-14.04}$ & $1.83^{+0.29}_{-0.24}$ \\
\vspace{0.1cm}
GRB190114C & 0.42 &116.4&\cmark&3& $1.48^{+0.54}_{-0.75} \times 10^{-5}$ & $6.17^{+0.99}_{-1.28}$ & $1.85^{+0.48}_{-0.41}$ \\
\vspace{0.1cm}
GRB221023A & \xmark &39.2& \xmark &10& $2.14^{+0.30}_{-0.30} \times 10^{-6}$ & $36.31^{+2.51}_{-3.34}$ & $2.15^{+0.15}_{-0.14}$ \\ \hline
\end{tabular}
\end{table*}

\begin{table*}[ht!]
\centering
\caption{Properties of GRBs in Sample-2, including the relative time bins analyzed, redshift (if available), $T_{90}^{\text{GBM}}$, availability of LLE data, and parameters derived from spectral analysis: spectral index ($p_{\rm index}$), synchrotron frequencies ($\nu_{\text{c}}$, $\nu_{\text{m}}$), bolometric flux, and the highest energy photon detected by \textit{Fermi}/LAT. GRBs marked with $*$ indicate the requirement for a high-energy cutoff.}\label{tab:Sample2}
\begin{tabular}{ l c c r c c c c c c}
\hline
\multirow{2}{*}{GRB name}  & Time-T$_{0}$ &  \multirow{2}{*}{z}    &\multirow{2}{*}{T$^{\rm GBM}_{90}$[s]}& \multirow{2}{*}{LLE} & \multirow{2}{*}{p$_{\rm {index}}$} & $\nu_{\text{c}}$        & $\nu_{\text{m}}$ & {Flux} ($\times10^{-6}$)  & E$_{\rm max}$ \\
          &     [s]      &           &        &         &  &   [keV]  &    [MeV]       &  [erg cm$^{-2}$ s$^{-1}$] &  [GeV] \\ \hline
          \vspace{0.1cm}
GRB 091031A & 0.0 - 6.4     &\xmark& 33.9 & \cmark & $3.08_{-0.23}^{+0.37}$& $117_{-26}^{+60}$ & $>0.2$ & $4.90_{-1.89}^{+0.08}$ & 0.3 \\
\vspace{0.1cm}
GRB 091003A & 0.0 - 21.1  &   0.8969   &   20.2        & \xmark & $4.13_{-0.39}^{+0.49}$ & $82_{-7}^{+7}$   & $0.61_{-0.12}^{+0.14}$ & $< 3.5$ & 2.7 \\

\vspace{0.1cm}
GRB 090323$^{*}$  & 0.0-130.0    &  3.57& 133.9        & \cmark &$3.22^{+0.04}_{-0.14}$ & $42^{-1.6}_{+1.8}$ & $1.18^{-0.22}_{+0.09}$ & $1.73_{-0.06}^{+0.04}$ & 0.4\\
\vspace{0.1cm}
GRB 100724A & 40.7 - 79.1&  1.288  &      114.7      & \cmark & $2.99_{-0.19}^{+0.25}$ & $37_{-5}^{+7}$   & $0.76_{-0.29}^{+0.47}$ & $1.02_{-0.79}^{+0.38}$ & 0.2 \\
\vspace{0.1cm}
GRB 110721A & 0.0 - 1.4 & \xmark &   21.8        & \cmark & $2.09_{-0.06}^{+0.08}$ & $27_{-3}^{+3}$      & $0.61_{-0.16}^{+0.25}$ & $3.44_{-2.97}^{+0.23}$ & 1.0 \\
\vspace{0.1cm}
GRB 120107A & 0.0 - 23.1 &  \xmark  &  23.0 & \xmark & $3.86_{-0.67}^{+0.75}$ & $24_{-5}^{+7}$     & $0.76_{-0.30}^{+0.51}$ & $4.48_{-3.92}^{+3.95}$ & 1.9 \\
\vspace{0.1cm}
GRB 120709A & 0.0 - 11.0 &  \xmark  &    27.3      &\xmark & $3.55_{-0.55}^{+0.81}$ & $45_{-12}^{+20}$   & $>0.9$ & $<14.1$ & 2.2 \\
\vspace{0.1cm}
GRB 130327B & 0.0 - 28.4 &  \xmark   &    31.2     &\xmark  & $4.48_{-0.53}^{+0.38}$ & $224_{-40}^{+196}$ & $0.24_{-0.04}^{+0.13}$ & $<9.75$ & 4.6 \\
\vspace{0.1cm}
GRB 140206B & 0.0 - 24.8 &  \xmark  &   146.7  &\xmark & $2.79_{-0.15}^{+0.26}$ & $14_{-1}^{+1}$     & $0.24_{-0.04}^{+0.04}$ & $1.35_{-1.05}^{+0.94}$ & 0.8\\
\vspace{0.1cm}
GRB 140523A & 0.0 - 15.0 &  \xmark  &   19.2&\xmark & $3.89_{-0.41}^{+0.58}$ & $37_{-6}^{+36}$    & $0.06_{-0.02}^{+0.04}$ & $<10$ & 2.4 \\
\vspace{0.1cm}
GRB 141028A & 0.0 - 18.0 &     2.33        & 31.5  & \cmark & $4.45_{-0.54}^{+0.38}$ & $29_{-8}^{+10}$  & $12.22_{-3.15}^{+2.73}$ & $7.94_{-0.79}^{+1.46}$ & 0.3 \\
\vspace{0.1cm}
GRB150523A &  0.0-34.9   &   \xmark  & 82.4  &\cmark& $3.43_{-0.98}^{+1.00}$ & $392_{-58}^{+123}$ & $>1.6$ & $1.41^{+0.04}_{-0.09}$ & 1.6 \\
\vspace{0.1cm}
GRB 160816A & 0.0 - 8.5 &   \xmark  & 11.1& \cmark & $4.25_{-0.41}^{+0.53}$ & $141_{-12}^{+49}$ & $0.15_{-0.01}^{+0.03}$ & $4.03_{-3.25}^{+0.71}$ & 1.1 \\
\vspace{0.1cm}
GRB 160905A & 0.0 - 37.4 &  \xmark   & 33.5 & \cmark  & $4.41_{-0.35}^{+0.33}$ & $>330$ & $>0.4$ & $4.03_{-3.25}^{+0.71}$ & 2.2 \\
\vspace{0.1cm}
GRB 170115B & 0.0 - 8.2 &   \xmark    &  44.3  & \cmark & $4.96_{-0.06}^{+0.03}$ & $1018_{-49}^{+238}$ & $1.07_{-0.07}^{+0.15}$ & $<6.21$ & 0.6 \\
\vspace{0.1cm}
GRB 171210A & 0.0 - 146.7 & \xmark &  143.1 & \xmark  & $4.52_{-0.34}^{+0.33}$ & $87_{-9}^{+68}$   & $0.10_{-0.01}^{+0.03}$ & $5.44_{-2.45}^{+1.77}$ & 0.7 \\
\vspace{0.1cm}
GRB 180703A & 0.0 - 22.3 &     0.6678        &  20.7 & \xmark  & $4.40_{-0.48}^{+0.40}$ & $97_{-9}^{+11}$    & $1.42_{-0.38}^{+0.50}$ & $1.59_{-1.20}^{+1.38}$ & 0.2 \\
\vspace{0.1cm}
GRB 181020A & 0.0 - 15.6&      2.938        & 15.1 & \xmark  & $3.77_{-0.31}^{+0.63}$ & $>200$ & $0.29_{-0.06}^{+0.22}$ & $3.00_{-2.36}^{+2.03}$ & 0.3 \\
\vspace{0.1cm}
GRB 190511A & 0.0 - 30.2 &  \xmark &  27.6 & \xmark  & $4.67_{-0.43}^{+0.24}$ & $110_{-13}^{+69}$  & $0.11_{-0.01}^{+0.04}$ & $7.58_{-7.57}^{+2.19}$ & 0.8 \\
\vspace{0.1cm}
GRB 210410A & 0.0 - 1.8 &   \xmark &  48.1 &\xmark  & $4.86_{-0.20}^{+0.10}$ & $850_{-83}^{+315}$  & $0.87_{-0.08}^{+0.19}$ & $3.76_{-3.76}^{+2.23}$ & 0.4 \\
\vspace{0.1cm}
GRB 210928A & 0.0 - 19.8&  \xmark  & 23.8  & \xmark  & $4.58_{-0.50}^{+0.30}$ & $178_{-27}^{+121}$ & $0.18_{-0.02}^{+0.08}$ & $4.03_{-3.25}^{+0.71}$ & 0.1 \\
\vspace{0.1cm}
GRB 211018A & 0.0 - 39.1 &  \xmark   & 123.9  &\xmark   & $3.58_{-0.86}^{+0.92}$ & $41_{-4}^{+6}$     & $0.48_{-0.16}^{+0.18}$ & $<4.74$ & 3.3 \\
\vspace{0.1cm}
GRB 220101A & 0.0 - 105.9 &    4.618       & 128.3 & \xmark   & $3.73_{-0.26}^{+0.41}$ & $58_{-6}^{+7}$    & $0.53_{-0.13}^{+0.17}$ & $8.25_{-3.04}^{+0.99}$ & 0.6 \\
\hline
\end{tabular}
\end{table*}

\subsection{Fermi-LAT data}
We used the \textsc{GTBURST} software from the official \textit{Fermi}-tools to extract and analyze the data from the GRBs listed in Table \ref{tab:Sample1} and \ref{tab:Sample2}/ Section \S\ref{sec:samples}. The high-energy data within the 0.1-10\,GeV energy band were extracted from a 12$^{\circ}$ region of interest (ROI) around the source location of the GRB. The source location (R.A., Dec.) and trigger time are obtained from the GBM online trigger catalog. A zenith angle cut of 100$^{\circ}$ was applied to minimize contamination from gamma-ray photons originating from the Earth's limb as a sanity measure.

The GRBs were selected ensuring a detection significance of more than 3 sigma in high-energy gamma-rays. The temporal bins for each GRB were chosen so that the total number of high-energy gamma-ray photons within a 5$^{\circ}$ radius around the GRB location is at least 15. Thus, ensuring a higher detection significance. 
We considered a spectral model of type "powerlaw2" for the analysis throughout along with "isotr\_template" and "template (fixed norm.)" for the particle background and the Galactic component, respectively. An "unbinned likelihood analysis" is considered with a minimum test statistic (TS$_{\rm min}$) of 10 which denotes the detection significance ($\sqrt{\rm TS}$) of more than 3. The selection of the instrument response function is made based on the duration. We used the response function "P8R3\_TRANSIENT010E\_V2" until 1000\,s from the trigger time following the analysis methods described in \citealt{2019ApJ...878...52A}.  The \textit{Fermi}/LAT data spectral products are produced through \textsc{GTBURST} using the standard ScienceTool\footnote{\url{ https://fermi.gsfc.nasa.gov/ssc/}} and {\tt gtbin} pipeline. In addition, we produce background counts and response files using {\tt gtbkg} and {\tt gtrspgen}, respectively (for details see \citealt{Ajello:2019zki}). A cut in the angle of the off-axis source of 90$^{\circ}$ has been considered to produce the spectral files in six energy bins between 100 MeV and 100 GeV. We fit the \textit{Fermi}/LAT spectrum on {\sc XSPEC} using Cash statistics. For each of the time bins, we further conduct the search for the highest-energy photon (E$_{\rm max}$, GeV) using ${gtsrcprob}$ tool which has probability of association with the GRB more than 0.9. We perform two independent analysis a) 0.1-100 GeV, and b) 100 MeV-E$_{\rm max}$.  
The spectral points presented in Figures 2, 4 and 5, are made following the strategy (a), whereas the spectral butter-fly plots are made with the analysis described in (b).

\subsection{Fermi-LLE data}

The GRBs listed in Samples-1 and -2 are detected by \textit{Fermi}/LAT during the prompt emission, thus leading to the possibility of detection in 30-100 MeV. We search for availability of the LLE data \cite{Pelassa:2010xp}. We downloaded data from the Fermi LAT Low-Energy Events Catalog\footnote{\url{https://heasarc.gsfc.nasa.gov/W3Browse/Fermi/fermille.html}}. LLE data are reduced using the same tools employed in the \textit{Fermi}/GBM analysis, selecting the source in the same time-bin to perform a simultaneous joint time-resolved analysis. In the spectral analysis, we excluded all channels outside the 30-100 MeV interval. For this dataset fit, we employed a Cash statistics (\texttt{cstat}). The column "LLE" in Tables \ref{tab:Sample1} and \ref{tab:Sample2} indicates the availability of the LLE data.

\section{Timing analysis}
To explore the origin of the HE emission observed by \textit{Fermi}/LAT, we empirically model the light curves in terms of the observed flux. 
In the standard forward shock model, excluding the effects of continuous energy injection, the bolometric light curve of the afterglow is expected to rise proportionally to t$^{\tau}$ until the deceleration time, after which it follows a power law decline t$^{-\gamma}$ \citep{Sari:1997qe}. To model the light curve empirically, we use the following function adopted from \cite{2010A&A...510L...7G}:
\begin{equation}
    {\rm F(t)} = \frac{{\rm A}_0 \left(\frac{\rm t}{{\rm t}_{\rm b}}\right)^{\tau}}{1+\left(\frac{{\rm t}}{{\rm t}_{\rm b}}\right)^{\tau+\gamma}}
    \label{eq:standard_aft}
\end{equation}
where ${\rm A}_{0}$ is the flux normalization, $t_b$ is the characteristic break time, and $\tau$ and $\gamma$ are the temporal rise and decay indices, respectively. In our calculation, we fix $\tau = 2$ and explore the posterior distribution of the model parameters. We derive the peak time as t$_{\rm p}$ = t$_{\rm b} \left(\frac{\tau}{\gamma}\right)^{\frac{1}{\tau+\gamma}}$. 

We built a likelihood based on the model function, comparing it with the observed \textit{Fermi}/LAT light curves. Logarithmic priors for the parameters A$_{0}$, t$_{\rm p}$, $\gamma$ are chosen in ranges $[-8.0,-4.0]$, $[-0.8,4.0]$, $[1.0,5.0 ]$, respectively. We perform a MCMC sampling using the Python package \textsc{emcee} \citep{Foreman-Mackey2013PASP..125..306F} to estimate the posterior distribution of the parameters.  
\begin{figure*} [ht!]
\centering
        \includegraphics[width=0.32\textwidth]{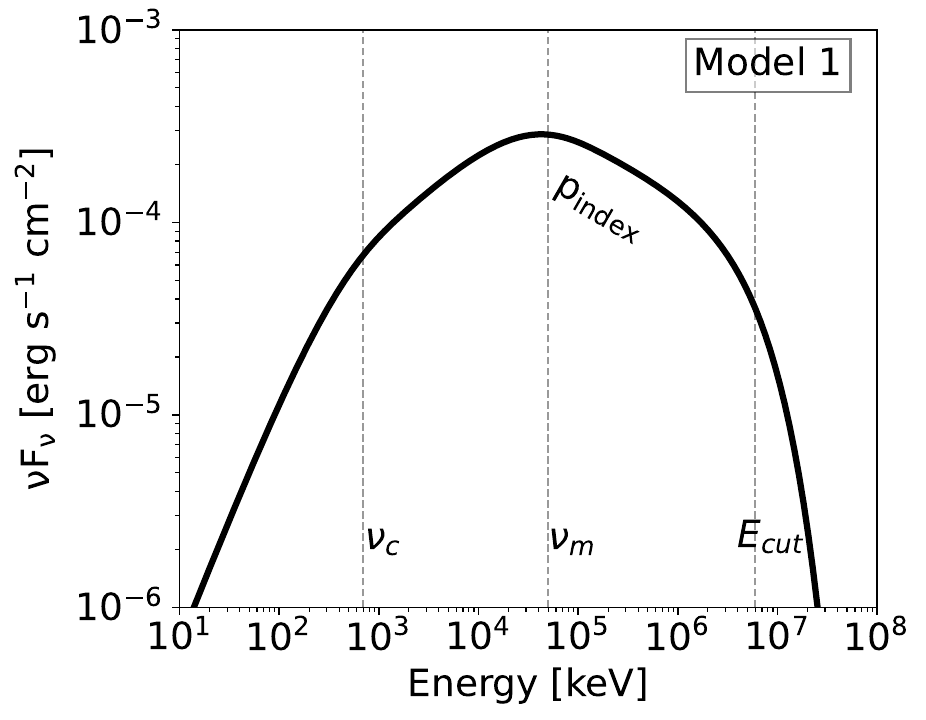}
        \includegraphics[width=0.32\textwidth]{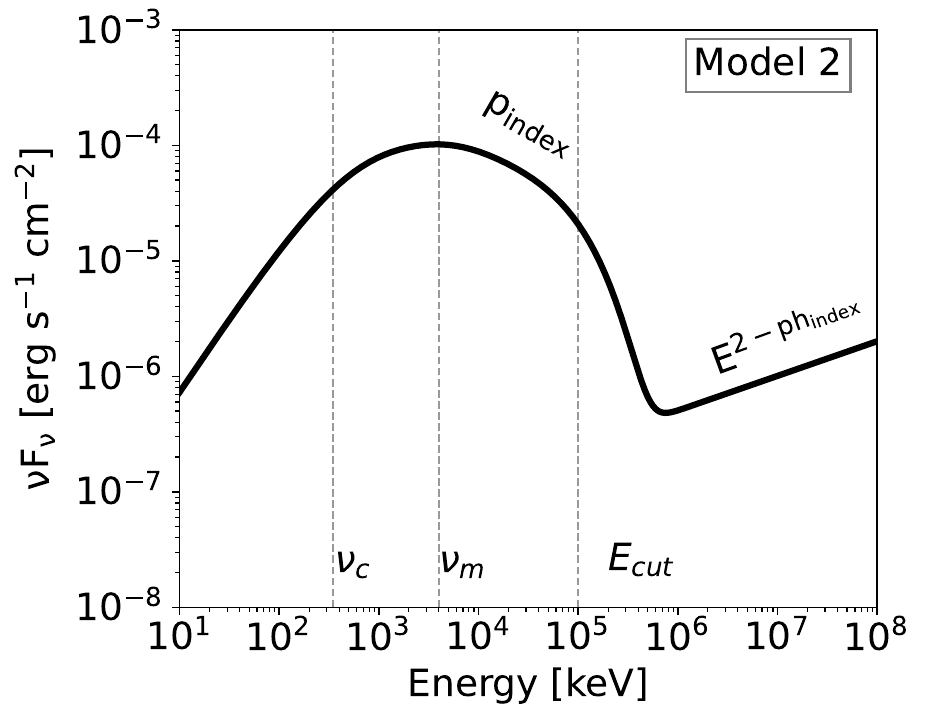}
        \includegraphics[width=0.31\textwidth]{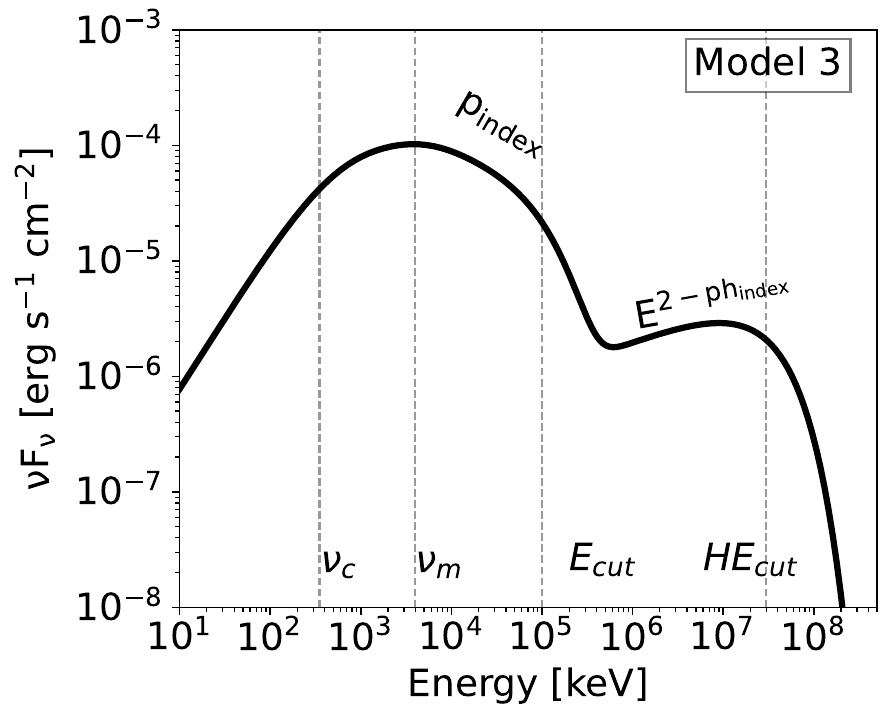}
        \caption{Model-1 considers only synchrotron emission. In case of Model-2, an additional power law is added. Model-3 is obtained adding to Model-2 a high energy cut off in the spectrum. The characteristic frequencies ($\nu_{c}$, $\nu_{m}$), and the cut-off energy (E$_{cut}$) are marked with the vertical lines. For detailed description see Section \S\ref{sec:model}.}\label{fig:model}
\end{figure*}

We define the best-fit values of the parameters as the median of their posterior distribution. The uncertainty for each parameter is expressed as the range between the 16th and 84th percentiles of the posterior distribution, representing the range within which the true value of each parameter is expected to lie with a probability of 68\%, given the model and the data. The best-fit values of the parameters and their uncertainties for each \textit{Fermi}/LAT light curve analyzed are reported in Table~\ref{tab:Sample1}. 
Some of the time bins selected through the method reported in Sect. \ref{sec:samples} are simultaneous with the relative prompt emission, as highlighted by \textit{Fermi}/GBM duration.
When fitting the afterglow light curve model to the LAT data, we find that some GRBs exhibit data points at early times in excess with respect to the best-fit afterglow light curve. The latter is mainly driven by the fit of late time data points, which are produced with a large temporal delay with respect to the GBM trigger, when the prompt emission component is largely sub-dominant. Therefore, the excess observed at early times might be caused by the superposition of both prompt and afterglow emissions. However, given the uncertainties of the ratio between the data and the model, it is difficult to distinguish the prompt and/or the afterglow contribution in the early GeV data. Hence, we performed a more detailed spectral analysis, described in the following section.

\section{Spectral analysis}\label{sec:model}

A temporal analysis alone is not sufficient to determine the origin of the early GeV emission.
In this study, we performed a detailed spectral analysis of GRB prompt emission, adopting a physical model based on the assumption that prompt emission is produced by synchrotron radiation. For the synchrotron part, we used a table model as described in \citealt{Oganesyan:2019fpa}. 
Synchrotron emission arises from non-thermal electrons distributed according to a power law \( \frac{dN_e}{d\gamma} \propto \gamma^{-p} \), where \(\gamma_m\) denotes the minimum Lorentz factor. The resultant spectrum is generated by combining the single-electron emission spectrum with the electron distribution after they have cooled via synchrotron radiation down to the cooling Lorentz factor \(\gamma_c\). 
The table spectra are constructed by exploring both the fast and slow cooling regime of synchrotron radiation. Importantly, the model assumes a fixed cooling frequency \(\nu_c = 1 \, \text{keV}\). 
Therefore, we apply the XSPEC convolution model \texttt{zmshift}, which adjusts the spectrum along the energy axis. 
The resulting parameters of the model are: p$_{\rm index}$, log($\gamma_{\rm m}/\gamma_{\rm c}$), \textit{zmshift} and the normalization \(N_\text{Sync}\). Spectra that reach GeV energies require an empirical high-energy cutoff to consider the potential spectral attenuation attributed to pair production \citep{Vianello:2017yfd, Mei:2022ics, Ravasio:2023vgh}. 
The energy resolution and sensitivity of LAT prevent an accurate characterization of the GeV spectrum. The addition of the HE cutoff allows for a proper description of the putative pair-attenuation, without strong assumptions on the spectral shape.
This table model with the additional high energy cutoff is named Model-1, hereafter. To account for a potential additional high-energy component, we test models with both a power law and a cut-off power law. For these cases, the added parameters include the power law photon index (ph$_{\rm index}$) and normalization. We term this as Model-2. For rare occasions, the data also require a high-energy cutoff. This third category is termed Model-3. 
The models and the relevant quantities are presented in Fig.~\ref{fig:model}. 
For comparison, we fit all GRBs in our sample using the Band model in the standard MeV range. This allows us to evaluate how the results differ when using a physical model instead of the empirical approach applied to most GRBs in our sample.

\subsection{Implementation of the models and spectral analysis}
We perform a joint spectral analysis using \textit{Fermi}/GBM, \textit{Fermi}/LLE, and \textit{Fermi}/LAT data, applying the three physical models described in Section \S\ref{sec:model} and shown in Fig.~\ref{fig:model}. The fitting process is carried out using \textit{XSPEC (v12.13.0c)}. In Model-1 (table synchrotron model), we combined an additional feature, the multiplicative component \textit{highecut}, that includes two parameters: $E_{b}$, fixed at 300 keV, and the fold energy $foldE$, which remains a free parameter. For Model-2 (synchrotron and power law combined) and Model-3 (synchrotron and cut-off power law components), in addition to Model-1, we add the built-in XSPEC models \textit{powerlaw} and \textit{cutoffpl}, respectively. These models introduce additional parameters: normalization, photon index (ph$_{\rm index}$), and, for the cut-off model, the folding energy (${foldE}$). In case of the conventional Band model, we use the \textit{grbm} model, with parameters comprising the two photon indices $\alpha$ and $\beta$, the characteristic energy $E_{c}$, and normalization.

\subsection{Model Comparison: The Akaike information criterion}

To select the best fit model among Model 1 and Model 2 for the GRBs in Sample 1, we use the Akaike Information Criterion  \citep[AIC;][]{Akaike:1974vps}. The AIC helps in evaluating and comparing different models. Specifically, AIC takes into account the number of degrees of freedom and the statistic values for each model fit. 
In practice, we evaluated the quantity 
\begin{equation}
    AIC = 2k + S
    \label{AIC}
\end{equation} 
for each model fitted to each spectrum, where $k$ is the number of model parameters and $S$ is the statistic value, which in this case consists of a combination of \texttt{pgstat} and \texttt{cstat}. We then compute the difference $\Delta AIC = AIC_{2} - AIC_{1}$, where $AIC_{2}$ is the more complex model, i.e. with more model parameters $k$.
We select the more complex model as the best-fit model if $\Delta AIC \geq 4$, which corresponds to a statistical improvement of $1\sigma$ \citep{burnham2004multimodel}.
We do not perform model comparison between the Band and synchrotron model. Band, being a phenomenological model, is able to easily accommodate the majority of the GRB spectra, without any insight on the physical processes behind. On the other hand, the synchrotron model allows to investigate the physical framework (with the same number of free parameters), but with less freedom in describing the spectral data.
Instead, our goal is to study whether the synchrotron emission can model the prompt spectrum up to HE with o without the addition of extra components.

\subsection{Parameter estimation: BXA}
Once the best-fit model is determined with the AIC test, we use the Bayesian X-ray analysis tool \textsc{BXA}\footnote{\url{https://johannesbuchner.github.io/BXA/}} together with PyXSPEC for precise parameter estimation. BXA is particularly advantageous for parameter estimation due to its integration of Bayesian inference with the nested sampling algorithm (like UltraNest), enabling thorough exploration of parameter space and delivering reliable estimates and uncertainties for model parameters. First, we implement the best-fit model, as identified by the AIC test, in PyXSPEC. We then define non-informative priors for the free parameters of the model.  Specifically, we apply logarithmic priors for the normalization of the synchrotron and power law components (where needed) and for the fold energy of the cutoff component, while using uniform priors for the $p$-index (2, 5), $log(\gamma_{m}/\gamma_{c})$ (-1, 2), zmshift (-0.999, 10), and the photon index of the power law $ph_{index}$. From the samples of the parameters $log(\gamma_{m}/\gamma_{c})$, \textit{zmshift} and the fold energy, we obtain "secondary" samples for the physical quantities of interest $\nu_{m}$, $\nu_{c}$, and E$_{cut}$ in the following way:

\begin{equation}
    \nu_{c} = \frac{1}{1 - zmshift}
    \label{nu_c}
\end{equation}

\begin{equation}
    \nu_{m} = \nu_{c}\left(\log\frac{\gamma_{m}}{\gamma_{c}}\right)^{2}
    \label{nu_m}
\end{equation}

\begin{equation}
    {\rm E}_{\rm cutoff} = 10^{\rm log(foldE)}
    \label{Ecut}
\end{equation}

We define the median of the posterior distribution as the mean value of each parameter and determine credible intervals as the range from the 16th to the 84th quantile. The values of the flux and the relative errors are computed directly from PyXspec. The resulting values of all the parameters for each spectra analyzed are reported in Table~\ref{tab:GRB080916} for GRB 080916C, and in the Appendix for the other GRBs.

\section{Results}

\begin{table*}[ht]
\centering
\caption{Synchrotron parameters for individual time bins of GRB 080916C, along with the bolometric flux and the highest energy photon detected by \textit{Fermi}/LAT.}\label{tab:GRB080916}
\begin{tabular}{c c c c c c c }
\hline
Time bin & \multirow{2}{*}{p$_{\rm index}$} & $\nu_{\text{c}}$    & $\nu_{\rm m}$  & E$_{\rm cutoff}$ & F$_{\rm Bol}$ ($\times$ 10$^{-6}$) & E$_{\text{max}}$ \\
(Time-T$_{0}$)[s] &        & [keV] &                        [MeV] &    [GeV] &   [erg cm$^{-2}$ s$^{-1}$] & [GeV] \\
\hline \\
\vspace{0.1cm}
0.0 - 5.27   & $2.94_{-0.15}^{+0.18}$ & $276_{-50}^{+258}$ &  0.49$_{-0.21}^{+0.46}$ & $>0.2$ & $8.32_{-0.08}^{+0.01}$ & 0.30 \vspace{0.1cm}\\ 
5.27 - 5.95   & $2.83_{-0.56}^{+0.84}$ & $38_{-7}^{+7}$ & $16.23_{-7.23}^{+10.0}$ & $>0.3$ & $15.2_{-0.03}^{+0.03}$ & 0.58 \vspace{0.1cm}\\
5.95 - 6.63   & $2.32_{-0.21}^{+0.34}$ & $19_{-5}^{+5}$ & $5.91_{-2.29}^{+4.21}$ & $>0.6$ & $14.8_{-0.02}^{+0.02}$ & 1.69 \vspace{0.1cm}\\
6.63 - 8.81   & $2.65_{-0.30}^{+0.38}$ & $49_{-7}^{+8}$ & $2.77_{-1.21}^{+1.96}$ & $>0.3$ & $6.25_{-0.02}^{+0.02}$ & 2.11 \vspace{0.1cm}\\
8.81 - 12.79  & $2.57_{-0.27}^{+0.28}$ & $47_{-6}^{+8}$ & $2.50_{-1.21}^{+1.66}$ & $>0.5$ & $4.62_{-0.06}^{+0.06}$ & 1.11 \vspace{0.1cm}\\
12.79 - 17.80 & $2.84_{-0.27}^{+0.31}$ & $58_{-7}^{+9}$ & $2.66_{-1.09}^{+1.38}$ & $>0.5$ & $4.35_{-0.04}^{+0.04}$ & 12.42 \vspace{0.1cm}\\
17.80 - 22.38 & $2.39_{-0.10}^{+0.13}$ & $100_{-22}^{+78}$ & $>0.11$ & $>0.6$ & $3.23_{-0.03}^{+0.03}$ & 2.57 \vspace{0.1cm}\\
22.38 - 27.02 & $2.79_{-0.29}^{+0.35}$ & $61_{-8}^{+13}$ & $1.17_{-0.59}^{+0.94}$ & $>0.2$ & $4.00_{-0.01}^{+0.01}$ & 2.50 \vspace{0.1cm}\\
27.02 - 36.75 & $2.97_{-0.23}^{+0.30}$ & $54_{-6}^{+8}$ & $1.33_{-0.49}^{+0.68}$ & $>0.4$ & $2.65_{-0.03}^{+0.03}$ & 6.72 \vspace{0.1cm}\\
36.75 - 45.79 & $2.91_{-0.25}^{+0.36}$ & $98_{-17}^{+35}$ & $0.95_{-0.61}^{+0.88}$ & $>0.5$ & $2.19_{-0.02}^{+0.02}$ & 27.43 \vspace{0.1cm}\\
\hline
\end{tabular}
\end{table*}

\begin{table*}[ht!]
\centering
\caption{Synchrotron parameters for each analyzed time bin of GRB 090902B, along with the bolometric flux and the highest energy photon detected by \textit{Fermi}/LAT.}\label{tab:090902B}
\begin{tabular}{c c c c c}
\hline
{Time bin} & E$_{\text{cutoff}}$  & \multirow{2}{*}{ph$_{\text{index}}$} & {Flux} ($\times 10^{-6}$) & E$_{\rm max}$ \\
(t-T$^{\rm GBM}_{0}$)[s]     & [MeV]                &     & [erg cm$^{-2}$ s$^{-1}$] & [GeV] \\
\hline \\
\vspace{0.1cm}
0.0 - 7.02 &  $0.50_{-0.01}^{+0.02}$ & $1.46_{-0.08}^{+0.07}$ & $9.57_{-0.46}^{+0.01}$ & 1.28 \vspace{0.1cm}\\
7.02 - 8.31 &  $1.01_{-0.03}^{+0.04}$ & $1.91_{-0.02}^{+0.02}$ & $3.41_{-0.08}^{+0.96}$ & 1.0 \vspace{0.1cm}\\
8.31 - 8.89 &  $1.01_{-0.05}^{+0.06}$ & $1.98_{-0.03}^{+0.03}$ & $3.95_{-0.61}^{+2.08}$ & 1.78 \vspace{0.1cm}\\
8.89 - 9.47  & $1.01_{-0.05}^{+0.06}$ & $1.98_{-0.02}^{+0.03}$ & $3.96_{-0.72}^{+3.39}$ & 3.10 \vspace{0.1cm}\\
9.47 - 10.33 & $0.61_{-0.02}^{+0.03}$ & $2.03_{-0.02}^{+0.03}$ & $3.14_{-0.57}^{+1.59}$ & 7.72 \vspace{0.1cm}\\
10.33 - 11.03 & $0.92_{-0.04}^{+0.05}$ & $2.01_{-0.03}^{+0.03}$ & $3.61_{-2.59}^{+0.23}$ & 2.10 \vspace{0.1cm}\\
11.03 - 13.18 & $0.80_{-0.11}^{+0.18}$ & $1.97_{-0.02}^{+0.02}$ & $1.72_{-1.07}^{+0.17}$ & 11.90 \vspace{0.1cm}\\
13.18 - 14.84 & $1.48_{-0.61}^{+1.43}$ & $1.85_{-0.03}^{+0.03}$ & $1.93_{-0.36}^{+1.33}$ & 14.20 \vspace{0.1cm}\\
14.84 - 15.93 & $0.64_{-0.05}^{+0.07}$ & $1.81_{-0.02}^{+0.02}$ & $3.09_{-1.86}^{+0.90}$ & 5.19 \vspace{0.1cm}\\
15.93 - 17.73 & $0.64_{-0.08}^{+0.11}$ & $1.85_{-0.03}^{+0.03}$ & $1.76_{-1.17}^{+0.60}$ & 2.53 \vspace{0.1cm}\\
17.73 - 21.78 & $>0.1$ & $>0.17$ & - & 5.36 \vspace{0.1cm}\\
\hline
\end{tabular}
\end{table*}

\begin{table*}[ht!]
\centering
\caption{Synchrotron parameters for each analyzed time bin of GRB 221023A, along with the bolometric flux and the highest energy photon detected by \textit{Fermi}/LAT.}\label{tab:221023A}
\begin{tabular}{c c c c c c c c }
\hline
Time bin & \multirow{2}{*}{p$_{\rm index}$} & $\nu_{\text{c}}$    & $\nu_{\rm m}$  & E$_{\rm cutoff}$ & \multirow{2}{*}{ph$_{\text{index}}$} & F$_{\rm Bol}$ ($\times$ 10$^{-6}$) & E$_{\text{max}}$ \\
(Time-T$_{0}$)[s] &        & [keV] &                        [MeV] &    [GeV] &       & [erg cm$^{-2}$ s$^{-1}$] & [GeV] \\
\hline \\
\vspace{0.15cm}
0.00 - 17.72 & $4.12_{-0.27}^{+0.27}$ & $115_{-5}^{+5}$ & $1694_{-164}^{+170}$ & $>0.2$ & - & $1.42_{-0.38}^{+0.25}$ & 5.64 \vspace{0.1cm}\\
17.72 - 21.37 & $3.49_{-0.27}^{+0.26}$ & $266_{-19}^{+25}$ & $1261_{-233}^{+222}$ & $>0.15$ & - & $3.76_{-1.31}^{+1.28}$ & 2.10 \vspace{0.1cm}\\
21.37 - 23.11 & $3.20_{-0.23}^{+0.24}$ & $271_{-33}^{+70}$ & $1008_{-457}^{+305}$ & $>0.25$ & - & $3.60_{-1.21}^{+2.11}$ & 4.05 \vspace{0.1cm}\\
23.11 - 24.41 & $3.20_{-0.15}^{+0.18}$ & $517_{-87}^{+355}$ & $657_{-157}^{+408}$ & $>0.37$ & - & $4.99_{-2.85}^{+3.02}$ & 0.96 \vspace{0.1cm}\\
24.41 - 28.45 & $3.92_{-0.39}^{+0.58}$ & $196_{-16}^{+18}$ & $1169_{-219}^{+272}$ & $>0.13$ & - & $3.10_{-4.67}^{+1.48}$ & 7.87 \vspace{0.1cm}\\
28.45 - 31.35 & $3.62_{-1.08}^{+0.95}$ & $86_{-10}^{+16}$ & - & $1.82_{-0.31}^{+1.58}$ (MeV) & $1.93_{-0.03}^{+0.04}$ & $1.16_{-0.10}^{+0.92}$ & 1.18 \vspace{0.1cm}\\
31.35 - 34.26 & $3.44_{-0.98}^{+1.03}$ & $38_{-9}^{+11}$ & $840_{-160}^{+313}$ & $1.90_{-0.05}^{+0.04}$ (MeV) & $1.90_{-0.05}^{+0.04}$ & $6.89_{-4.99}^{+1.50}$ & 6.67 \vspace{0.1cm}\\
34.26 - 37.65 & $2.22_{-0.14}^{+0.16}$ &  $382_{-77}^{+102}$ & - & $6.41_{-3.27}^{+2.58}$ & - & $4.33_{-1.22}^{+0.07}$ & 2.24 \vspace{0.1cm}\\
37.65 - 41.38 & $2.19_{-0.11}^{+0.12}$ & $14_{-3}^{+4}$ & $237_{-73}^{+101}$ & $>0.5$ & - & $3.91_{-0.15}^{+0.18}$ & 9.63 \vspace{0.1cm}\\
41.38 - 44.24 & $2.09_{-0.06}^{+0.10}$ & - & $161_{-30}^{+41}$ & $>0.8$ & - &  $2.34_{-0.05}^{+0.47}$ & 2.19 \vspace{0.1cm}\\
\hline
\end{tabular}
\end{table*}

\begin{figure*}[ht!]
\caption{The time resolved spectral analysis of GRB~080916C. The upper-panel of the top left plot shows the fit of LAT light curve in 0.1-10 GeV with the model described in Eq. \ref{eq:standard_aft} along with the GBM light curves (BGO and  NaI). A comparison between the inferred spectral index ($\phi$) and the observed spectral index is given in the middle panel. The bottom panel shows the degree of deviation of the light curve from the model prediction. The rest of the panels in the plot depict the spectral fit in a time resolved manner. The details of the time resolved spectral fit is given in Table \ref{tab:GRB080916}.}\label{fig:080916}
    \includegraphics[width=0.65\textwidth, height = 8cm]{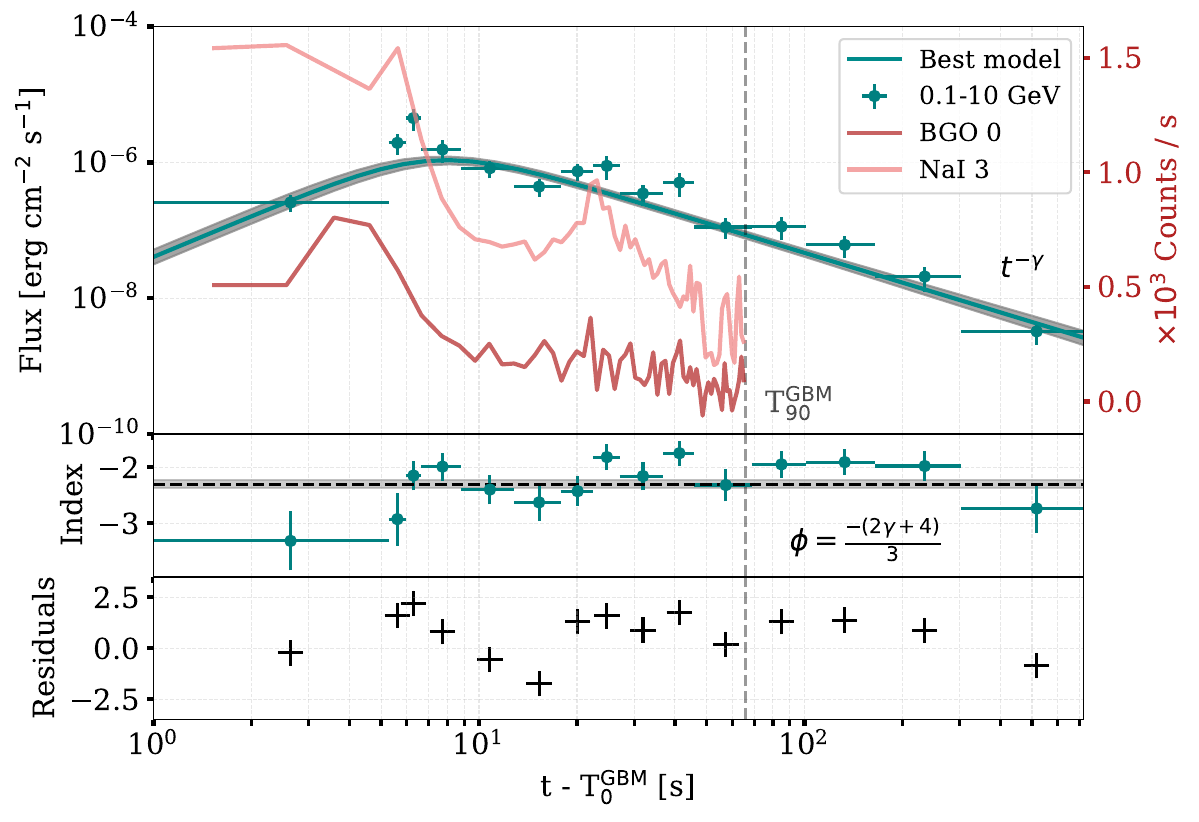}
    \includegraphics[width=0.33\textwidth]{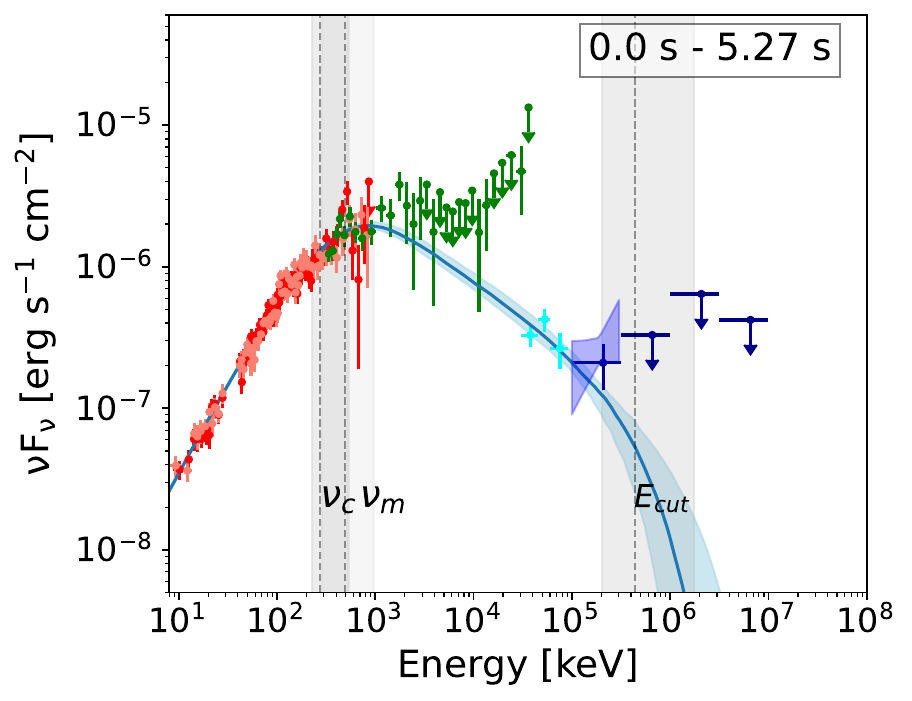}
    \includegraphics[width=0.33\textwidth]{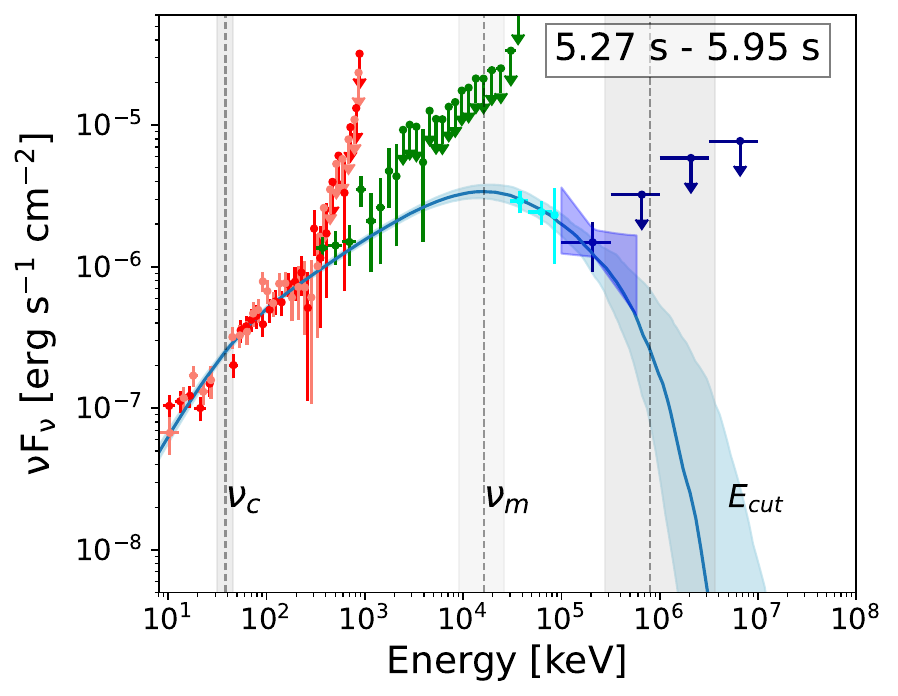}
    \includegraphics[width=0.33\textwidth]{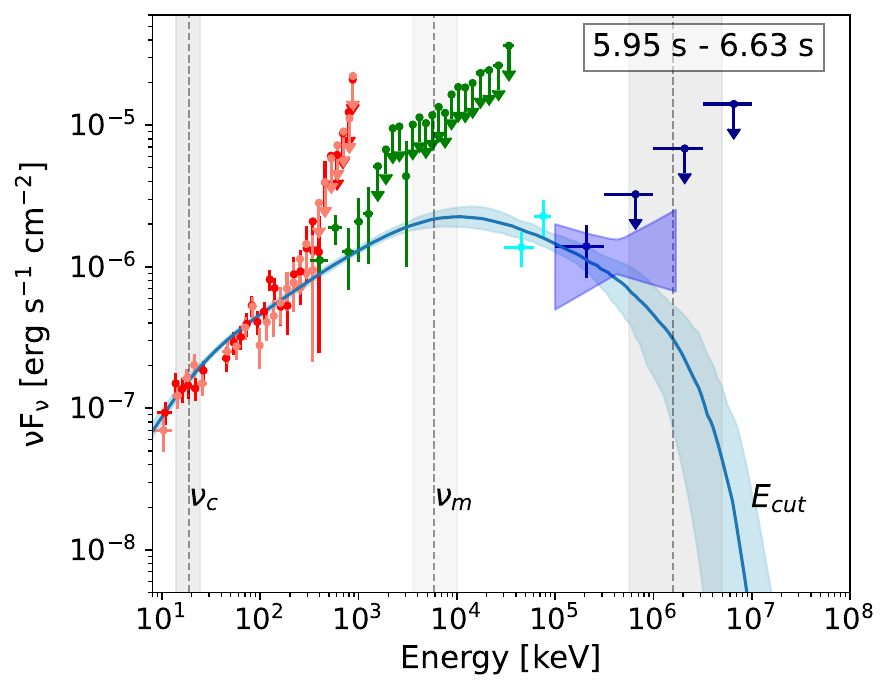}
    \includegraphics[width=0.33\textwidth]{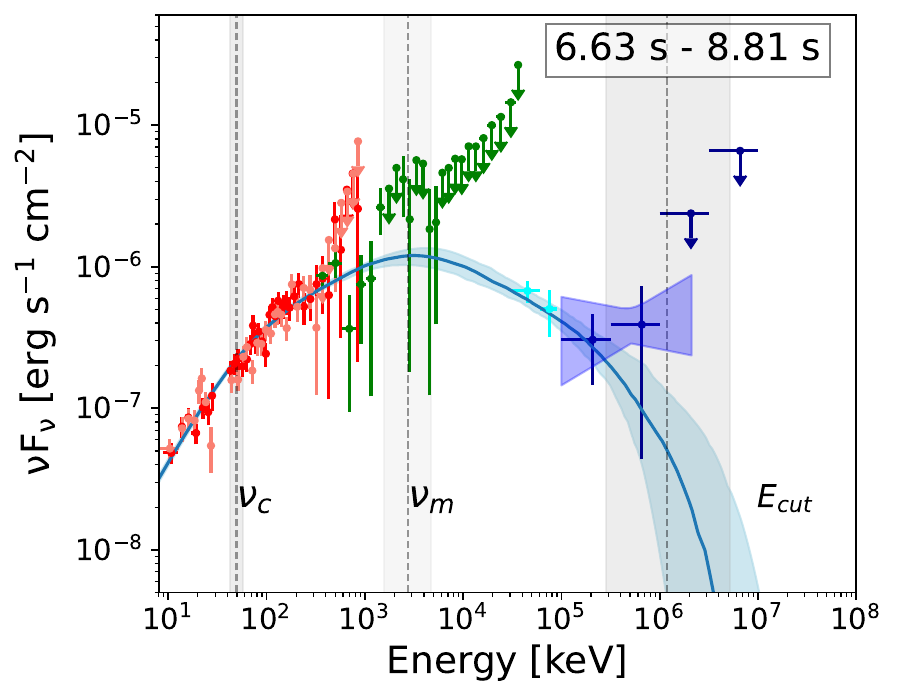}
    \includegraphics[width=0.33\linewidth]{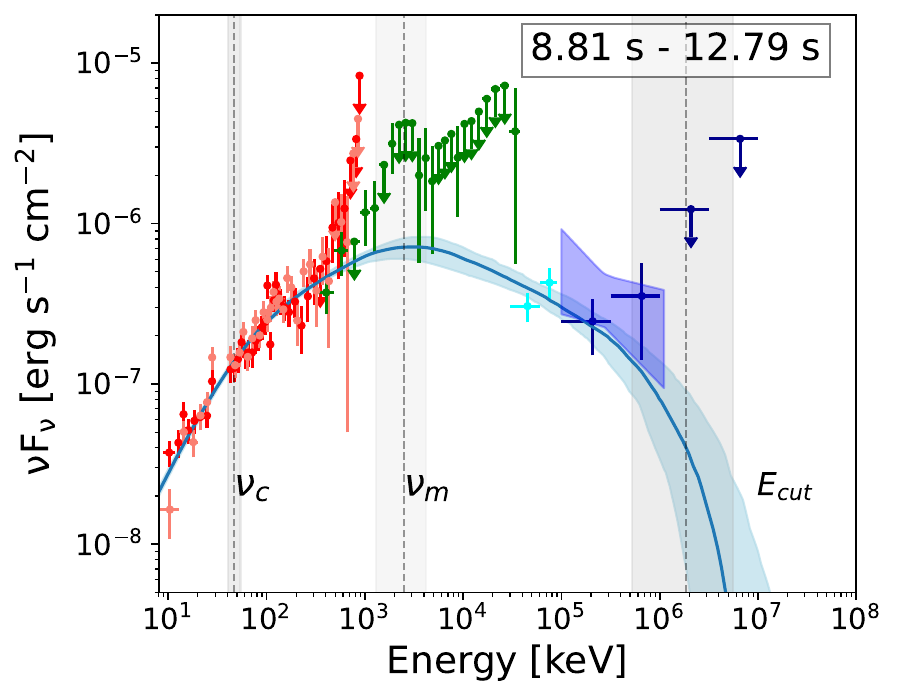}
    \includegraphics[width=0.33\linewidth]{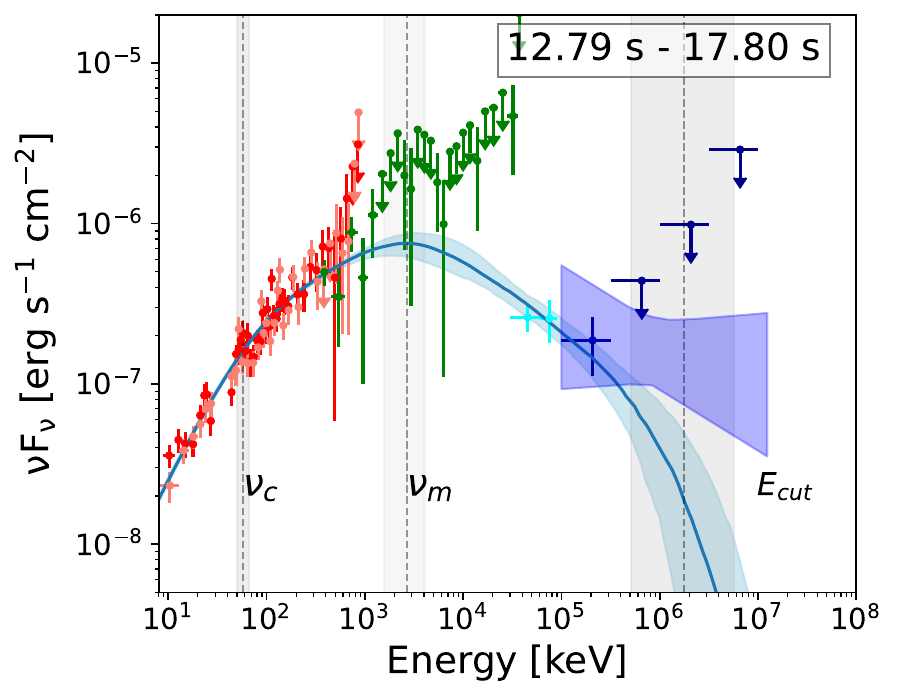}
    \includegraphics[width=0.33\linewidth]{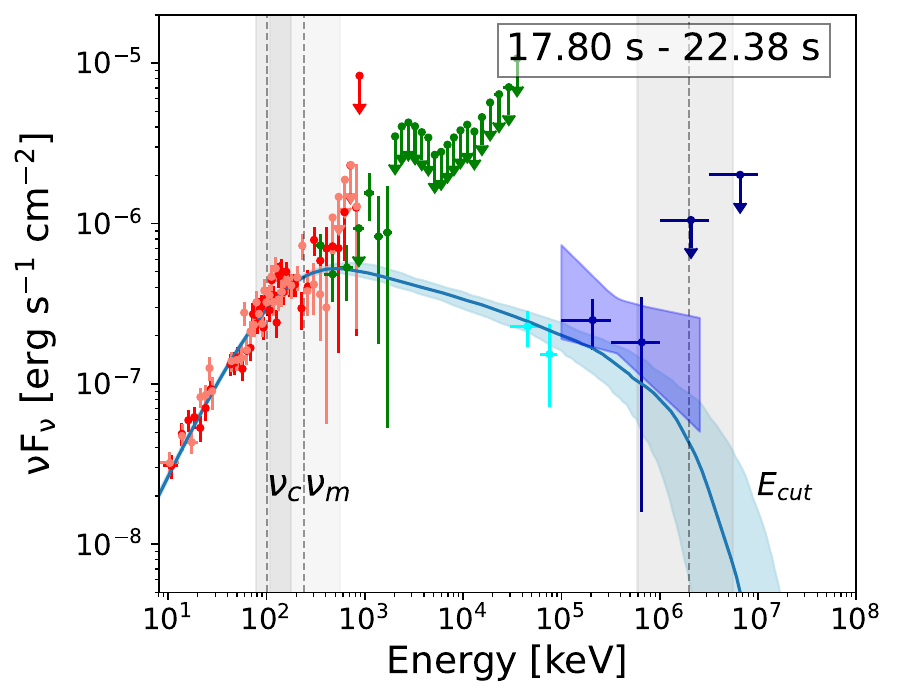}
    \includegraphics[width=0.33\linewidth]{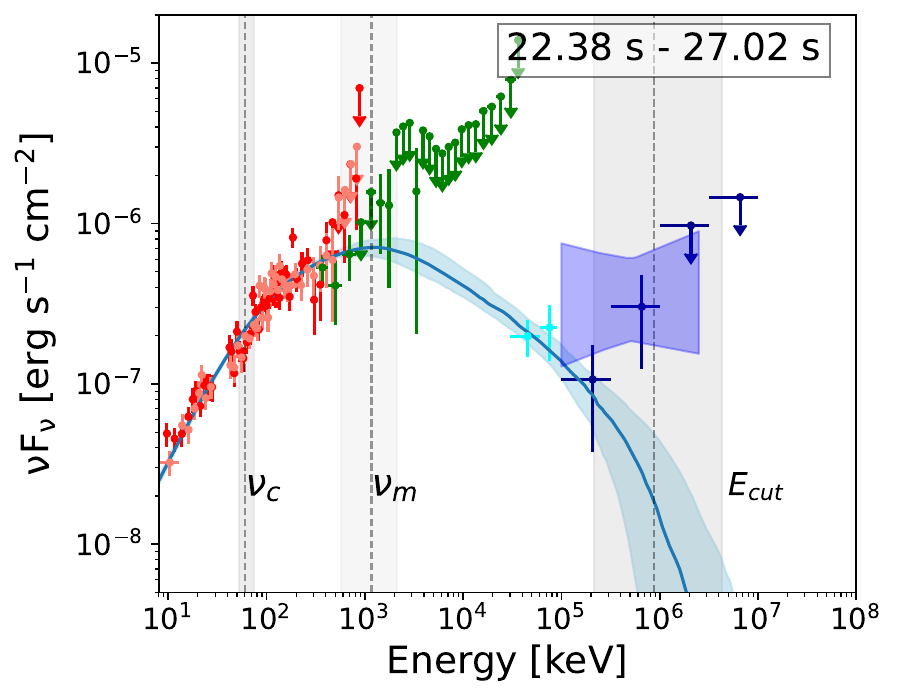}
    \includegraphics[width=0.33\linewidth]{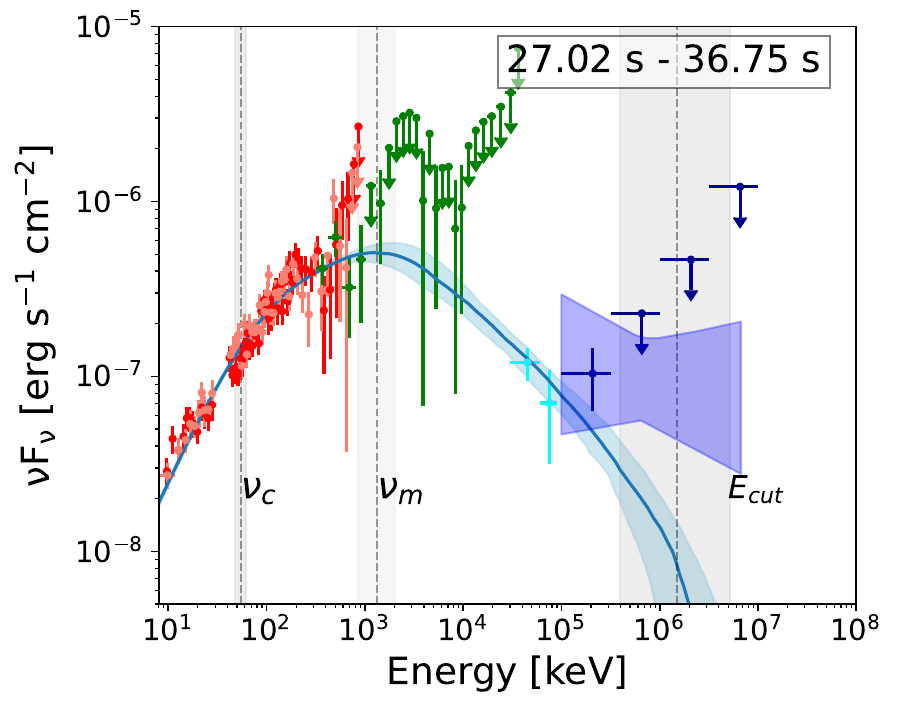}
    \includegraphics[width=0.33\linewidth]{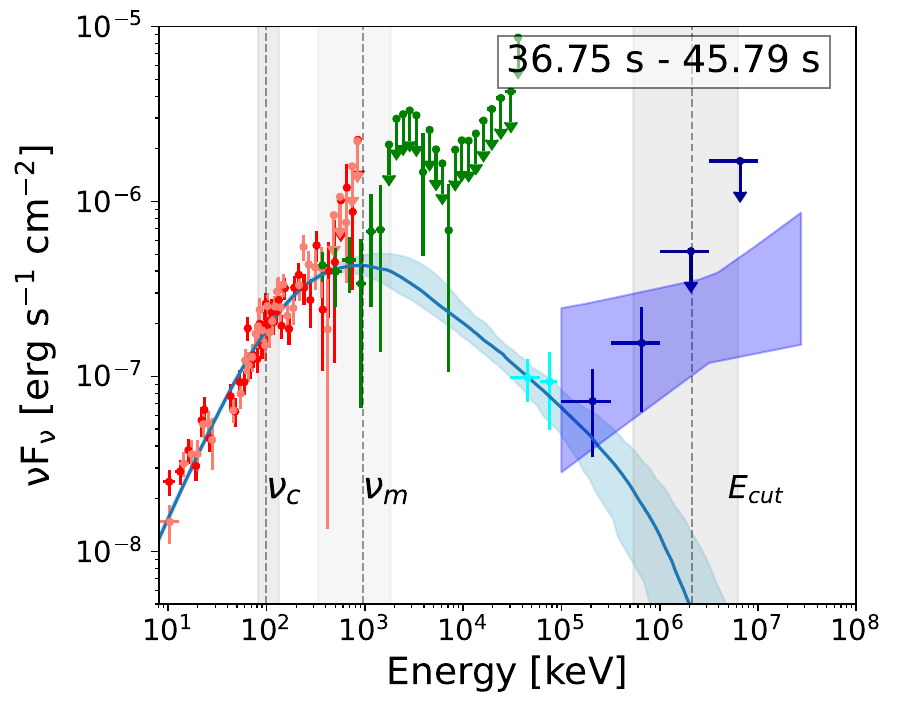}   
\end{figure*}

\begin{figure*}[ht!]
    \centering    
    \caption{Same as Fig. \ref{fig:080916} but for GRB 090902B. See Table \ref{tab:090902B} for the spectral fit parameters.}
    \label{fig:090902B}
    \includegraphics[width=0.65\linewidth, height=8cm]{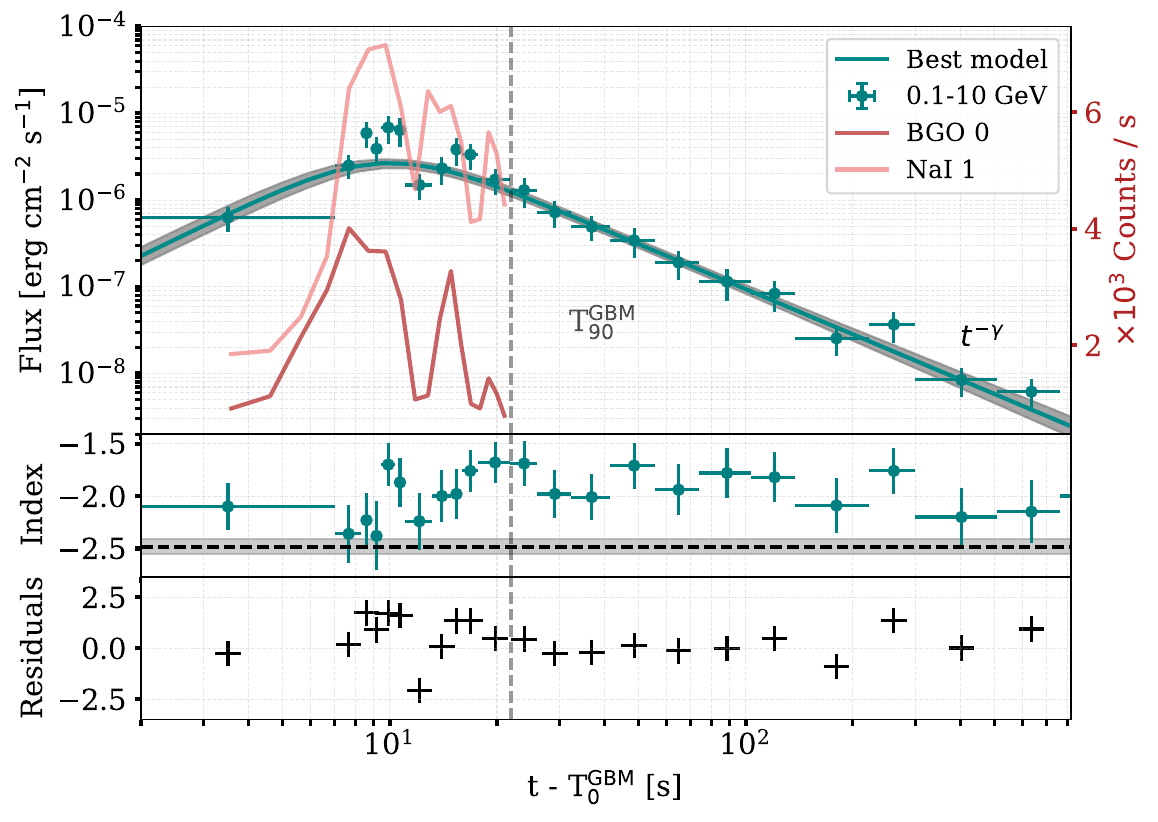}
    \includegraphics[width=0.33\linewidth]{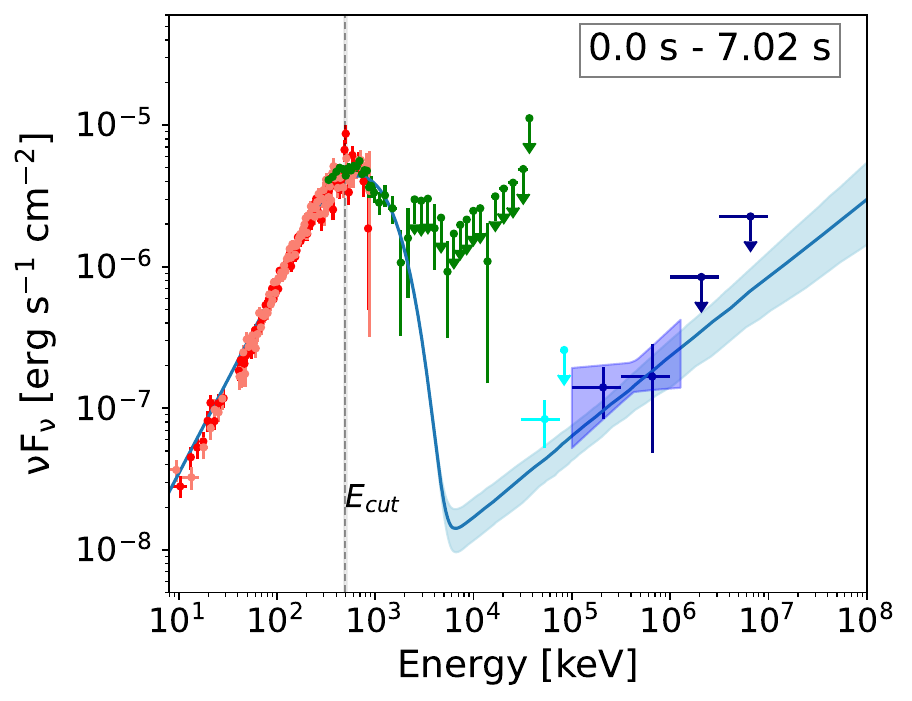}
    \includegraphics[width=0.33\linewidth]{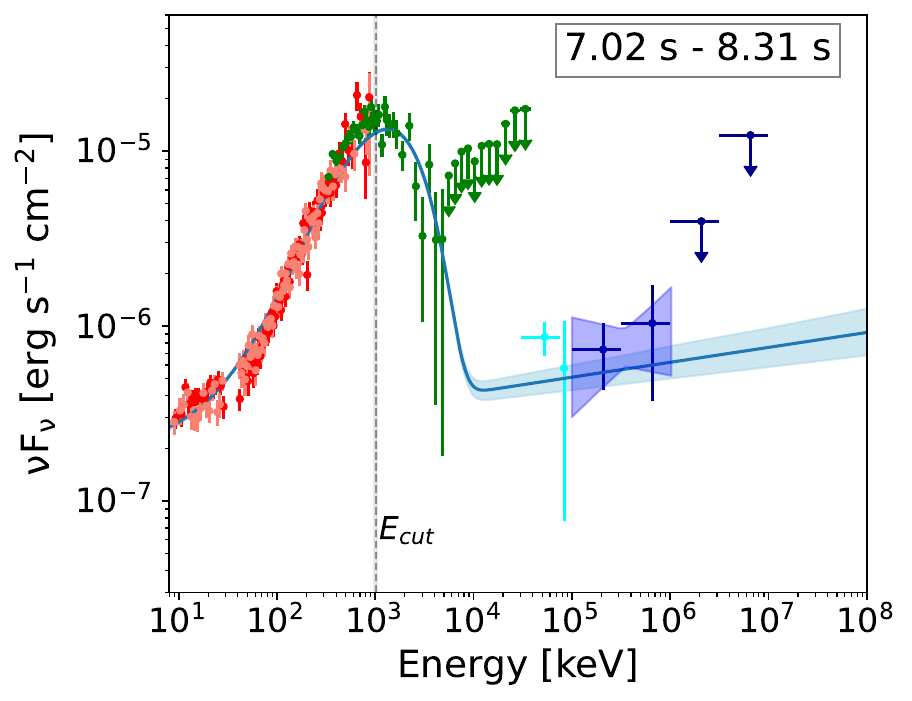}
    \includegraphics[width=0.33\linewidth]{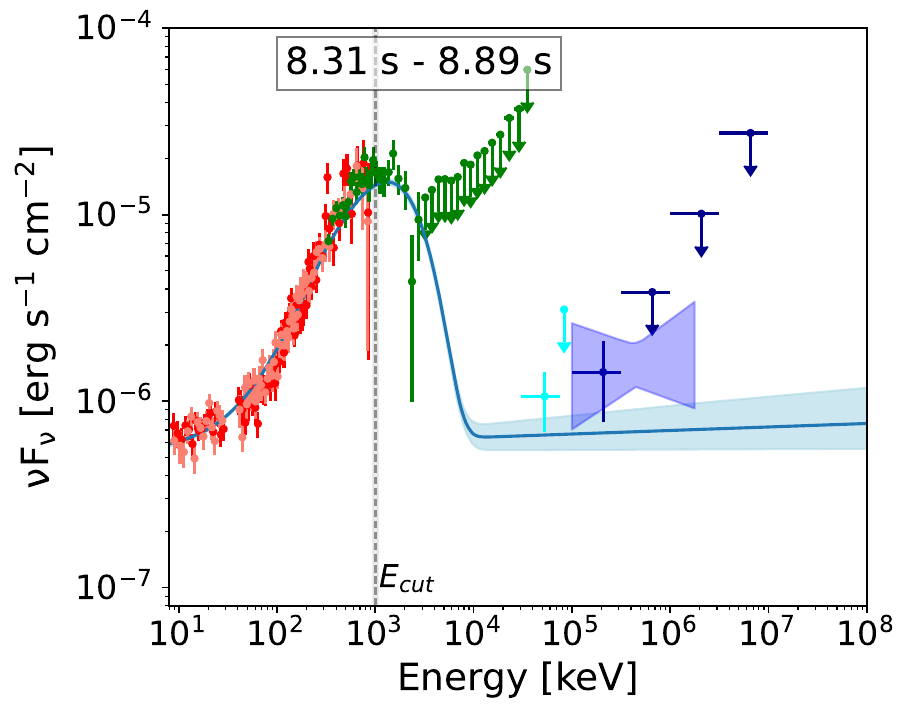}
    \includegraphics[width=0.33\linewidth]{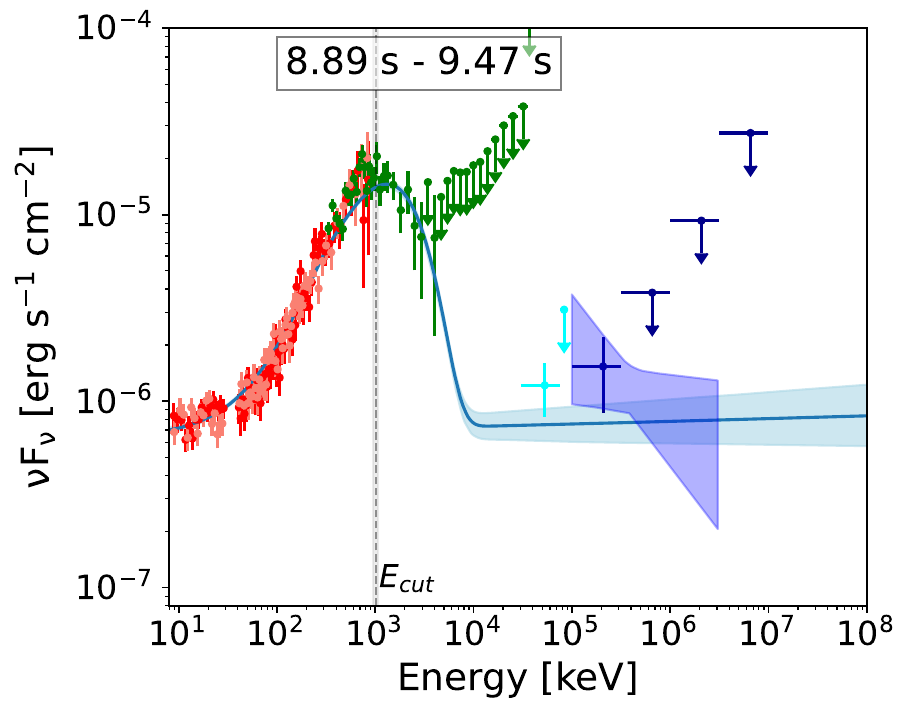}
    \includegraphics[width=0.33\linewidth]{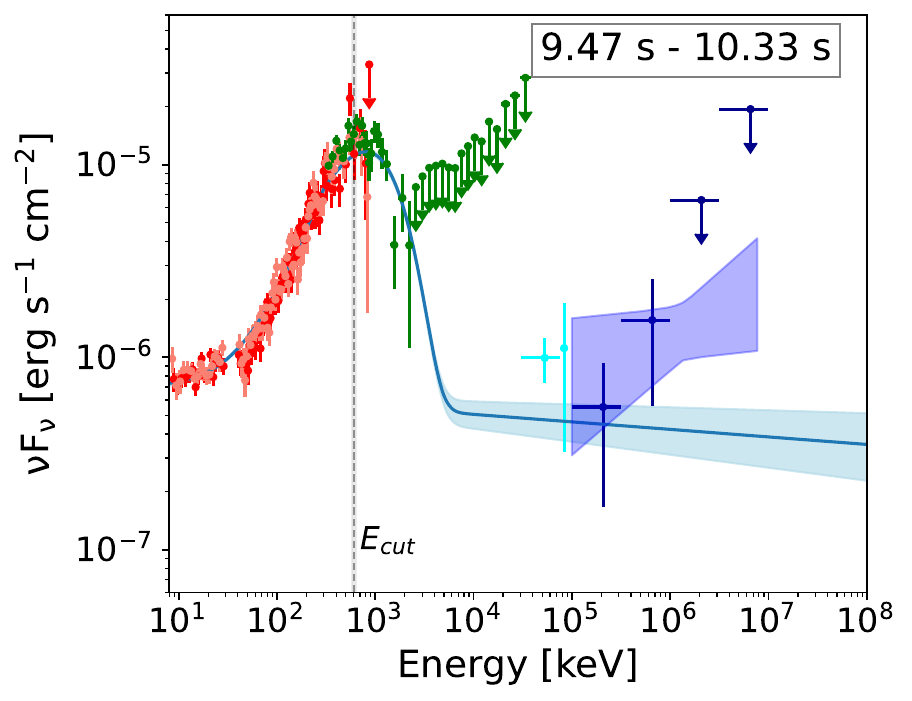}
    \includegraphics[width=0.33\linewidth]{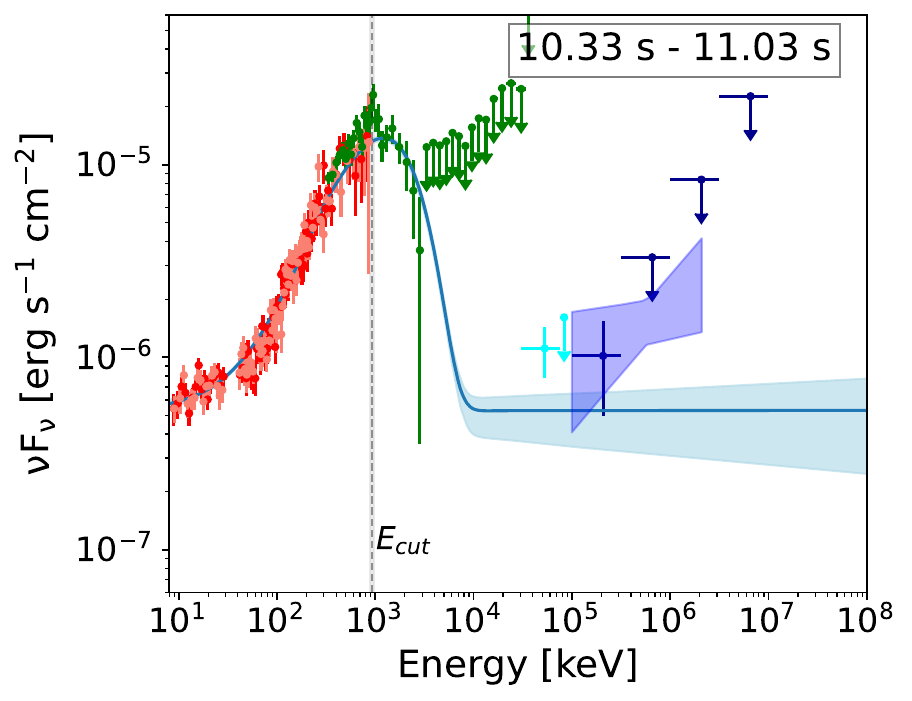}
    \includegraphics[width=0.33\linewidth]{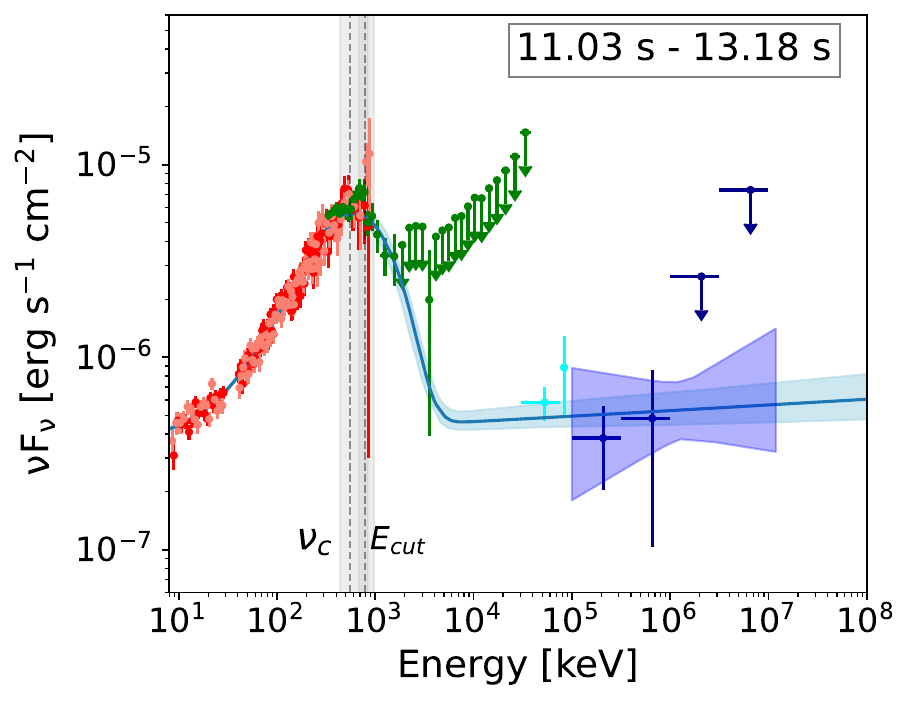}
    \includegraphics[width=0.33\linewidth]{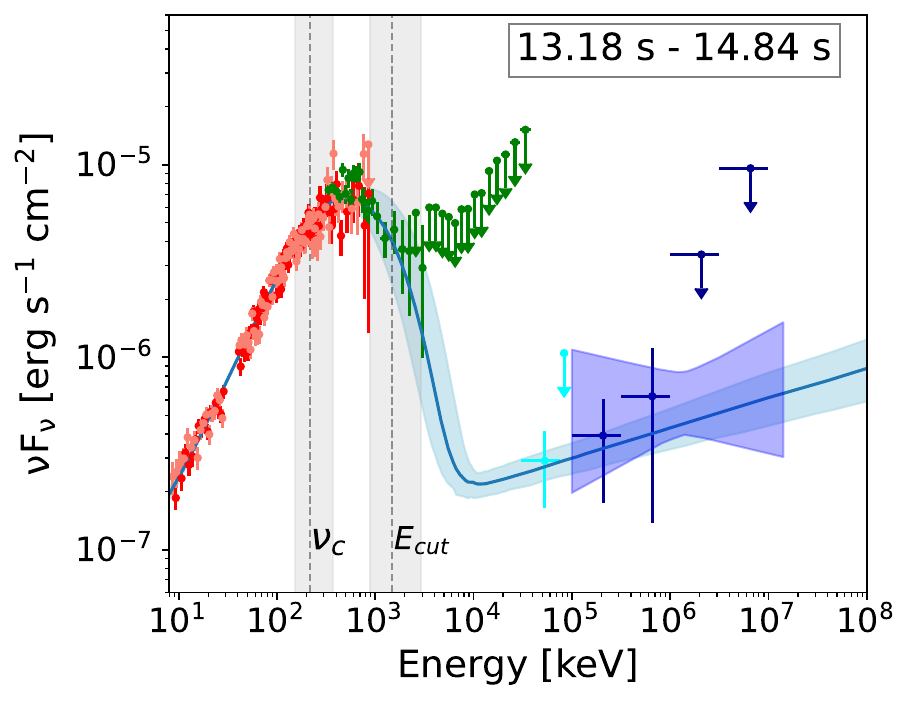}
    \includegraphics[width=0.33\linewidth]{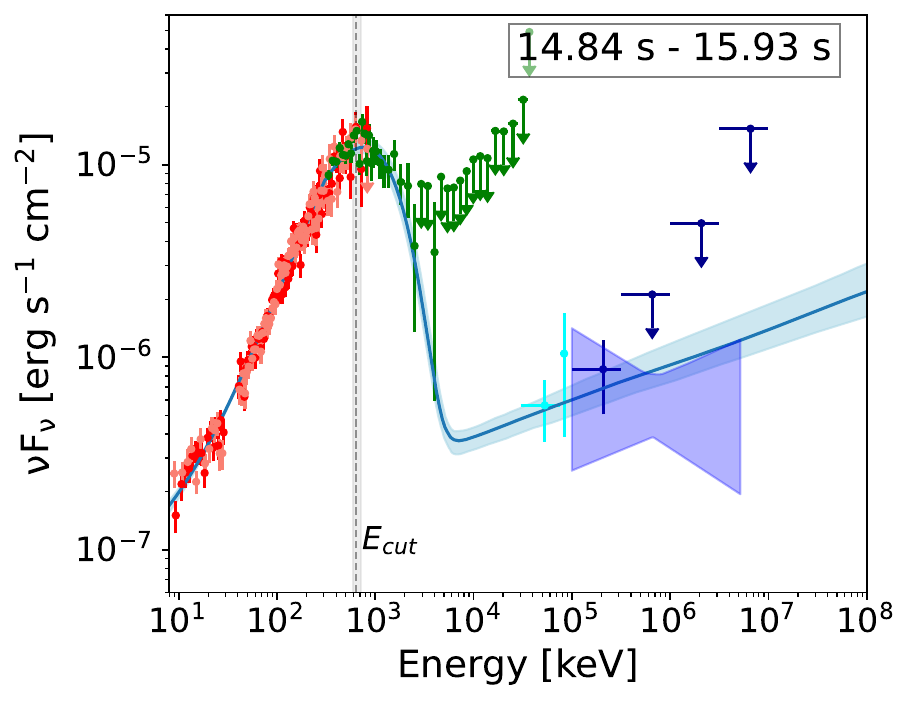}
    \includegraphics[width=0.33\linewidth]{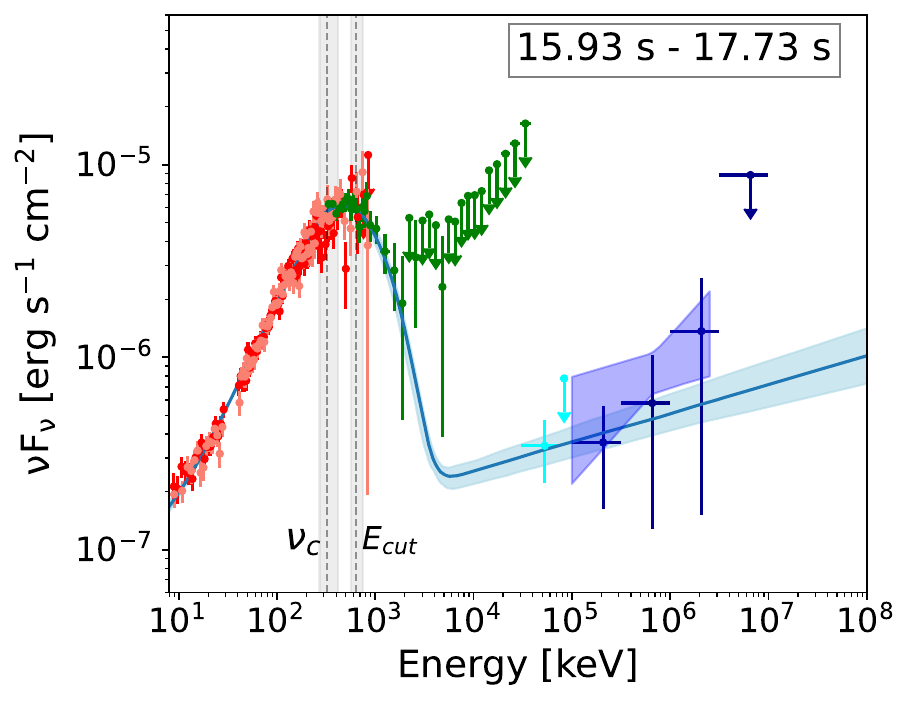}

\end{figure*}

\begin{figure*}[ht!]
    \centering
    \caption{Same as Fig. \ref{fig:080916} but for GRB 221023A. See Table \ref{tab:221023A} for the spectral fit parameters.}
    \label{fig:220123}
    \includegraphics[width=0.65\textwidth, height = 8cm]{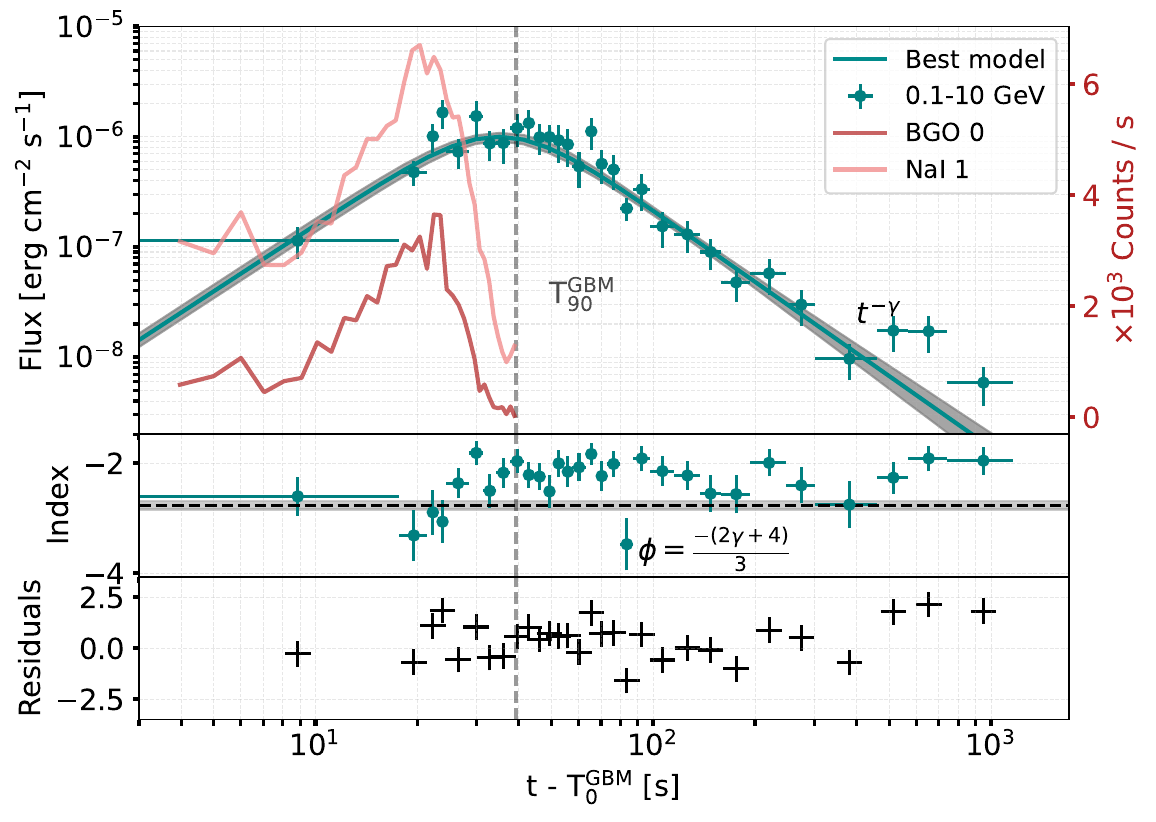}
    \includegraphics[width=0.33\linewidth]{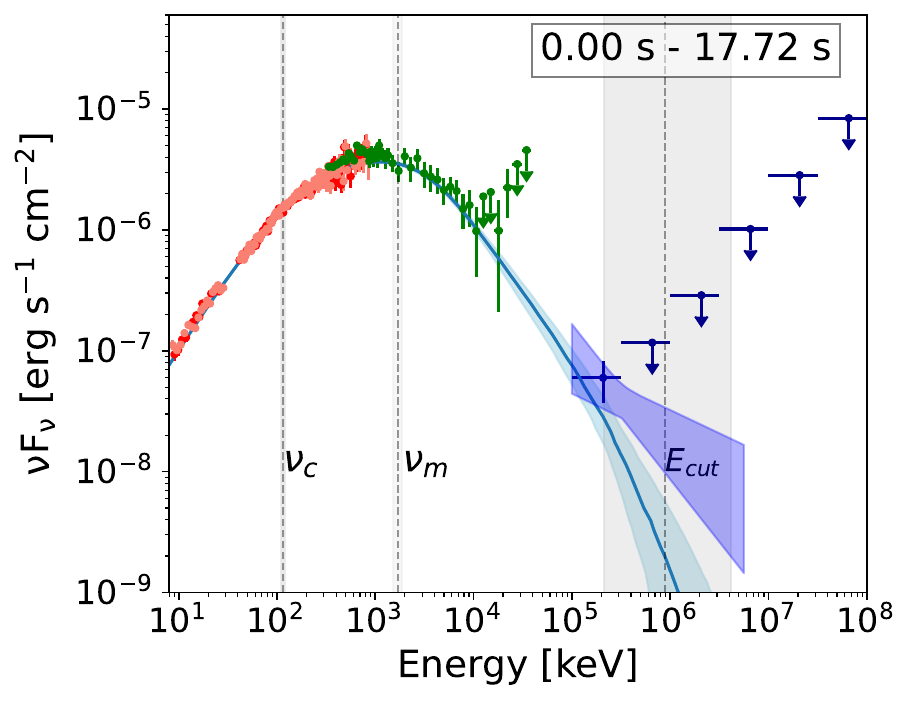}
    \includegraphics[width=0.33\linewidth]{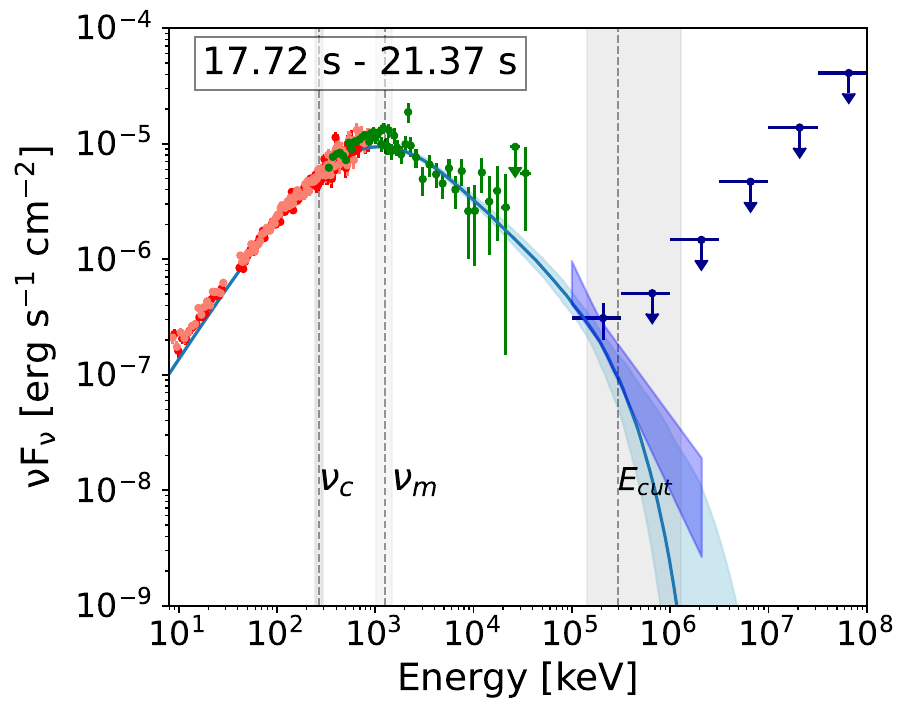}
    \includegraphics[width=0.33\linewidth]{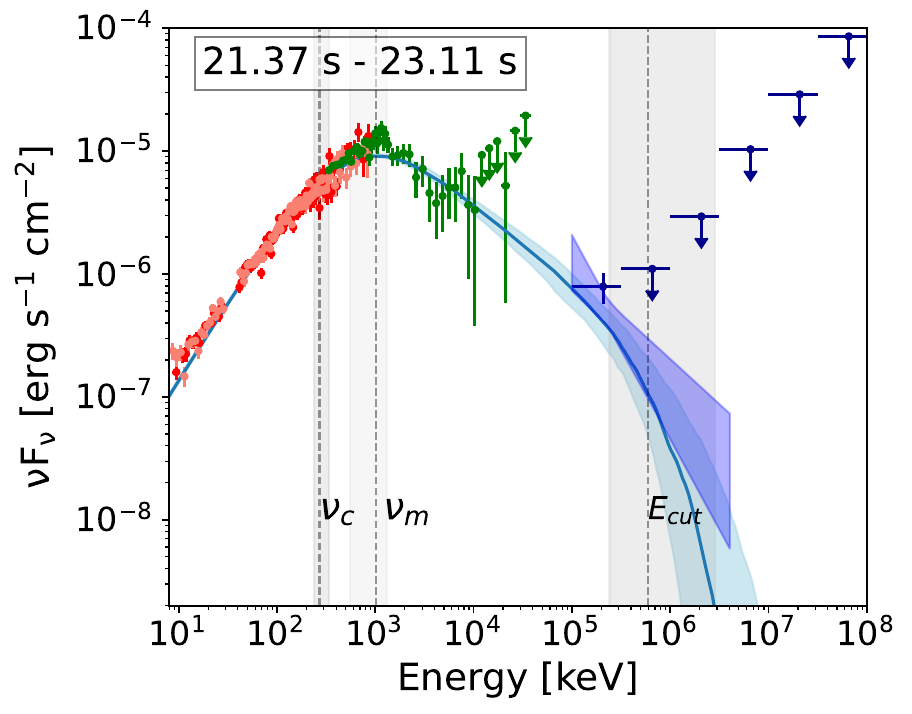}
    \includegraphics[width=0.33\linewidth]{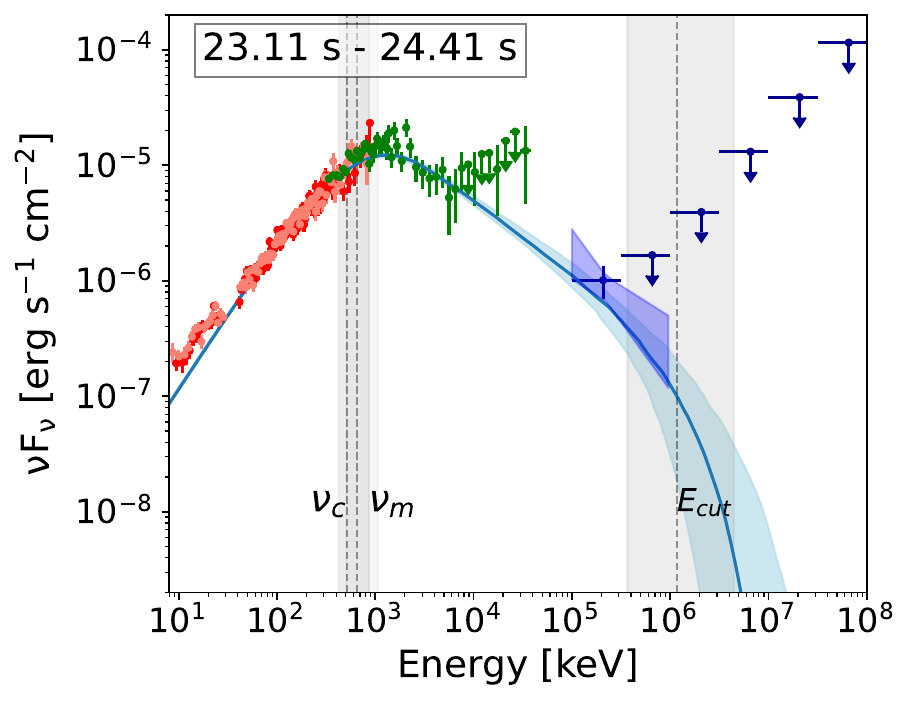}
    \includegraphics[width=0.33\linewidth]{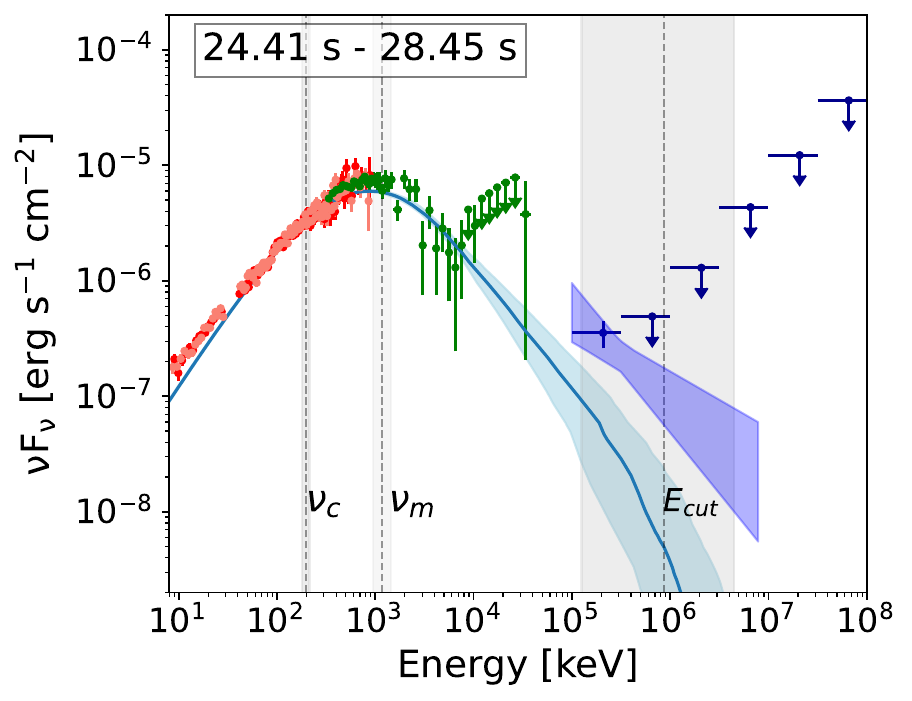}
    \includegraphics[width=0.33\linewidth]{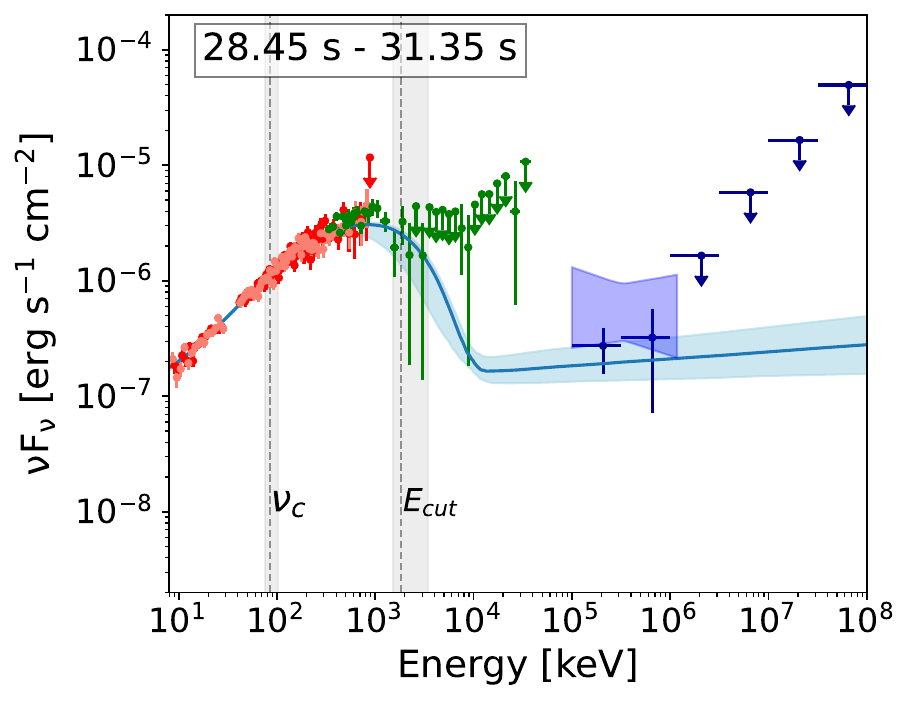}
    \includegraphics[width=0.33\linewidth]{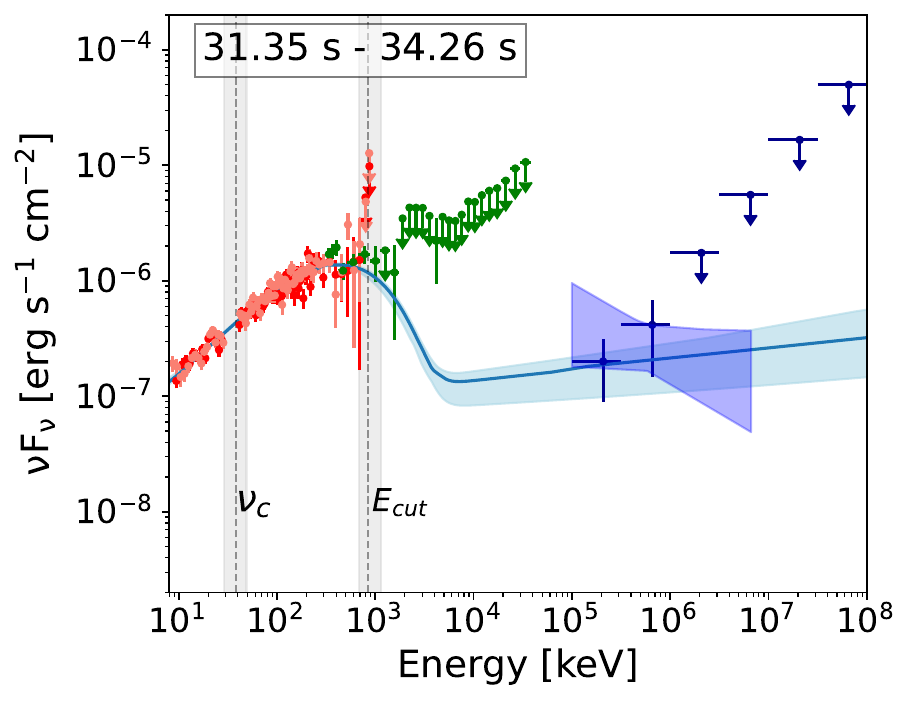}
    \includegraphics[width=0.33\linewidth]{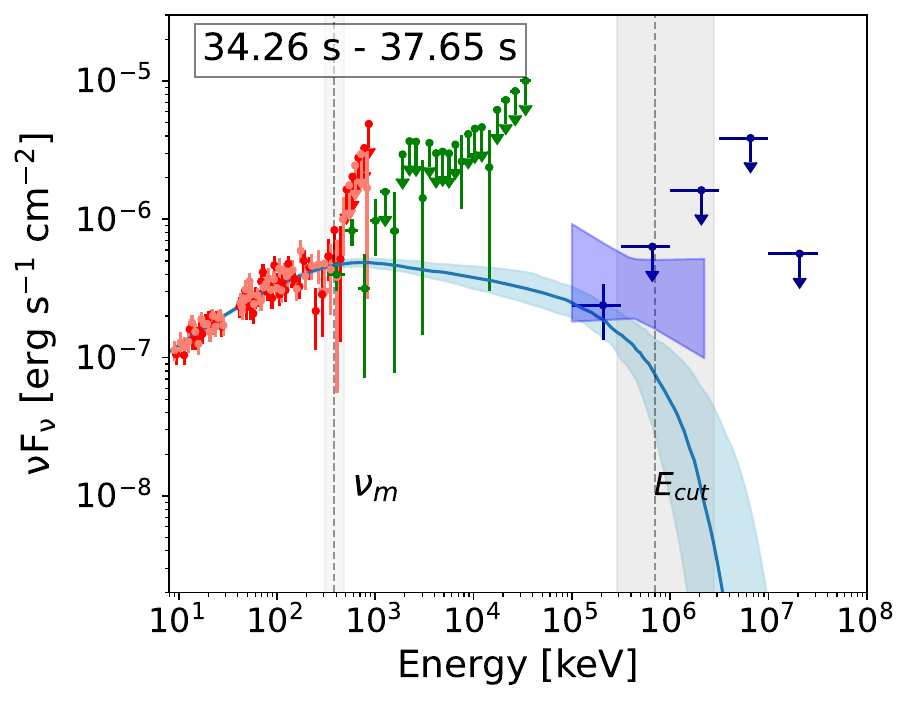}
    \includegraphics[width=0.33\linewidth]{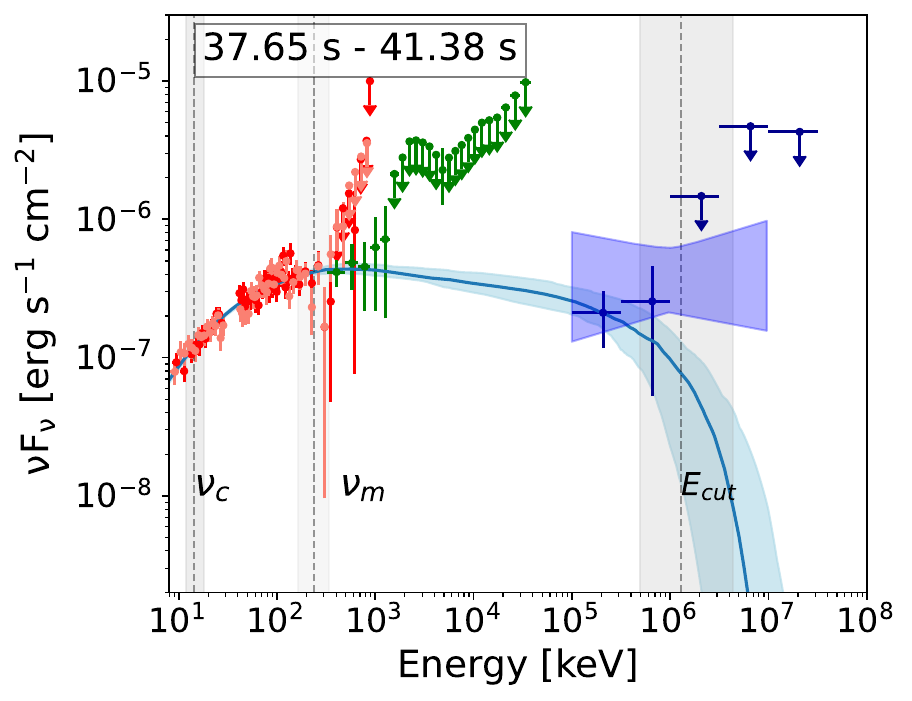}
    \includegraphics[width=0.33\linewidth]{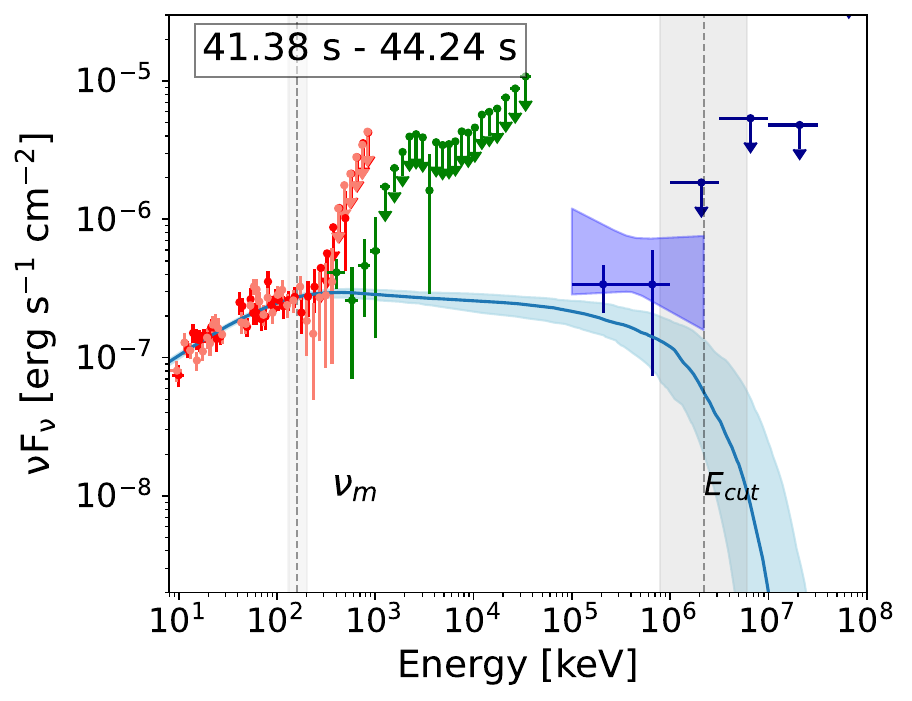}

\end{figure*}

\begin{figure*}[ht]
    \centering
    \caption{Same as Fig. \ref{fig:080916} but for GRB 190114C. See Table \ref{tab:190114C} for the spectral fit parameters.}
    \label{fig:all_plots_190114}
    \includegraphics[width=0.7\textwidth]{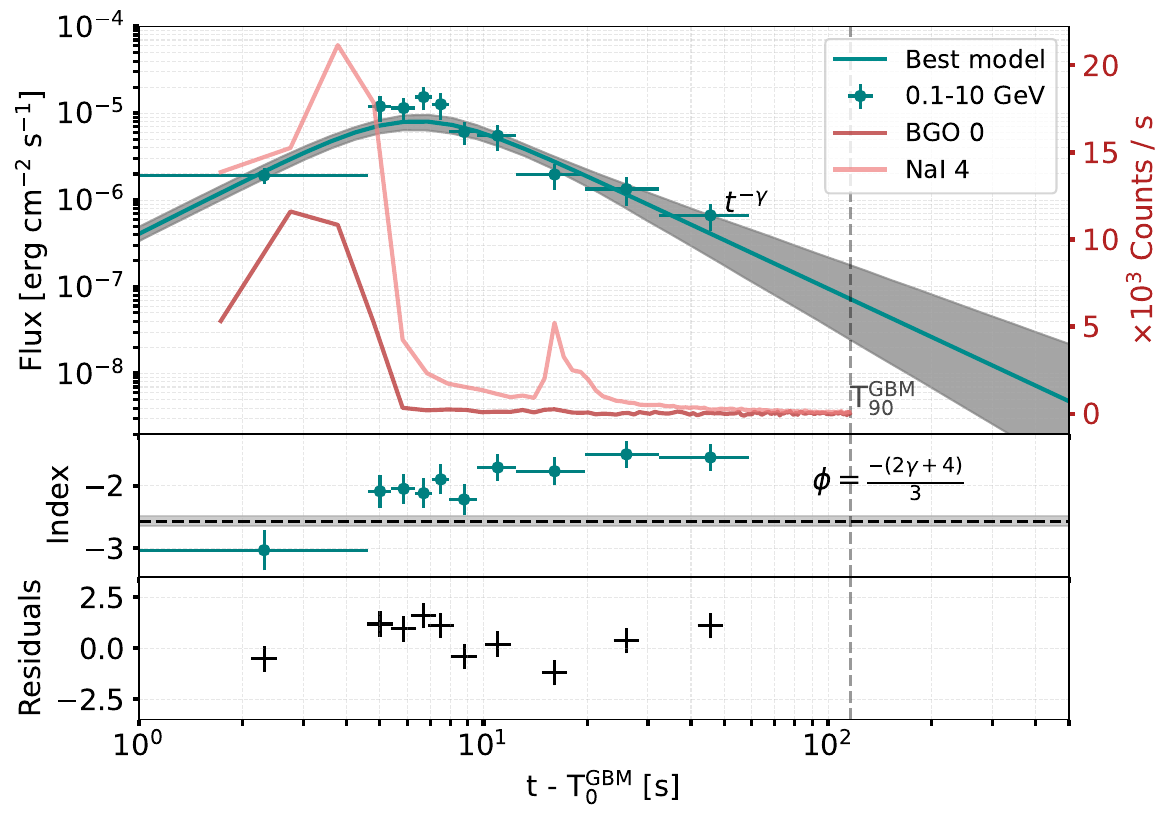}
    \includegraphics[width=0.33\linewidth]{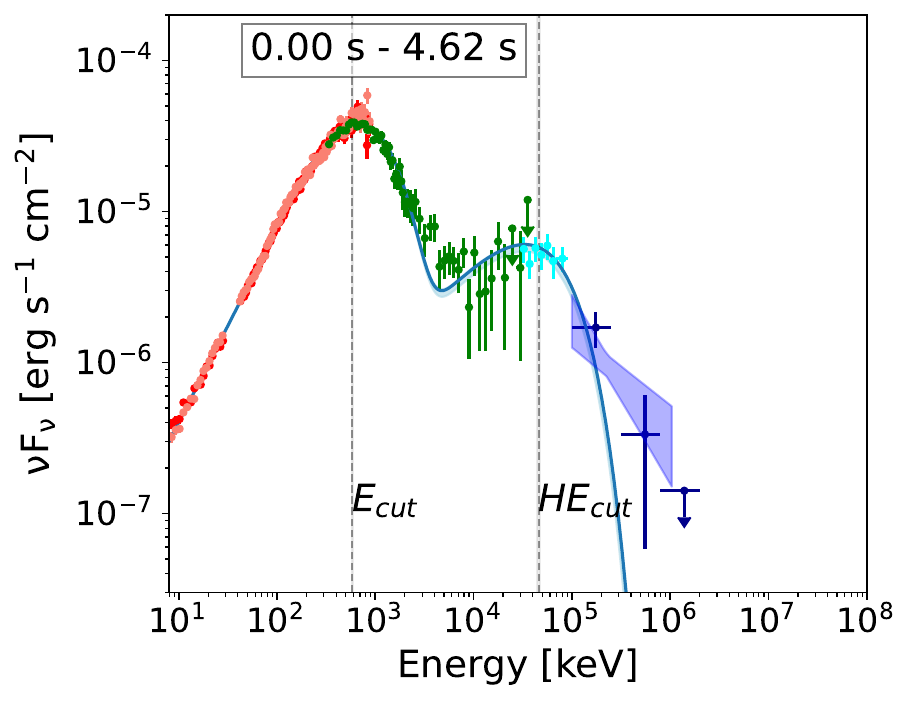} 
    \includegraphics[width=0.33\linewidth]{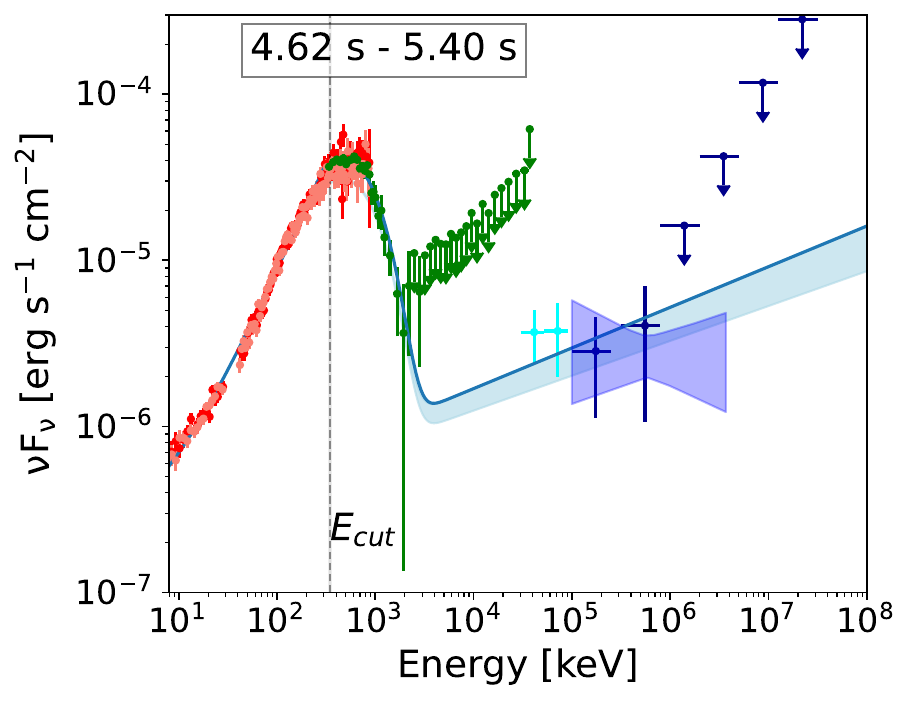}
    \includegraphics[width=0.33\linewidth]{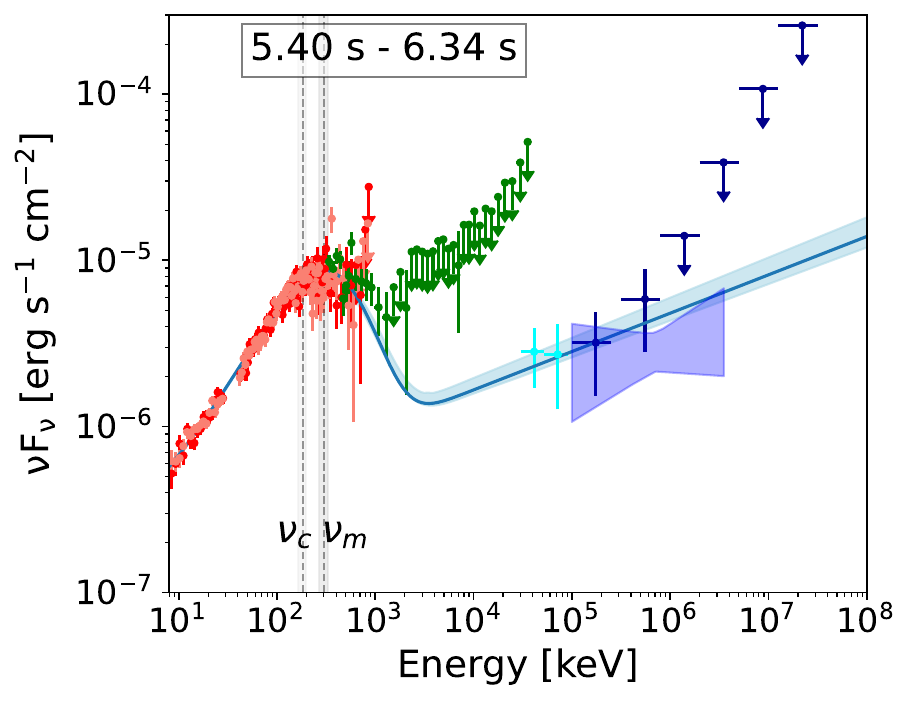}    

\end{figure*}

In this work, we performed a temporal analysis of the \textit{Fermi}/LAT lightcurves of 12 GRBs and a time-resolved spectral analysis of 35 GRBs, encompassing a total of 90 spectra. 

\subsection{Temporal analysis}

For Sample-1, we conducted a timing analysis, to study the evolution of \textit{Fermi}/LAT light curves. Our analysis shows that, in most cases, the empirical model (Eq. \ref{eq:standard_aft}) adequately represents the \textit{Fermi}/LAT light curve, with some exceptions. Fig.~\ref{fig:080916} presents the results of this analysis, using GRB 080916C as an illustrative example. In the top panel, we display the afterglow model fitted to the data; the central panel shows the LAT photon index; and the bottom panel presents the residuals.
The synchrotron process within the external shock model establishes a connection between the temporal decay index ($\gamma$) of the light curve and the energy spectral index ($\phi$), referred to as the closure relations \citep{Kumar:2009vx, Ghisellini:2009rw}. These relations provide a convenient method to verify whether the observed radiation originates from the external shock. In particular, high energy emission is often associated with the synchrotron radiation from the forward shock accelerated electrons beyond the synchrotron cooling frequency. If so, then the spectrum should be $\propto \nu^{-p/2}$, while the lighcurve in the external shock scenario should decay as $\propto t^{-(3p-2)/4}$ \citep{Panaitescu:2000bk, Beniamini:2015eaa, Beniamini:2016hzc}. From this, we can derive the closure relation $\phi = -(2\gamma + 4)/ 3$.
Therefore, from the fit of the decay temporal index $\gamma$, we derive the predicted spectral index $\phi$, which is shown with its uncertainties as a shaded gray area in the central panel of Fig.~\ref{fig:080916}. 
The majority of data points which show an excess are clustering around the peak of the \textit{Fermi}/LAT light curve, and within the duration of the burst (T$_{90}$ measured by the \textit{Fermi}/GBM) ensuring the emission in GeV energies during the prompt phase. However, we did not observe any significant excess over $5\sigma$. We computed also the quantity t$_{\rm peak}$/T$_{90}$, which gives information about which fraction of the duration is happening during prompt and which one is more likely afterglow-related. The median values for t$_{\rm peak}$/T$_{90}$ and $\gamma$ found in our sample are $0.51^{+0.82}_{-0.09}$ and $1.76^{+0.55}_{-0.30}$ respectively.

\subsection{Spectral analysis}

To better understand the nature of the HE emission, we performed a spectral analysis.
For Sample-1, which includes time-resolved spectra, Model-1 (only synchrotron emission and high energy cut-off) and Model-2 (a combination of synchrotron and power law) are employed. When Model-2 is selected as best-fit model by the Akaike test, further evaluation is performed using Model-3 (synchrotron and power law with high-energy cut-off). Meanwhile, for Sample-2, due to the absence of significant GeV excess beyond synchrotron emission, we limit our analysis to Model-1, without the addition of a high energy cutoff. 
The subsequent sections highlight the most suitable model category and are illustrated with examples.

\subsubsection{Model-1: Synchrotron-dominated spectra}
In total, 32 GRBs are best-fitted by a pure synchrotron with a high-energy cut-off (Model-1), resulting in a total of 75 spectra for which the HE emission is well explained by synchrotron radiation in a marginally fast-cooling regime.
In particular, in the illustrative case of GRB 080916C, the cooling regime shows an evolution: during the first time-bin, at the onset of the prompt emission, the synchrotron frequencies $\nu_{m} = 276_{-50}^{+258}$ keV and $\nu_{c} = 0.49_{-0.21}^{+0.46}$ MeV are particularly close (see Fig.~\ref{fig:080916} and Table~\ref{tab:GRB080916} for details). However, in subsequent time bins, they become distinct and well-constrained. Additionally, the post-peak spectrum hardens over time, with the spectral index $p$ evolving from $2.85$ in the $0.0-5.27$\,s time bin to $2.18$ in the $5.95 - 6.63$ s time bin, before softening again at later times. Moreover, the model provides tight constrains on the cut-off energy, E$_{\rm cutoff}$, in all the time bins analyzed (see Table~\ref{tab:GRB080916}).

The inclusion of \textit{Fermi}/LAT and LLE data significantly improved the constraints on the $p$ index. To show this, we fitted three physical models all based on the synchrotron emission to all the GRBs in both samples, with and without \textit{Fermi}/LAT data (Fig.~\ref{p_hist_GBM}). The improvement in constraints is shown in Figure \ref{p_hist_GBM}. We observe that the fit of GRB spectra without high-energy data returns unconstrained values of p$_{\rm index}$ (gray bins). In contrast, the inclusion of MWL data between 30 MeV and 10 GeV helps to constrain the parameter, returning a median value around $p \simeq2.7$ .

\subsubsection{Model-2: Synchrotron + Power Law dominated spectra}

From our analysis, we identified a total of 14 spectra requiring the addition of a secondary power law component superimposed to the main synchrotron model. In this section, we report two GRBs as examples, namely GRB 221023A and GRB 090902B.

In case of GRB 221023A, the power law component emerges at later times, around $\sim 25$ s after the burst, and subsequently it fades away. Interestingly, the onset of the power law coincides with a softening of the synchrotron emission, which exhibits a particularly soft electron distribution index $p \simeq 3.5$. In these cases, as well as in the case of GRB 190114C, descibed in Section \S\ref{sec:model3}, the synchrotron radiation alone is not capable to describe the curvature around the peak of the spectrum. 
To fully characterize the spectral softening, we include an additional cut-off component to the synchrotron component in the energy range 100 keV - 1 MeV. 

In the case of GRB 090902B, the power law component is present throughout the entire duration of the prompt emission (Fig.~\ref{fig:090902B}). It extends up to lower energies, influencing the synchrotron spectrum even at few keV with a larger impact with respect to the case of GRB 221023A. Moreover, in GRB 090902B, the flux of the power law component 
closely tracks the time evolution of the synchrotron one, indicating a possible connection between the two emission processes. For GRB 221023A, the power law component appears only in few time bins, making it difficult to probe its time evolution and to establish a consistent physical scenario.

\subsubsection{Model-3: Synchrotron + cut-off Power Law dominated spectra}\label{sec:model3}

In the case of GRB 190114C, during its first time bin (0 - 4.6 s) the spectrum clearly requires the addition of a HE cutoff to the power law component superimposed to the synchrotron spectrum, described in Model-2. This new model, named Model-3, allows us to constrain the peak of the power law component at higher energies, thanks to the availability of data from LLE and LAT. In all the time bins, the synchrotron component appears to be narrow, with both $\nu_{m}$ and $\nu_{c}$ unconstrained. Indeed, all the spectra show a strong softening observed by BGO. Previous studies also showed that a cut-off in the MeV energies is required to explain the observed spectra \citep{Ajello:2019avs}.

The second spectral component represented by a cutoff power law in addition to the synchrotron component peaks at approximately 46 MeV, representing a rare instance where the high-energy peak can be identified. However, in the subsequent time bins, the high-energy cut-off is not required, resulting in a harder power law index (Fig. ~\ref{fig:all_plots_190114}).

\begin{table*}[ht!]
\centering\caption{Synchrotron parameters for each analyzed time bin of GRB 190114C, along with the bolometric flux and the highest energy photon detected by \textit{Fermi}/LAT.}\label{tab:190114C}
\begin{tabular}{c  c c c c c}
\hline

{Time bin} &  E$_{\text{cutoff}}$ & \multirow{2}{*}{ph$_{\text{index}}$} & HE$_{\text{cutoff}}$ & {Flux} ($\times 10^{-6}$) & E$_{\rm max}$ \\
(t-T$^{\rm GBM}_{0}$)[s] & [MeV]                &     & [MeV]                & [erg cm$^{-2}$ s$^{-1}$] & [GeV] \\
\hline \\
\vspace{0.1cm}
0.0 - 4.62 & $0.58_{-0.07}^{+0.07}$ & $1.28_{-0.01}^{+0.01}$ & $46.03_{-1.59}^{+1.38}$ & $109.6_{-0.7}^{+24}$ & 1.04 \vspace{0.1cm}\\
4.62 - 5.40 & $0.35_{-0.01}^{+0.01}$ & $1.77_{-0.02}^{+0.02}$ & - & $127.4_{-6.7}^{+11.8}$ & 3.70 \vspace{0.1cm}\\
5.40 - 6.34 & $1.40_{-0.29}^{+0.15}$ & $1.77_{-0.01}^{+0.02}$ & - & $65.3_{-5.9}^{+7.4}$ & 3.52 \vspace{0.1cm}\\
\hline
\end{tabular}
\end{table*}

\section{Discussion}
\subsection{Temporal properties}

In this paper, we analyzed GRBs jointly detected by \textit{Fermi}/LAT and \textit{Fermi}/GBM. The selected GRBs are divided into two subgroups, Sample-1 and Sample-2. 

GRBs in Sample-1 have more than one temporal bin with significant detection in GeV energies, hence allowing us to perform a temporal analysis of the HE light curve (0.1-10\,GeV).  We fit an analytical function describing an afterglow light curve (F$_{\rm 0.1-10\,GeV} \propto t^{-\gamma}$) to the \textit{Fermi}/LAT dataset, which extends up to more than 1000\,s beyond \textit{Fermi}/GBM prompt duration, T$_{90}^{\rm GBM}$. During the early prompt phase, the data points deviatew from the afterglow model. However, this excess is not significant enough to rule out an afterglow origin. Thus, the origin of the HE emission is rather inconclusive from the temporal fitting of the light curve to claim an association with prompt or afterglow emissions.

However, the temporal decay index $\gamma$ yields an average value of approximately 1.5. To produce photons with $E > 0.1$ GeV, the associated electrons must be in the fast-cooling regime. Following the deceleration time, the afterglow flux is expected to decay as $\gamma = -(3p-2)/4$, where $p$ is the slope of the electron distribution function, as described in \cite{Nava:2016onl}. Using the inferred temporal index, a spectral index can be estimated using $\phi = -p/2 -1$. The derived spectral index most often is in accordance with the values observed in LAT spectra. However, there are some GRBs (e.g., GRB 090902B, see Fig.~\ref{fig:090902B}), where the observed spectral index at GeV energies deviates from the predicted one derived from the temporal index. In these cases, the observed spectral indices at GeV energies are harder than the ones expected (see Figs. \ref{fig:090902B} and \ref{fig:220123}). At early times, such hardening is consistent with the emergence of a power law component, as observed in GRB~090902B and GRB~221023A. At later times, spectral hardening could suggest the presence of a potential SSC component of the afterglow. The energy range covered in our analysis, up to $\sim$100 GeV, does not identify the presence of the peak of this SSC component mostly due to the low-sensitivity of LAT. This further suggests that the appearance of the peak of the SSC component might be expected beyond 100 GeV (in the VHE gamma-rays). In addition, we also find that in many cases the HE photons seem to be delayed with respect to the keV-MeV emission. This delay was reported by many authors for different GRBs \citep[i.e.][]{ackermann2013first, Castignani:2014gaa, 2019ApJ...878...52A}.

\subsection{Spectral analysis} 
\begin{figure}[ht!]
    \centering \includegraphics[width=\linewidth]{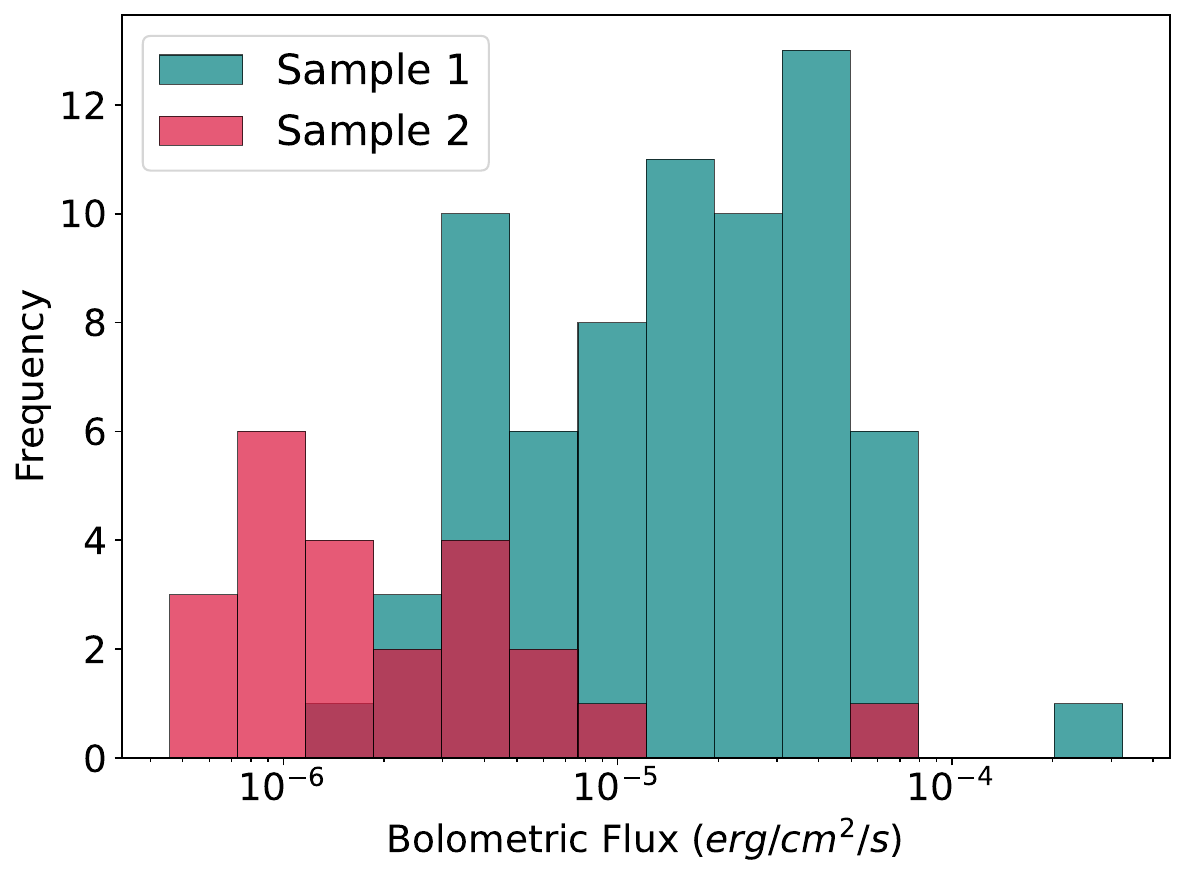}
    \caption{Histogram of the bolometric flux of GRBs. The bolometric fluxes are calculated in the energy band of 10 keV-10 GeV. The histogram hints towards the fact that the GRBs considered in Sample-1 are predominantly brighter than Sample-2. Therefore, Sample-1 is associated with significant data in both LLE and LAT.}
    \label{fig:hist_flux}
\end{figure}

We investigated the nature of early high-energy emission by performing a detailed time-resolved spectral analysis for GRBs in Sample-1. Several studies previously argued that GRB spectra cannot be explained by synchrotron radiation from a distribution of non thermal electrons \cite[e.g.,][]{Beloborodov:2012ys, Axelsson:2014ora, Burgess:2017eek}, since GRB prompt spectra are in general much narrower than what predicted by synchrotron processes. However, our analysis extends the spectral coverage to energies exceeding 30 MeV by including LLE and LAT data. Such large energy range (10\,keV up to $\sim$ 10\,GeV) reveals a unique perspective: most of the prompt GRB spectra we analyzed appear significantly broader than previously expected, making them consistent with a synchrotron origin. Specifically, in Sample-1, 52 spectra (10 GRBs) are well described by a pure synchrotron model that incorporates a high-energy cut-off, which effectively describes the high-energy emission. In contrast, the spectra of 23 GRBs in Sample-2 are well-fitted with a synchrotron model without requiring a high-energy cut-off. This is because the data for Sample-2 exhibit a lower bolometric flux compared to Sample-1 (as illustrated in Fig.~\ref{fig:hist_flux}), returning spectral parameters not well-constrained. 
However, \textit{Fermi}/LAT data remain consistent with the synchrotron model across all the analyzed spectra. In addition, the availability of significant BGO and LLE data significantly impacts the estimation of the slope of the high-energy emission. This effect is more prominent in Sample-2, where rarely we have GRBs which are bright in BGO energy range, and only 7 GRBs have available LLE data. For these reasons, the $p$ index in this sample is poorly constrained.

It should be noted that, in most cases, synchrotron emission is in an intermediate cooling regime, where the values of the characteristic frequencies $\nu_{m}$ and $\nu_{c}$ are close, or even nearly identical ($1<\nu_{m}/\nu_{c}<3$).  This is clearly demonstrated in Fig.~\ref{fig:alpha_vs_ratio}, where most of the data points show a frequency ratio $\nu_{m}/\nu_{c}\sim1$. Specifically, when the two frequencies are approximately equal, the slope of the low-energy portion of the spectrum (i.e. before the break) still matches the expected behavior of a synchrotron spectrum. There are 15 cases (3 GRBs) in Sample-1 where the spectra appear significantly narrower, with the electron spectral index $p>4$, consistent with findings reported in previous works (see, e.g., \citealt{Burgess:2017eek}). In these cases, an additional power law component is necessary to describe the high-energy data. In order to fully explain such spectra within the synchrotron model framework, the introduction of a cutoff at lower energies is required, typically around a few MeV. In particular, this cut-off is consistently observed whenever a power law component is present in the prompt spectrum. Interestingly, a cut-off around 5 MeV is physically anticipated in the pair-loading afterglow scenario, as proposed by \cite{Beloborodov:2005nd}.

In particular, we notice peculiarity in the time resolved spectra of three GRBs: GRB 090902B, GRB 190114C, and GRB 221023A. For GRB 090902B, the power law component persists throughout the entire prompt phase, exhibiting spectral evolution that appears consistent with the variability observed in the GBM data. This suggests that the GeV component is most likely associated with the prompt emission. In contrast, for GRB 221023A, the power law emerges only at later times, specifically during the final stages of the prompt emission, when the GBM flux is comparatively lower. This indicates that, in this case, the power law is more likely linked to the afterglow, which begins to overlap with and influence the prompt phase. For the spectrum of GRB 190114C during the time interval from 0.0 to 4.62 seconds, where a secondary component peaking at approximately 46 MeV is observed, our findings align with those reported by \citealt{Ajello:2019zki}. However, we interpret the first spectral component as synchrotron emission with a MeV cutoff. The second component is distinctly identifiable only in the initial time bin analyzed, when its peak energy is low enough to appear at high energies, before evolving into a power law in subsequent bins. This evolution suggests that the peak shifts to higher energies while the keV-MeV component decreases in flux, eventually adopting a power law-like appearance. This behavior indicates that the HE emission in this case may originate from the afterglow, which begins to dominate during the latter part of the prompt phase. Consequently, the entire keV-GeV emission could represent a transition from the prompt to the afterglow phase, as suggested by \citealt{Ravasio:2019yhd}. However, due to the sensitivity limitations of \textit{Fermi}/LAT and the resulting poor spectral resolution, we are unable to precisely determine the nature and temporal evolution of this distinct power law component.
\begin{figure*}[h!]
    \begin{center}
        \includegraphics[width=12cm, height=12cm]{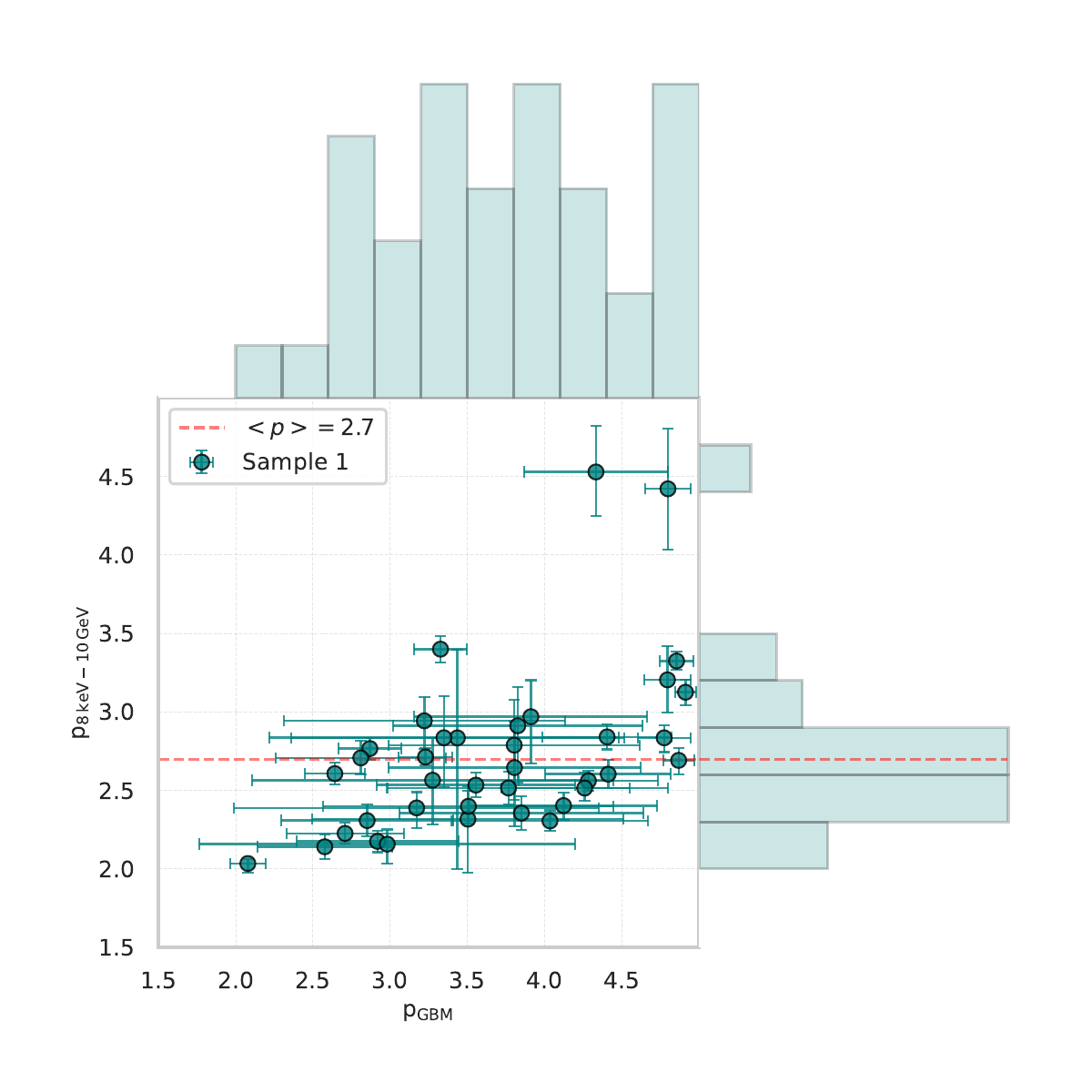}
        \caption{Comparison between the $p$ index obtained from the spectral analysis between 8 keV-10 GeV and using only the GBM data for Sample-1 and their corresponding uncertainties. The histogram on the sides provide a distribution of the indices for each cases. The index while adding the GeV data provides significant constrains.}
        \label{p_hist_GBM}    
    \end{center}
\end{figure*}
To compare our findings with previous studies, we also fitted the synchrotron model within the keV–MeV range, therefore using only GBM data. In this limited range, the $p$ index distribution is much broader, yielding less precise constraints. The use of LLE data was essential to have a better picture of the prompt spectrum after the peak. In fact, this methodology allowed us to place tighter constraints on the electron spectral index $p$, which exhibits a distribution that peaks around 2.7, as shown in Fig.~\ref{p_hist_GBM}. This aligns well with theoretical predictions from the particle acceleration mechanisms \citep[for a recent review]{Sironi:2015oza}. 

It is interesting to note the evolution of the frequencies $\nu_{m}$ and $\nu_{c}$ for the GRBs in Sample-1. Since all the analyzed spectra appear to be in a fast-cooling regime, $\nu_{m}$ consistently represents the spectral peak energy. In some cases, such as GRB 080916C, $\nu_{m}$ and $\nu_{c}$ are well-separated ($\nu_{m}/\nu_{c}>1$)  with the peak energy closely following the flux evolution, in other words, shifting to higher energies as the flux increases. However, in other cases like GRB 090926A (see Fig.~\ref{fig:090926A}), the two frequencies are indistinguishable, and the peak energy remains constant over time. In these cases, the high-energy slope becomes harder as the flux increases.
Thus, there is no universal pattern for the evolution of the peak energy in the prompt spectrum, as the behavior varies case to case.
\begin{figure}
    \includegraphics[width=\linewidth]{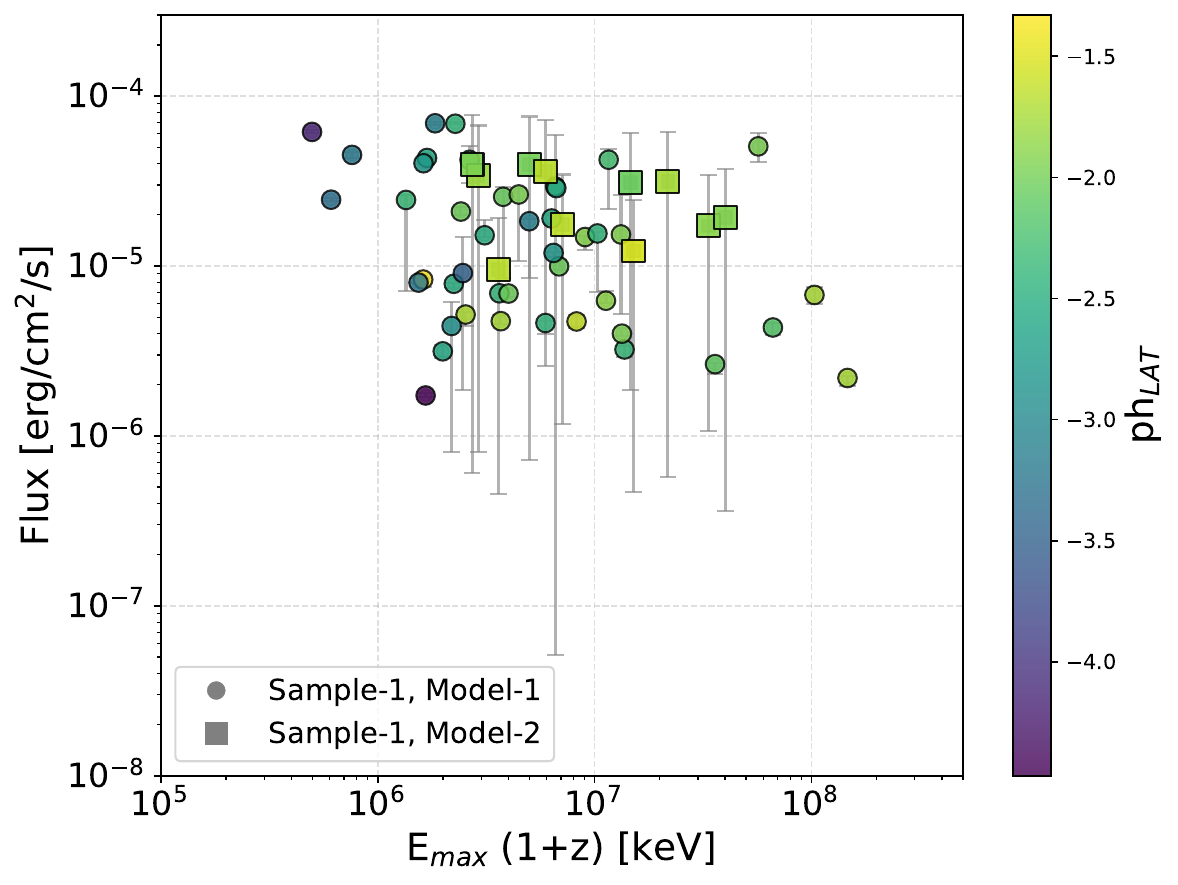}
    \caption{Bolometric flux vs. E$_{max}$ for the GRBs in Sample-1. Each entry in the plot represents a spectrum of Sample-1. 
    The best fitted models (Model-1 or Model-2) of the individual spectrum are indicated with different markers.
    The color bar indicates the corresponding spectral index for each spectrum (represented by each point in the graph) in the high-energy gamma-rays detected with LAT. The highest energy photons are represented in the co-moving frame. For each spectrum the corresponding LAT photon index in the energy between 0.1-10 GeV is indicated by a color gradient.}
    \label{fig:FluxvsEmax}
\end{figure}
We calculated the bolometric flux within the energy range spanning 10\,keV to 10\,MeV for the time-resolved spectra of GRBs in Sample-1, specifically when the spectra are optimally fitted using Model-1 and Model-2. Figure \ref{fig:FluxvsEmax} illustrates the relationship between bolometric flux and the maximum photon energy in the co-moving frame in those temporal bins.  Spectra with higher flux levels tend to exhibit a softer LAT index, which indicates an early prompt emission phase, and consequently shows a lower maximum photon energy because of the spectral cut-off previously described. Conversely, spectra with lower bolometric flux lead to the production of photons with higher energies. This could be due to the emergence of the afterglow component around the end of the prompt emission. However, the association of this HE emission, exhibiting harder photon indices, with the afterglow is particularly challenging, given the poor sensitivity of \textit{Fermi}/LAT.

\begin{figure}[ht!]
    \centering \includegraphics[width=\linewidth]{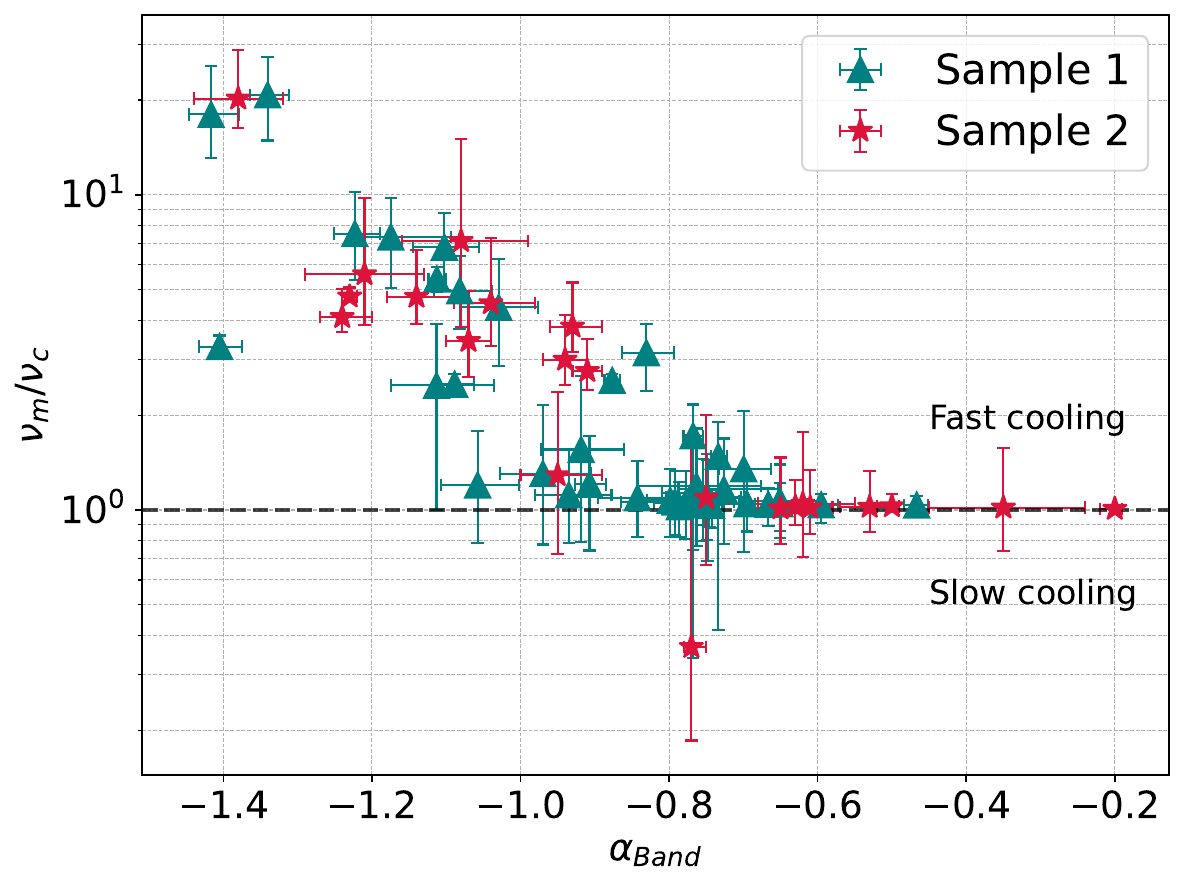}
    \caption{Spectral index $\alpha$ from the Band model compared with the ratio of two characteristic frequencies $\nu_{m}$, and $\nu_{c}$ derived from the synchrotron model. The ratio $\nu_{m}/\nu_{c}$ indicates the cooling regime, where a value grater and less than 1 represents the fast and slow cooling regime, respectively. The majority of the spectra are found to be in the fast- or marginally fast cooling regime.}
    \label{fig:alpha_vs_ratio}
\end{figure}

\subsection{Comparison with Band model}
We investigated the impact of physical models in describing prompt emission spectra. To do so, we fitted our dataset not only with a synchrotron model, but also with the empirical Band function, which was already adopted in previous studies. We focused on the comparison between the Band function's spectral index before the peak, $\alpha$, with the frequencies ratio $\nu_{m}/\nu_{c}$. This comparison is illustrated in Fig.~\ref{fig:alpha_vs_ratio}.

We observe that when the ratio $\nu_{m}/\nu_{c}>1$, the corresponding $\alpha$ tends to be softer, i.e.  $\alpha < -0.8$ . The larger $\nu_{m}/\nu_{c}$ is, the more $\alpha$ approaches the value of $\sim-1.5$, which is the slope expected in a fast-cooling regime. In contrast, when $\nu_{m}/\nu_{c}\sim1$, $\alpha\sim-0.6$, typical of the slow-cooling regime. These findings are consistent with previous studies both involving the Band function \citep{Band:1993eg} and exploring the synchrotron mechanism \citep{Mei2024}, further confirming the robustness of our approach. Additionally, in some spectra, the peak energy shifts to higher energies when fitted with a synchrotron model. However, when $\nu_{m}/\nu_{c}\sim 1$, the spectrum is well-described by the Band function, since the exhibits a single energy break which corresponds to the peak energy. In these cases, the synchrotron model and the Band one lead to similar results. 

We also compare the synchrotron spectral index $p$ with the high-energy spectral index of the Band model, $\beta$. When fitted with the synchrotron model, the GRBs in our sample predominantly show a spectrum in the fast-cooling regime, when the photon index after the peak depends on $p$ as $\beta = -p/2 -1$. 
Our analysis in most cases shows the consistency between these two parameters. 
Although, the uncertainties obtained from the Band model are significantly larger, which do not lead to constraints on the spectra.

\section{Conclusions}

In this study, we performed a detailed analysis of gamma-ray bursts (GRBs) detected simultaneously by \textit{Fermi}/LAT and \textit{Fermi}/GBM, focusing on the temporal and spectral features of their high-energy emission within the synchrotron scenario.
The temporal analysis of the \textit{Fermi}/LAT data was conducted under the assumption of an afterglow origin for the late-time HE component. We compared the spectral index inferred from the analysis of the light curves and the observed spectral index in the GeV band, to examine their consistency with the afterglow modeling. Our results show that the early LAT emission, particularly during the prompt phase, deviate from the expectations of standard afterglow scenarios.
To further investigate these discrepancies, we performed multi-wavelength spectral modeling based on synchrotron emission, deriving key physical parameters such as the characteristic frequencies, $\nu_m$ and $\nu_c$, and the slope of the electron distribution, $p$. These findings provide important insights into the origin of the early GeV emission observed in these bursts. Below, we summarize our strategy and the key findings:

\begin{itemize}
\item  \textit{GRBs selected in this work:}
We searched for GRBs jointly detected by \textit{Fermi}/GBM and \textit{Fermi}/LAT, localized by Swift/BAT, with and without redshift up to year 2023. Our binning is driven by LAT time-bins, for whose we require a test statistic value that exceeds 10. We examined two distinct samples: 1) GRBs with at least two consecutive time bins jointly detected (Sample-1), 2) GRBs with single-bin detections (Sample-2). For Sample-1, we have a total of 12 GRBs; for Sample-2, we have 23 GRBs, resulting in a total sample of 35 GRBs and 90 spectra, analyzed in the energy range from 10 keV up to the highest energy photon detected ($\sim$ GeV). 

\vspace{0.3cm}
\item  \textit{Modeling of the high energy light curve and spectral prediction:}
For GRBs in Sample-1, the evolution of their LAT light curve can typically be modeled as an afterglow with a temporal decay index $\gamma\sim 1.5$. This value is roughly consistent with predictions of afterglow lightcurves from synchrotron radiation above the cooling frequency. However, deviations from the afterglow model, particularly near the peaks of the \textit{Fermi}/LAT light curves show potential contributions from the prompt phase. Nonetheless, these deviations are not statistically significant enough to confirm the presence of prompt emission contamination. The spectral indices inferred during the temporal decay are generally consistent with those observed in the LAT energy range, reinforcing the link between the temporal and spectral properties. In some cases, a spectral hardening is observed at later times, suggesting the possible emergence of a synchrotron self-Compton (SSC) component in the afterglow phase, which may peak beyond the sensitivity range of the LAT instrument. \par

\vspace{0.3cm}
\item  \textit{Modeling of MWL spectra with synchrotron emission and inferences of the characteristic frequencies:} We assumed a synchrotron origin for the prompt emission and fit all spectra in our sample using a synchrotron model. The spectral analysis provides distinct insights from those revealed by the temporal analysis: while the temporal features suggest an afterglow origin for the GeV photons, the spectral properties indicate that they are more likely associated with the prompt emission. This discrepancy shows the critical importance of combining spectral and temporal analyses, as relying solely on temporal data may lead to incorrect conclusions about the origin of high-energy photons. Our spectral investigation reveals that most GRB spectra align with synchrotron emission extending into the GeV range. A significant portion of GRBs in Sample-1 (52 spectra) is well-represented by a synchrotron model featuring a high-energy softening. For Sample-2, where LAT data quality is comparatively lower, the spectra also remain consistent with synchrotron emission, though constraints on parameters like the $\rm p_{\rm index}$ are less precise. This highlights the importance of high-energy data in refining physical models.

\vspace{0.3cm}
\item  \textit{Best fitted models to characterize the observed spectra:} 
We find that the majority of GRB spectra collected in this work can be described with a synchrotron model with a high-energy cutoff due to pair annihilation (Model-1). 
The two characteristic frequencies $\nu_{m}$ and $\nu_{c}$ are well constrained in the majority of the cases, and in a large number of spectra seem to be very close to each other or almost identical, pointing towards a marginally fast cooling synchrotron regime. However, for some GRBs, we find that the model needs to be accompanied by additional features. For example, in some cases, and in particular GRB~090902B, the spectra requires both a cut-off in the spectra around few MeV and an additional broad-band power law component that are necessary to describe the high-energy behavior (Model-2). 
The power law component is rather constrained by the low-energy data below 50 keV as the emission can not be higher than the flux observed at those energies. Moreover, for GRB 190114C, the data required an additional cut-off in the high-energy gamma-rays defining the peak of an additional component around 50 MeV. 
It is important to state that in all the cases which require an additional power law, the spectra require a MeV suppression (which is predicted in the pair-loaded afterglow scenario) in order to be consistent with the synchrotron emission.

 \vspace{0.3cm}
\item  \textit{Observation of a broader synchrotron component:} Synchrotron fits return a value of the electron spectral index $p\simeq2.7$, consistent with theoretical predictions for shock acceleration and reconnection processes \citep{Sironi:2015oza}. The inclusion of \textit{Fermi}/LLE and \textit{Fermi}/LAT data significantly improves the constraints on $p$ compared to fits limited to the \textit{Fermi}/GBM range. The temporal and spectral properties of the high-energy emission provide strong evidence for synchrotron radiation as the dominant mechanism during the prompt phase of most GRBs. The spectral broadening observed when including LAT data reconciles previous discrepancies between observations and synchrotron model predictions, which were based on narrower energy ranges.

\vspace{0.3cm}
\item  \textit{Emergence of a power law component at high-energies:} For a subset of GRBs, an additional power law component (Model-2) is necessary to explain the observed spectra. When this component is included, the synchrotron spectral component often exhibits a MeV cutoff, potentially linked to a pair-loaded afterglow. This extra component is notably present in GRBs such as GRB 090902B, GRB 221023A, and GRB 190114C, suggesting the involvement of an additional high-energy mechanism. In some cases, this component extends to lower energies, potentially contaminating the synchrotron spectrum. However, due to limited high-energy spectral data quality, the exact nature of this power law component remains uncertain. It may originate from the prompt phase in some bursts and from the afterglow phase in others.

\vspace{0.3cm}
\item  \textit{Inference from the highest photon energy during prompt emission:}
Photons with the highest energies observed in the GRBs reported in the work range between 0.1 and 100 GeV. We observed that the initial bright emission phases (higher bolometric flux) are associated with softer spectra in the high-energy gamma-rays and hence lower photon energies. On the contrary, the states with low bolometric fluxes are associated with the later time emission (close to the end of the prompt emission) and most often represented by a harder (around spectral index of $-2$) spectral index in the high-energy gamma-rays. Observation of a high-energy photon from a spectrum, in particular beyond the high-energy spectral cut-off, might indicate an afterglow contamination. However, the poor sensitivity in LAT does not allow us to identify the presence of an afterglow emission component, if present. This is particularly relevant for the observation of the GRBs in prompt phases with current and future ground based telescope, such as MAGIC and CTAO-LST. An early observation of any very-high-energy component can confirm the origin of this high energy component. 

\vspace{0.3cm}
\item  \textit{Comparison with Band modeling:}
We fit all the GRBs spectra in our sample with both the empirical Band model and the physical synchrotron model. The Band model is a simple analytical model that consists of two power laws smoothly connected at a characteristic frequency, which represents the peak energy in the  $\nu F_{\nu}$ representation. In contrast, the synchrotron model, together with the high-energy cut-off, is able to identify the presence of spectral breaks. These breaks, together with the addition of the high-energy data (above 30 MeV), implies a comprehensive characterization of multi-wavelength prompt spectra. This helps us to identify the origin of the high-energy emission, whether it is a separate spectral component (i.e., an additional power law) or an extension of the synchrotron emission. Moreover, the high-energy part of the spectrum, while fitted with synchrotron, provides directly constraints on the particle distribution.

\vspace{0.3cm}
\item \textit{Detection of prompt emission in VHE Gamma-Rays:} 
Ground-based Imaging Atmospheric Cherenkov Telescopes (IACTs) like MAGIC, H.E.S.S., and the forthcoming Cherenkov Telescope Array Observatory (CTAO) need at least 20 seconds to reposition and target the transient discovered by localizing keV/MeV telescopes (such as, Swift/BAT, \textit{Fermi}/GBM). The operations of IACTs are further constrained by a duty cycle of about 10-15\%. In addition, the time needed for inter-telescope communication (usually in the range of tens of seconds) relies on the triggering instrument. For instance, the \textit{Fermi}/GBM identifies a large number of bursts (around 240 GRBs per year), but the localization (larger than 100 deg$^{2}$ considering systematic and statistical uncertainties) it provides at trigger time, or via follow-up notices, is considerably larger than the field of view (FoV) of the IACTs (e.g., $\sim$5 deg$^{2}$). One strategy that is potentially important involves conducting a tiling observation of the GRB localization to capture the VHE transient signal. In a conservative scenario, the expected flux due to synchrotron emission within the 0.03-1.0 TeV energy range is approximately 10$^{-12}$ erg cm$^{-2}$ s$^{-1}$. The sensitivity of the CTAO/LST for brief observations lasting about 10 seconds (or nearly 100 seconds) is roughly 10$^{-9}$ (or 10$^{-10}$) erg cm$^{-2}$ s$^{-1}$. Nonetheless, instances of a power law component emerging in several cases in our sample offer a promising opportunity to detect VHE gamma-ray emission either during ongoing prompt emission or in the early afterglow phase. This suggests that the most favorable scenario for detecting the VHE component of GRBs from the onset of prompt emission is achievable only with telescopes that have a larger field of view and higher duty cycles (about more than 80\%). Consequently, LHAASO and HAWC are prime candidates for discovering the VHE counterpart during the prompt emission phase. However, a limitation of these instruments is their high energy threshold, which is above a few hundreds of GeV (unlike the IACTs, which have a threshold of approximately tens of GeV), and their limited sensitivity at these higher energies.

\end{itemize}
In conclusion, the temporal and spectral properties of the high-energy emission provide strong evidence for synchrotron radiation as the dominant mechanism for the prompt phase of most GRBs. The requirement for an additional power law component in some GRBs points to the potential presence of additional physical processes (such as inverse Compton scattering, prompt or afterglow related). Further investigation of these cases, especially with multi-wavelength observations extending beyond the LAT energy range, at very high energies, is necessary to clarify the nature of these components. Overall, our analysis highlights the importance of extending GRB spectral studies to higher energies and emphasizes the need for simultaneous multi-wavelength observations to disentangle the complex interplay between the prompt and afterglow emissions.

\begin{acknowledgements}
BB and MB acknowledge financial support from the Italian Ministry of University and Research (MUR) for the PRIN grant METE under contract no. 2020KB33TP. The authors thank the Director and the Computing and Network Service of the Laboratori Nazionali del Gran Sasso (LNGS-INFN). This research used resources of the LNGS HPC cluster realized in the framework of Spoke 0 and Spoke 5 of the ICSC project - Centro Nazionale di Ricerca in High Performance Computing, Big Data and Quantum Computing, funded by the NextGenerationEU European initiative through the Italian Ministry of University and Research, PNRR Mission 4, Component 2: Investment 1.4, Project code CN00000013 - CUP I53C21000340006. We acknowledge the CINECA award under the ISCRA initiative, for the availability of high-performance computing resources and support

\end{acknowledgements}
\bibliographystyle{aa} 
\bibliography{references}

\clearpage

\appendix

\renewcommand{\thefigure}{A\arabic{figure}} 
\setcounter{figure}{0} 
\section{GRBs in Sample-1}

In this section, we present the results of our analysis for the time-resolved GRB sample. The physical parameters derived from our fits are summarized in Table A1. Additionally, the spectra for each time interval, along with the LAT and GBM light curves (as shown in Figure \ref{fig:080916} of the main text), are displayed in Figures A1–A8.

\begin{table*}[h!]
\centering
\begin{tabular}{l c c c c c c c }
\hline
\multirow{2}{*}{GRBs}&{Time bin}      & \multirow{2}{*}{$p_{\text{index}}$} & $\nu_{\text{c}}$ &  $\nu_{\text{m}}$& $E_{\text{cutoff}}$& {Flux} ($\times10^{-6}$) & \boldmath$E_{\text{max}}$ \\
    & (t-T$^{\rm GBM}_{0}$) [s]&                    &   [keV]          &  [keV]           &  [GeV]             &   [$ \text{erg/cm}^{2}/\text{s}$] &  [GeV] \\ 
\hline 
\vspace{0.1cm}
\multirow{10}{*}{GRB 090926A}&0.00 - 5.73   & $3.32_{-0.06}^{+0.06}$ & $221_{-6}^{+8}$ & $237_{-15}^{+29}$ & $>1.7$ & $32.7_{-2.94}^{+0.06}$ & 1.61 \\  \vspace{0.1cm}
&5.73 - 6.61   & $2.53_{-0.08}^{+0.08}$ & $169_{-16}^{+75}$ & $193_{-32}^{+91}$ & $>0.6$ & $32.6_{-2.9}^{+3.0}$ & 0.43 \\ \vspace{0.1cm}
&6.61 - 7.42   & $2.83_{-0.08}^{+0.09}$ & $205_{-15}^{+71}$ & $221_{-25}^{+65}$ & $12.1_{-9.6}^{+3.8}$ & $20.5_{-1.0}^{+0.7}$ & 1.22 \\ \vspace{0.1cm}
&7.42 - 9.71   & $3.13_{-0.08}^{+0.08}$ & $144_{-6}^{+27}$ & $154_{-12}^{+25}$ & $>2.7$ & $41.8_{-5.1}^{+0.1}$ & 3.32 \\ \vspace{0.1cm}
&9.71 - 10.31  & $2.18_{-0.07}^{+0.07}$ & $97_{-10}^{+43}$ & $114_{-23}^{+67}$ & $>0.6$ & $30.6_{-3.6}^{+0.1}$ & 0.85 \\ \vspace{0.1cm}
&10.31 - 11.26 & $2.14_{-0.08}^{+0.08}$ & $42_{-7}^{+26}$ & $61_{-20}^{+51}$ & $ >0.3$ & $18.4_{-1.2}^{+0.1}$ & 1.44 \\ \vspace{0.1cm}
&11.26 - 11.97 & $2.69_{-0.08}^{+0.09}$ & $101_{-6}^{+26}$ & $109_{-11}^{+26}$ & $>11.8$ & $21.2_{-1.0}^{+0.1}$ & 2.04 \\ \vspace{0.1cm}
&11.97 - 13.18 & $2.56_{-0.06}^{+0.06}$ & $88_{-7}^{+29}$ & $96_{-11}^{+32}$ &  $>1.4$ & $20.5_{-1.0}^{+0.1}$ & 0.78 \\ \vspace{0.1cm}
&13.18 - 14.74 & $2.31_{-0.07}^{+0.06}$ & $50_{-7}^{+29}$ & $62_{-15}^{+44}$ & $>0.9$ & $5.7_{-1.3}^{+0.1}$ & 2.21 \\ \vspace{0.1cm}
&14.74 - 17.33 & $2.40_{-0.08}^{+0.09}$ & $51_{-5}^{+21}$ & $57_{-8}^{+27}$ & $>2.1$ & $3.5_{-0.3}^{+0.1}$ & 2.66 \\ 
\hline \vspace{0.1cm}
\multirow{2}{*}{GRB 090510} &0.50 - 0.77 & $2.88_{-0.46}^{+0.69}$ & $876.44_{-385.00}^{+843.21}$ & $>1574$ & $>0.6$ & $20.1_{-5.1}^{+2.7}$ & 3.46 \\ \vspace{0.05cm}
 &0.77 - 0.91 & $3.62_{-1.04}^{+0.92}$ & $427.64_{-177.19}^{+339.71}$ & $>20454$ & $>0.1$ & $62.0_{-1.3}^{+1.2}$ & 29.9 \\ 
\hline
\multirow{2}{*}{GRB 110731A}&0.00 - 5.52   & $2.84_{-0.08}^{+0.08}$ & $116_{-12}^{+58}$ & $126_{-17}^{+60}$ & $>0.6$ & $4.7_{-0.2}^{+0.4}$ & 0.96 \\ \vspace{0.05cm}
&5.52 - 7.85   & $2.36_{-0.11}^{+0.11}$ & $109_{-24}^{+78}$ & $>110$ & $>0.3$ & $11.9_{-0.6}^{+0.2}$ & 0.83 \\ 
\hline
\multirow{3}{*}{GRB 130427A}&11.11 - 14.63  & $3.40_{-0.08}^{+0.09}$ & $57_{-1}^{+2}$ & $379_{-29}^{+26}$ & $>1.2$ & $42.2_{-0.2}^{+0.7}$ & 8.67 \\ \vspace{0.05cm}
&14.63 - 17.53  & $2.77_{-0.07}^{+0.07}$ & $22_{-1}^{+1}$ & $136_{-15}^{+14}$ & $>1.3$ & $15.3_{-0.1}^{+0.8}$ & 9.88 \\ \vspace{0.05cm}
&17.53 - 22.61  & $2.61_{-0.07}^{+0.07}$ & $9_{-1}^{+1}$ & $103_{-9}^{+10}$ & $>0.8$ & $6.8_{-0.7}^{+0.2}$ & 77.10 \\
\hline
\multirow{5}{*}{GRB 131108A}&0.29 - 2.65   & $2.51_{-0.10}^{+0.08}$ & $181_{-22}^{+83}$ & $205_{-36}^{+113}$ & $>0.3$ & $42.3_{-0.5}^{+0.4}$ & 0.66 \\ \vspace{0.05cm}
&2.65 - 4.95   & $2.60_{-0.09}^{+0.09}$ & $222_{-30}^{+116}$ & $261_{-51}^{+161}$ & $>0.5$ & $7.8_{-0.4}^{+0.4}$ & 1.07 \\ \vspace{0.05cm}
&4.95 - 6.37   & $2.52_{-0.10}^{+0.10}$ & $174_{-28}^{+105}$ & $233_{-67}^{+200}$ & $>0.5$ & $6.9_{-0.4}^{+0.5}$ & 1.18 \\ \vspace{0.05cm}
&6.38 - 7.95   & $2.40_{-0.10}^{+0.11}$ & $99_{-18}^{+66}$ & $140_{-46}^{+134}$ & $>0.5$ & $6.9_{-0.3}^{+0.6}$ & 0.75 \\ \vspace{0.05cm}
&7.95 - 11.97  & $2.16_{-0.09}^{+0.12}$ & $40_{-9}^{+23}$ & $244_{-181}^{+225}$ & $>0.3$ & $5.2_{-0.7}^{+0.2}$ & 0.59 \\
\hline
\multirow{3}{*}{GRB 160509A}&14.77 - 17.53 & $2.71_{-0.06}^{+0.05}$ & $147_{-7}^{+41}$ & $158_{-14}^{+42}$ & $0.2_{-0.04}^{+0.1}$ & $24.6_{-0.6}^{+0.1}$ & 0.28 \\ \vspace{0.05cm}
&17.53 - 18.28 & $2.03_{-0.03}^{+0.06}$ & $105_{-9}^{+13}$ & $1039_{-378}^{+428}$ & $0.11_{-0.02}^{+0.02}$ & $45.1_{-0.2}^{+0.2}$ & 0.35 \\ \vspace{0.05cm}
&18.28 - 24.26 & $2.23_{-0.07}^{+0.07}$ & $72_{-9}^{+56}$ & $105_{-35}^{+83}$ & $0.1_{-0.02}^{+0.03}$ & $7.9_{-0.1}^{+0.5}$ & 2.33 \\
\hline
\multirow{6}{*}{GRB 160625B}&188.71 - 189.72 & $3.39_{-0.27}^{+0.25}$ & $>430$ & $1019_{-529}^{+48}$ & $0.1_{-0.03}^{+0.05}$ & $68.8_{-0.5}^{+0.2}$ & 0.95 \\
&189.72 - 191.25 & $3.32_{-0.25}^{+0.22}$ & $>232$ & $758_{-473}^{+334}$ & $>0.1$ & $69.2_{-0.1}^{+0.2}$ & 0.50 \\
&191.25 - 193.83 & $3.54_{-0.09}^{+0.12}$ & $>200$ & $253_{-39}^{+153}$ & $>0.5$ & $28.7_{-0.1}^{+0.3}$ & 2.77 \\
&193.83 - 198.00 & $3.95_{-0.08}^{+0.08}$ & $315_{-14}^{+91}$ & $334_{-23}^{+72}$ & $>0.5$ & $43.2_{-0.4}^{+0.9}$ & 0.70 \\
&198.00 - 201.02 & $3.02_{-0.06}^{+0.06}$ & $343_{-14}^{+79}$ & $368_{-30}^{+82}$ & $0.08_{-0.02}^{+0.02}$ & $61.53_{-0.5}^{+0.2}$ & 0.60 \\
&201.02 - 203.75 & $3.55_{-0.06}^{+0.06}$ & $285_{-19}^{+127}$ & $297_{-22}^{+102}$ & $>0.5$ & $40.3_{-0.6}^{+0.1}$ & 0.68 \\
\hline
\multirow{3}{*}{GRB 170214A}&0.00 - 62.83 & $3.14_{-0.11}^{+0.11}$ & $190_{-13}^{+95}$ & $197_{-14}^{+79}$ & $0.1_{-0.03}^{+0.06}$ & $4.54_{-0.25}^{+0.11}$  & {0.60} \\
&62.83 - 65.52 & $2.31_{-0.10}^{+0.10}$ & $210_{-27}^{+103}$ & $248_{-50}^{+175}$ & $0.2_{-0.06}^{+0.1}$ & $9.16_{-0.42}^{+0.16}$ & 0.34 \\
&65.52 - 71.76 & $2.71_{-0.11}^{+0.11}$ & $148_{-22}^{+110}$ & $188_{-47}^{+148}$ & $>0.3$ & $4.51_{-0.43}^{+0.12}$ & 0.62 \\
\hline
\end{tabular}
\label{table:timeresolvedSample}
\caption{This table presents the synchrotron parameters obtained from our analysis of the full time-resolved GRB sample, listed for each analyzed time interval. For each time bin, the table includes the derived synchrotron parameters, the bolometric flux, and the maximum energy photon.}
\end{table*}

\begin{figure*}[ht]
    \centering
    \includegraphics[width=0.65\linewidth, height=8cm]{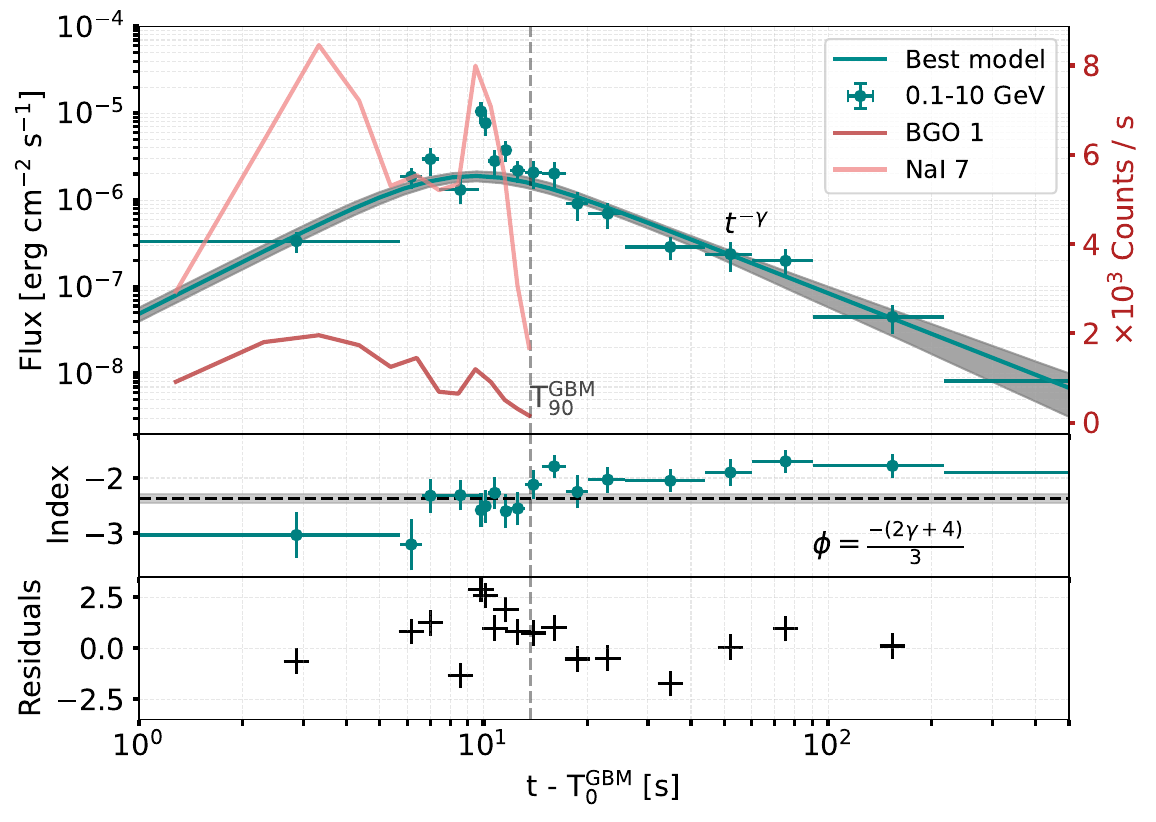}
    \includegraphics[width=0.33\linewidth]{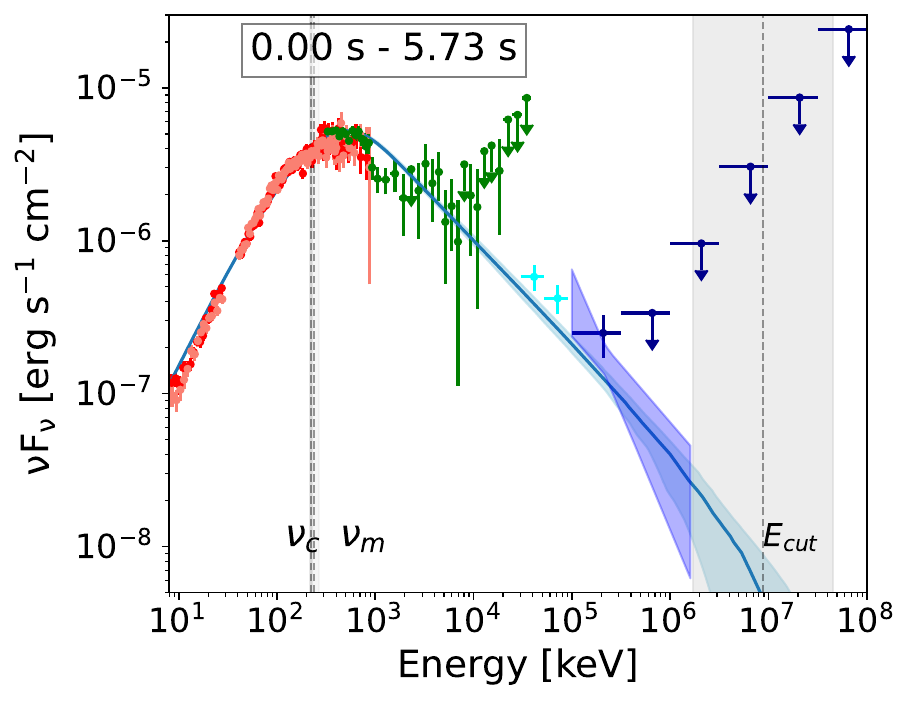}
    \includegraphics[width=0.33\linewidth]{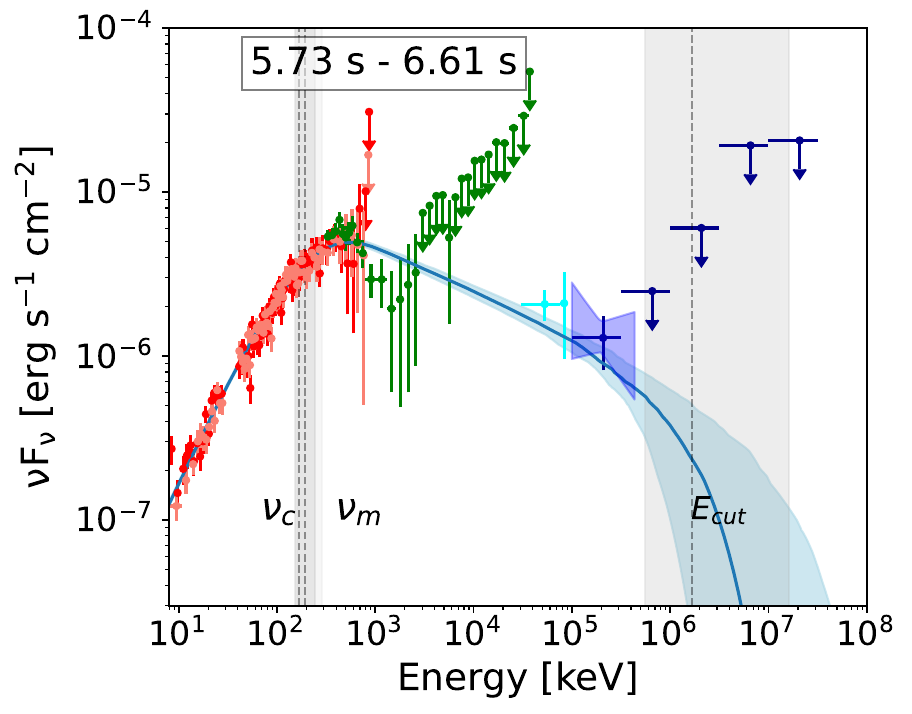}
    \includegraphics[width=0.33\linewidth]{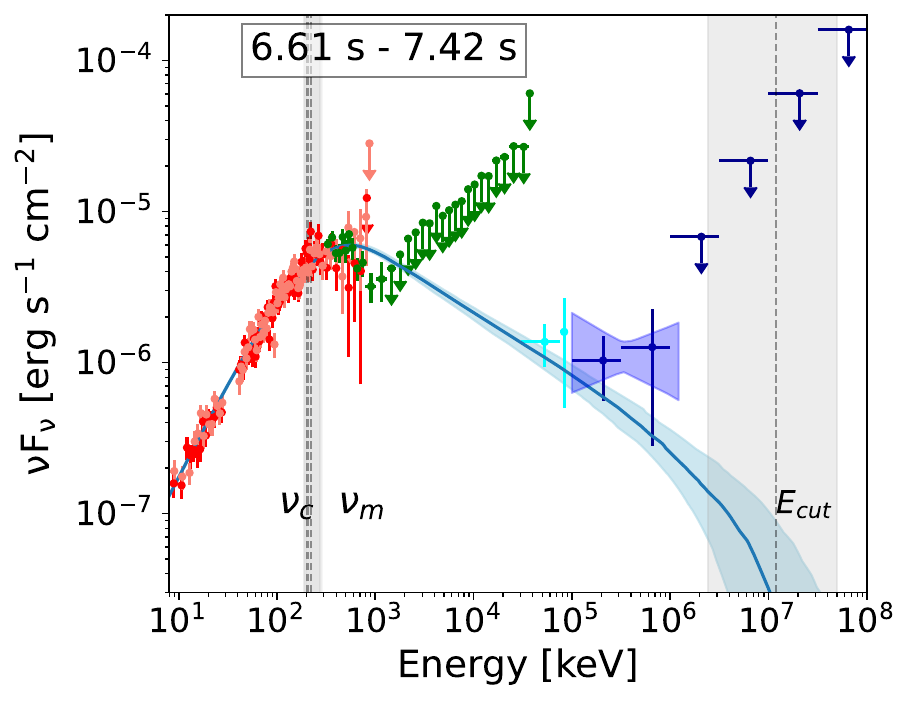}
    \includegraphics[width=0.33\linewidth]{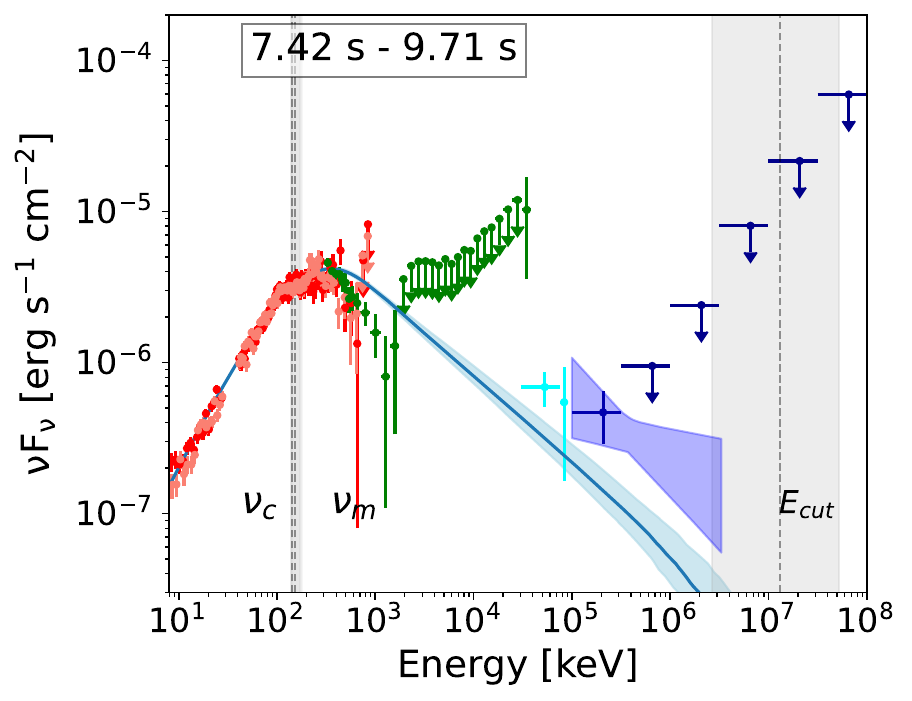}
    \includegraphics[width=0.33\linewidth]{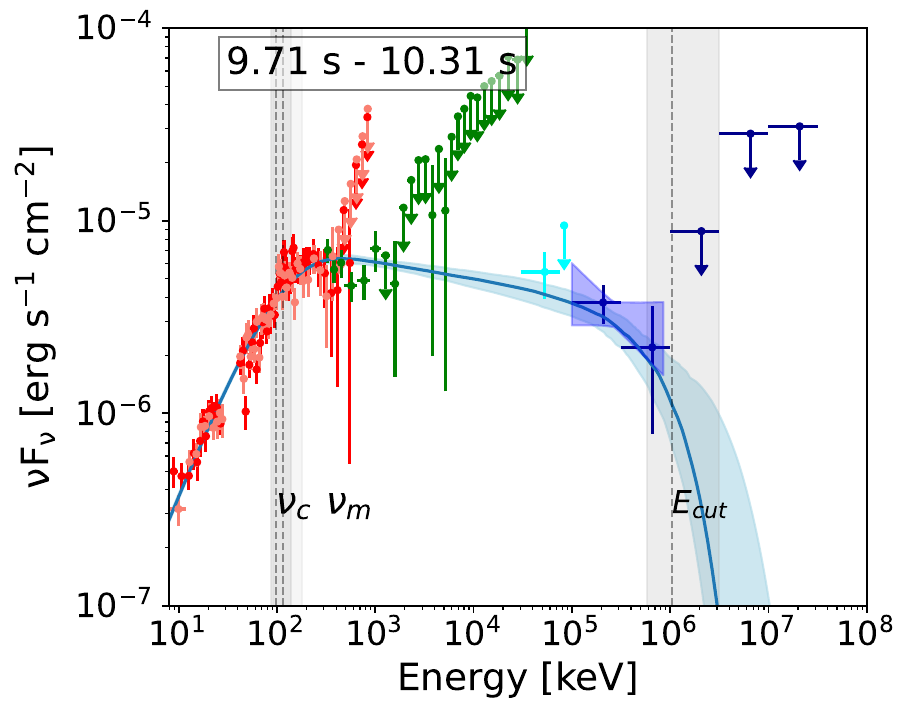}
    \includegraphics[width=0.33\linewidth]{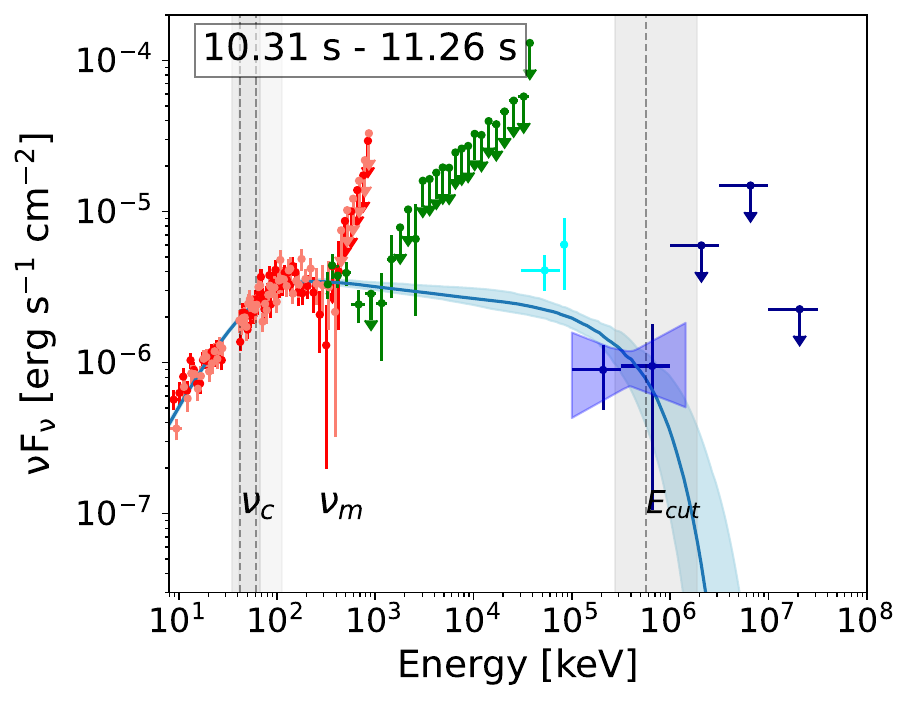}
    \includegraphics[width=0.33\linewidth]{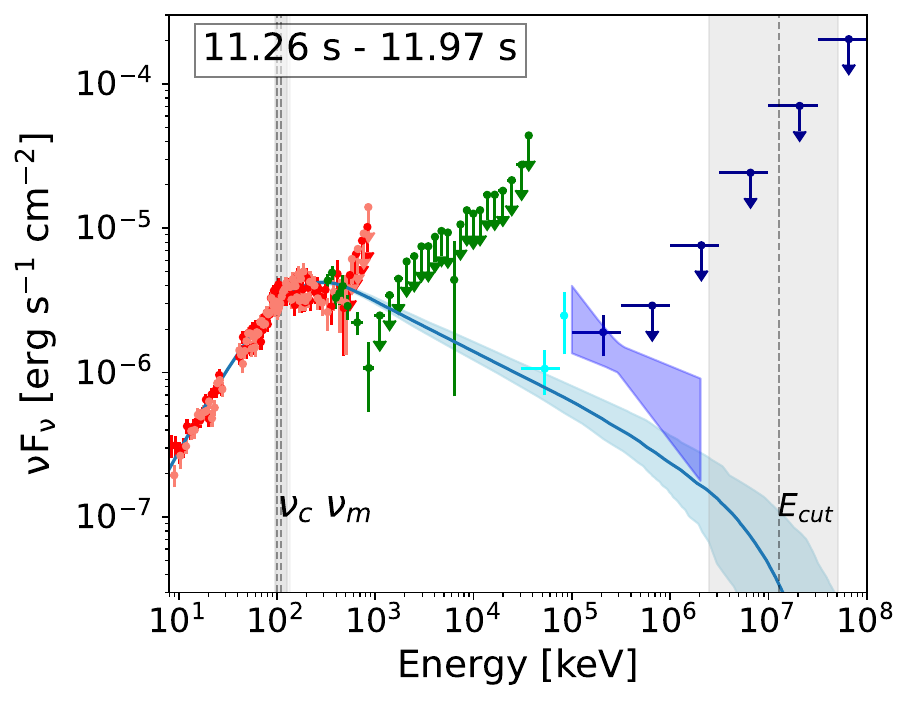}
    \includegraphics[width=0.33\linewidth]{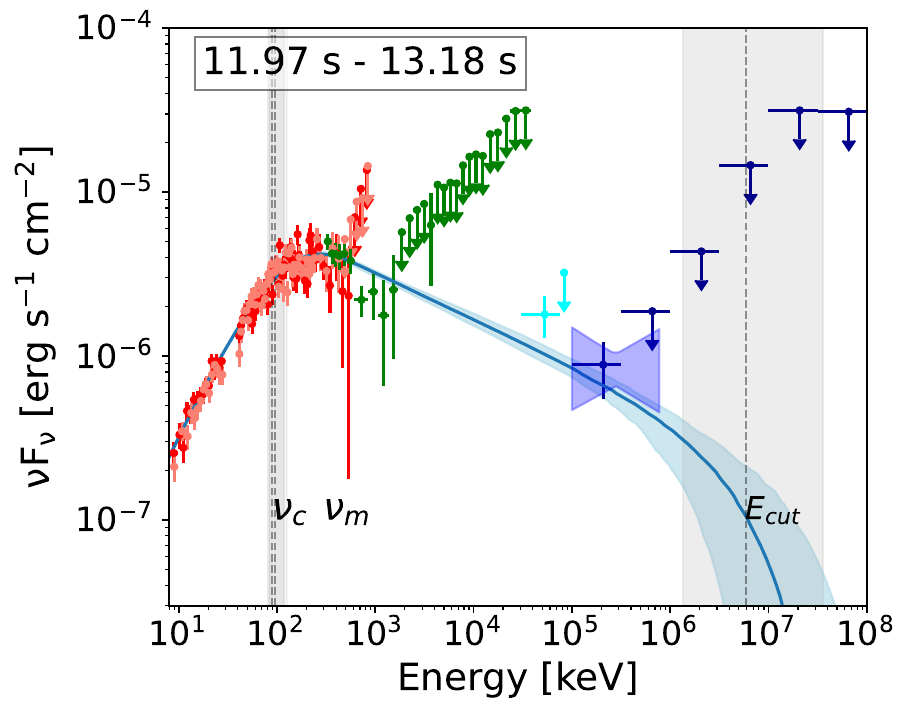}
    \includegraphics[width=0.33\linewidth]{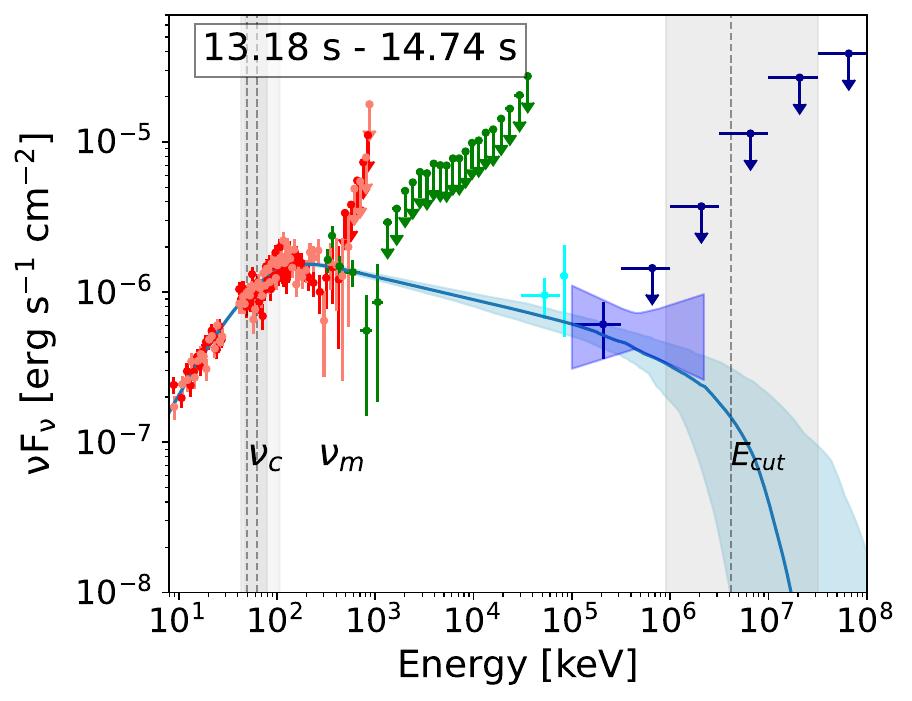}
    \includegraphics[width=0.33\linewidth]{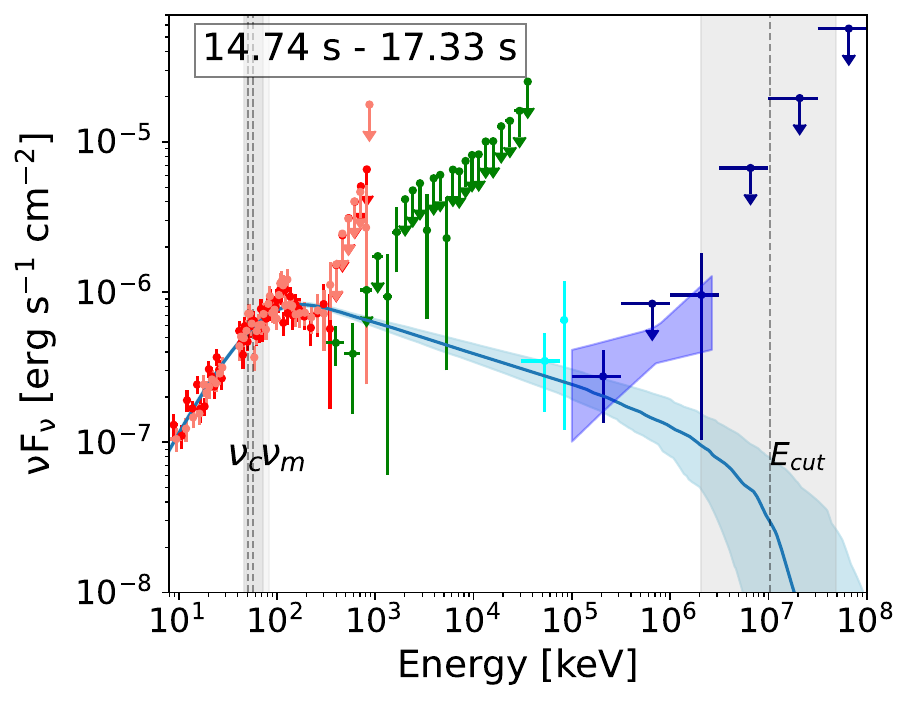}
    \caption{Spectrum of GRB 090926A.}
    \label{fig:090926A}
\end{figure*}

\begin{figure*}[ht]
    \centering \caption{Spectrum 090510.}
    \label{fig:090510}
    \includegraphics[width=0.65\linewidth, height=8cm]{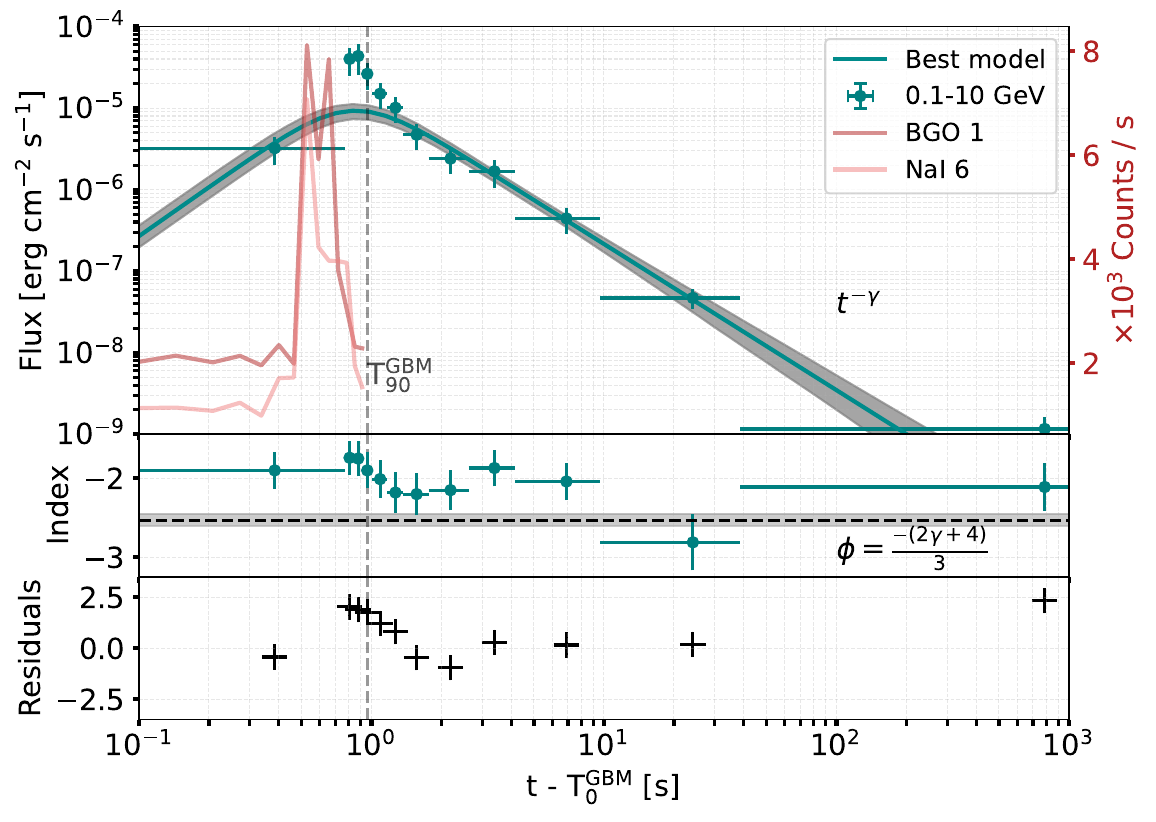} \newline
    \includegraphics[width=0.33\linewidth]{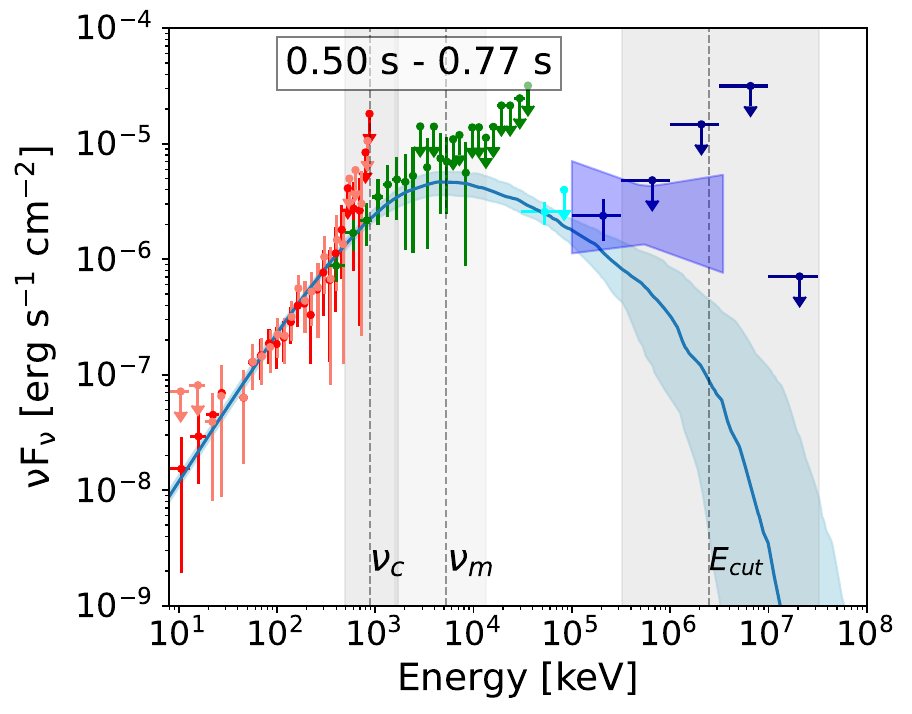}
    \includegraphics[width=0.33\linewidth]{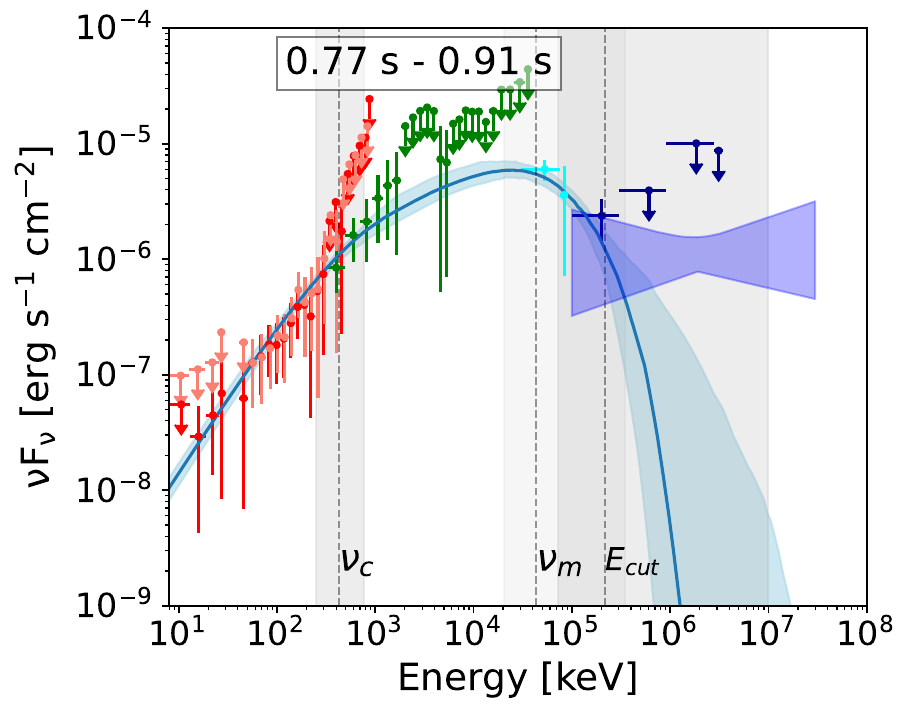}
\end{figure*}

\begin{figure*}[ht]
    \centering \caption{Spectrum GRB 110731A.}
    \label{fig:110731A}
    \includegraphics[width=0.65\linewidth, height=8cm]{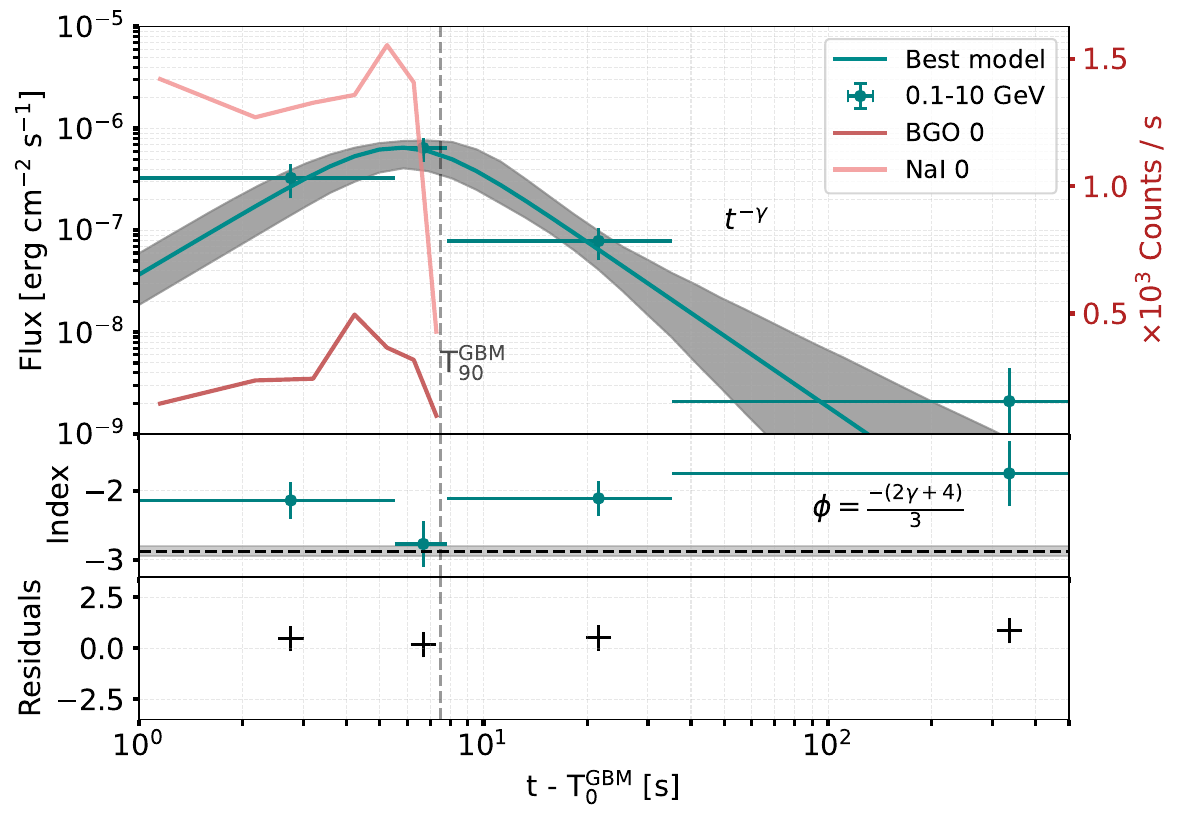} \newline
    \includegraphics[width=0.33\linewidth]{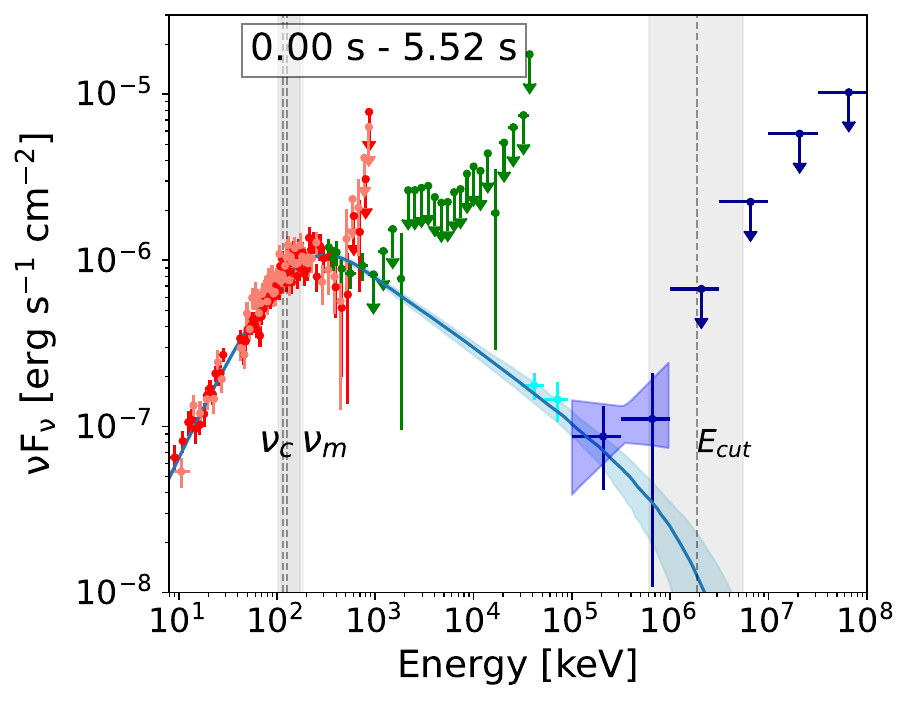}
    \includegraphics[width=0.33\linewidth]{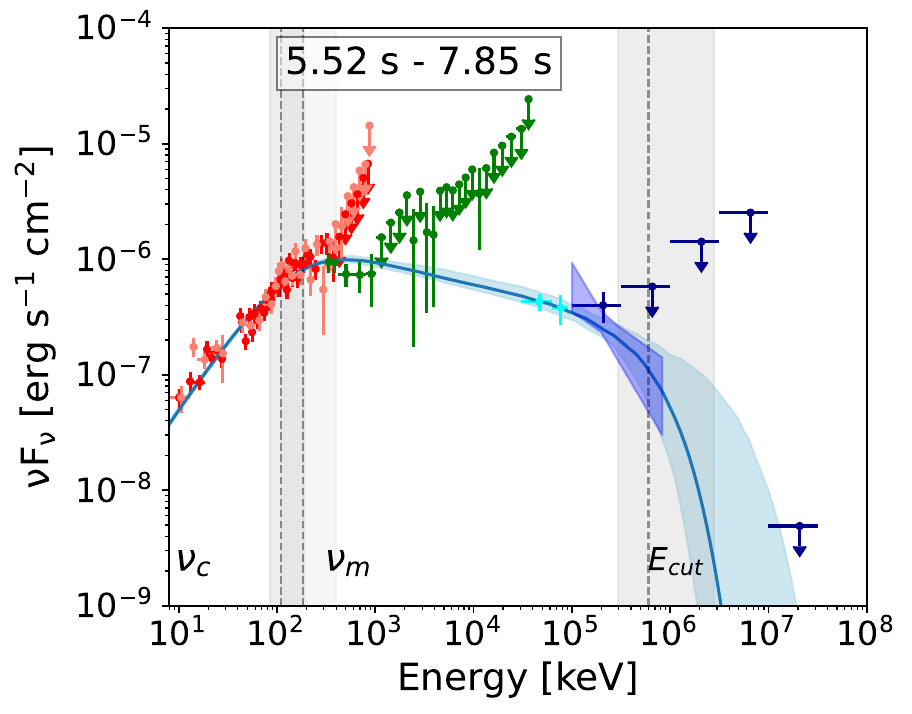}
\end{figure*}

\begin{figure*}[ht]
    \centering \caption{Spectrum GRB 130427A.}
    \includegraphics[width=0.66\linewidth]{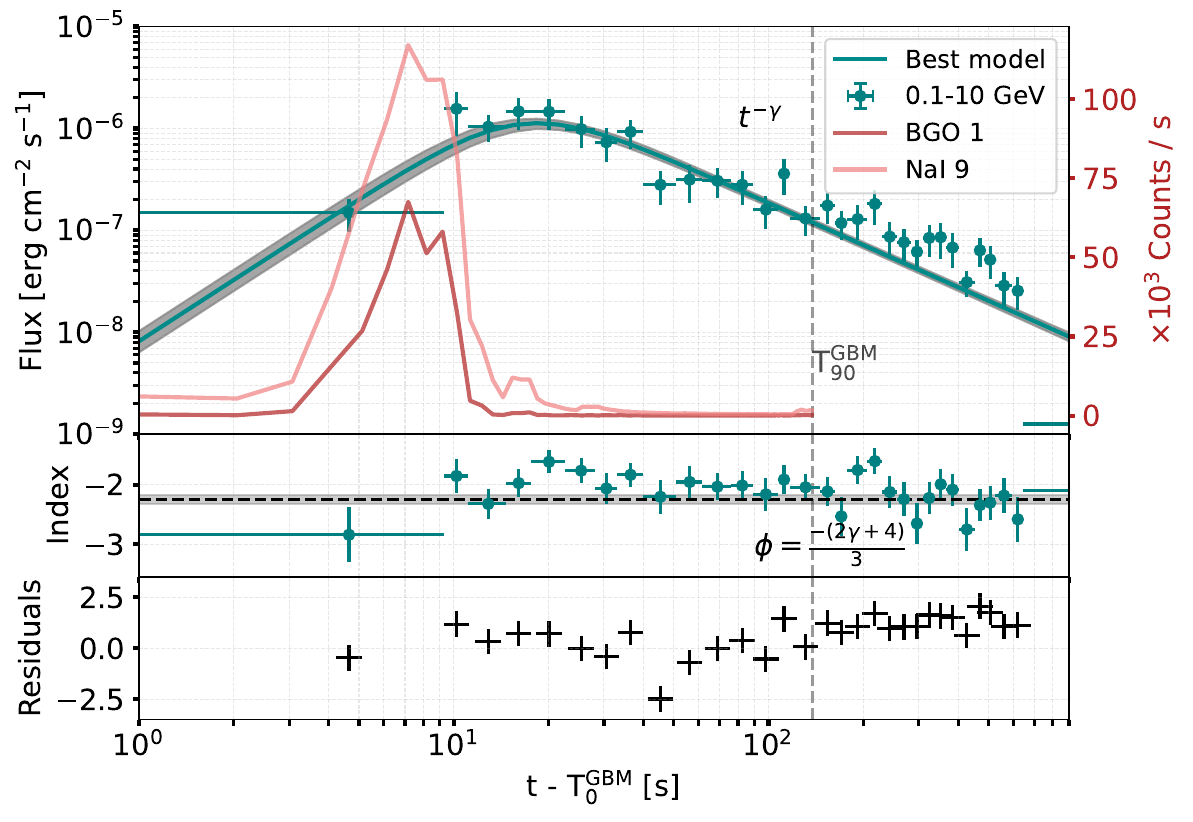} \newline
    \includegraphics[width=0.33\linewidth]{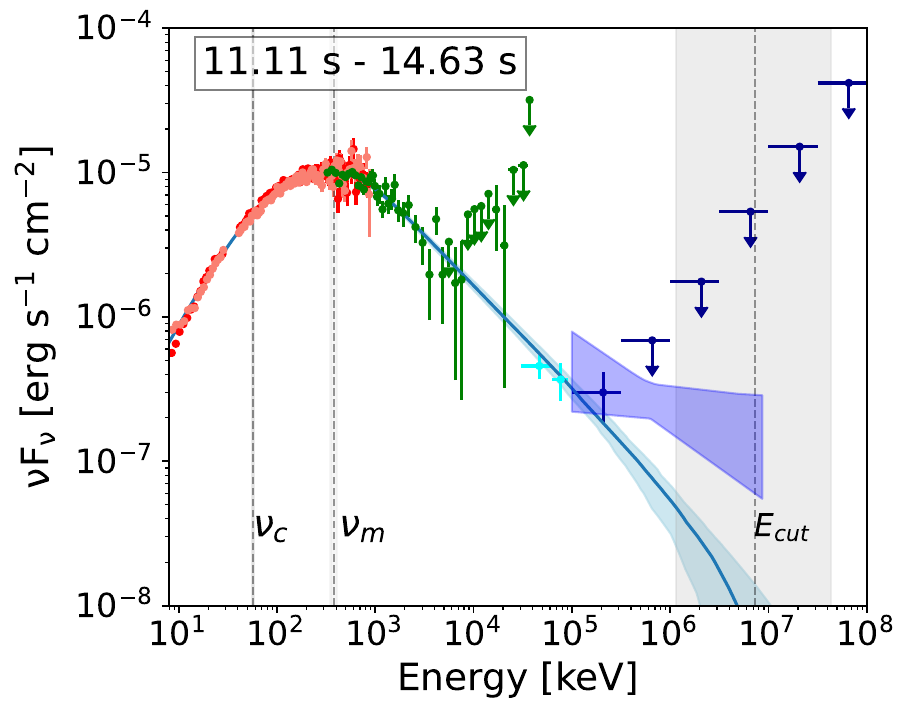}
    \includegraphics[width=0.33\linewidth]{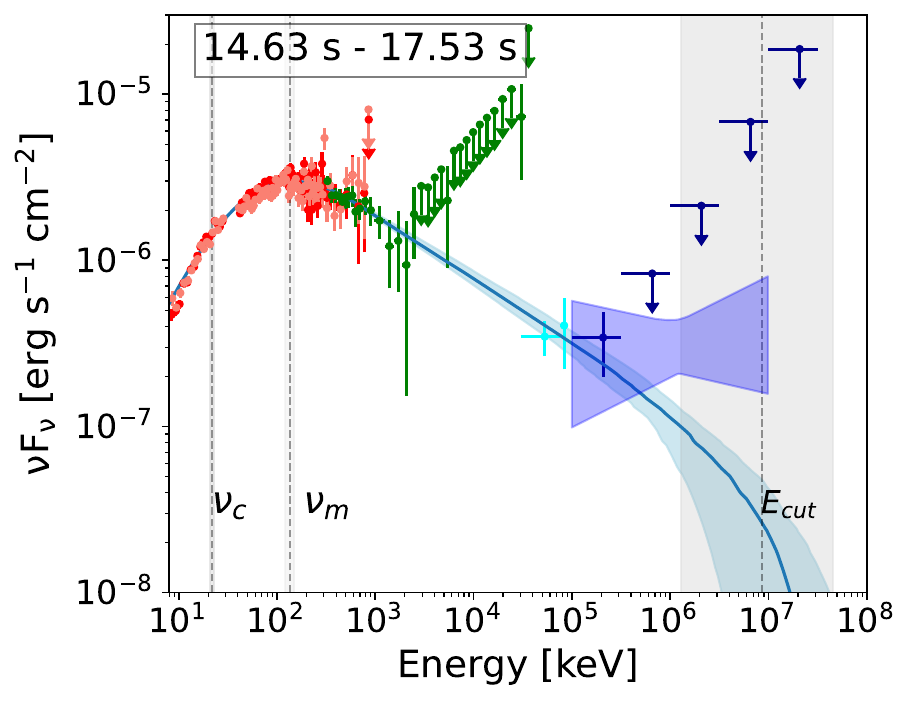}
    \includegraphics[width=0.33\linewidth]{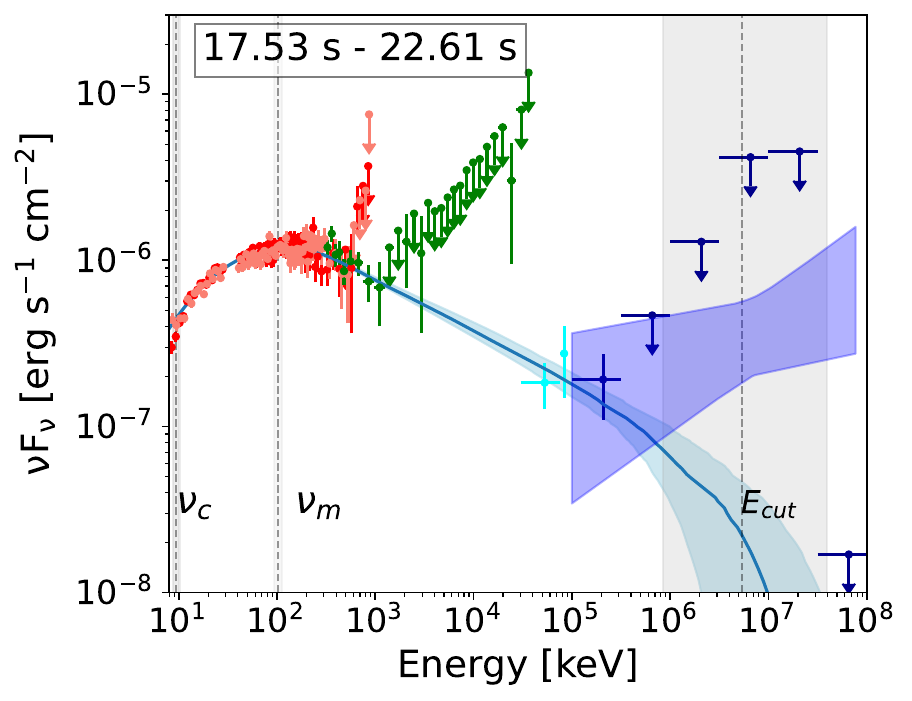}
\end{figure*}

\begin{figure*}[ht]
    \centering \caption{Spectrum GRB 131108A.}
    \includegraphics[width=0.66\linewidth]{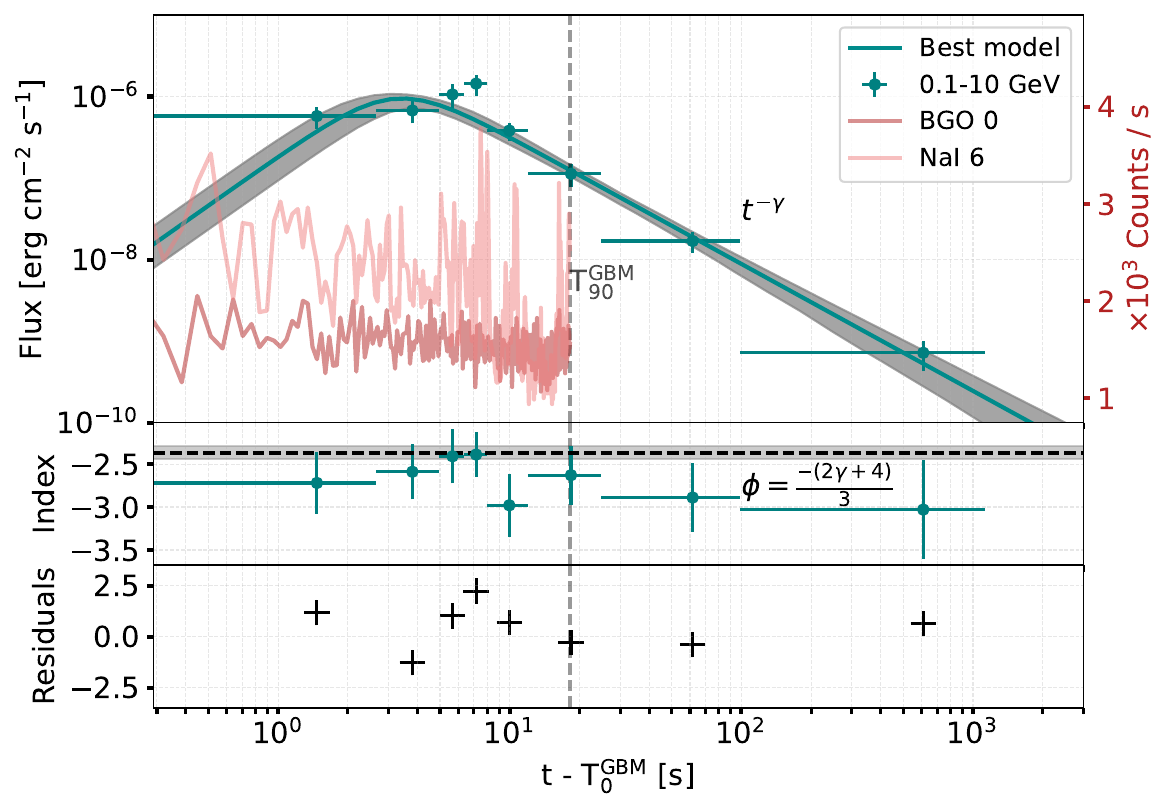} \newline
    \includegraphics[width=0.33\linewidth]{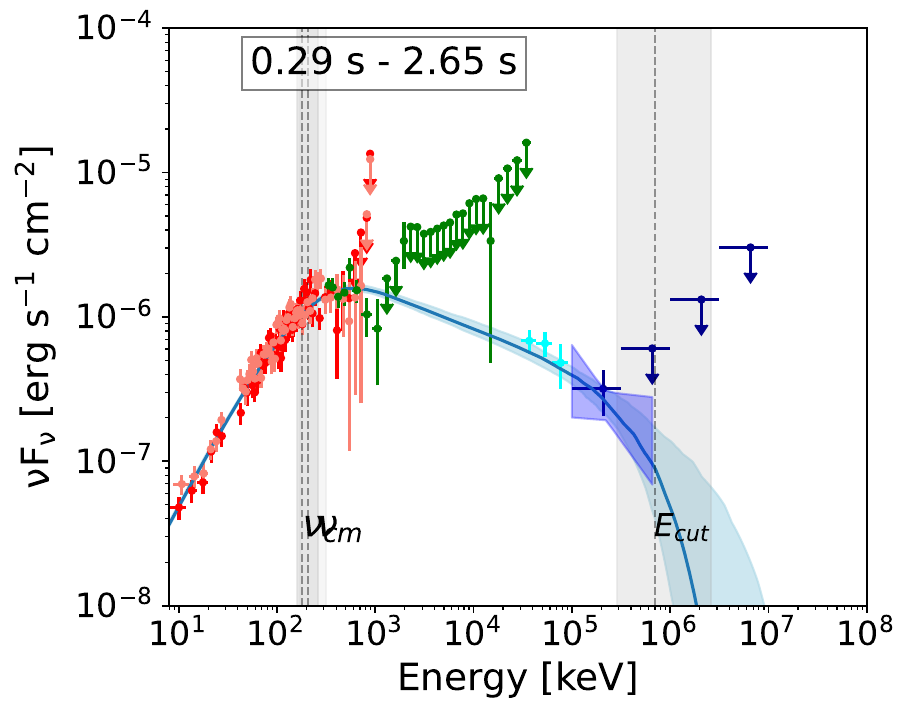}
    \includegraphics[width=0.33\linewidth]{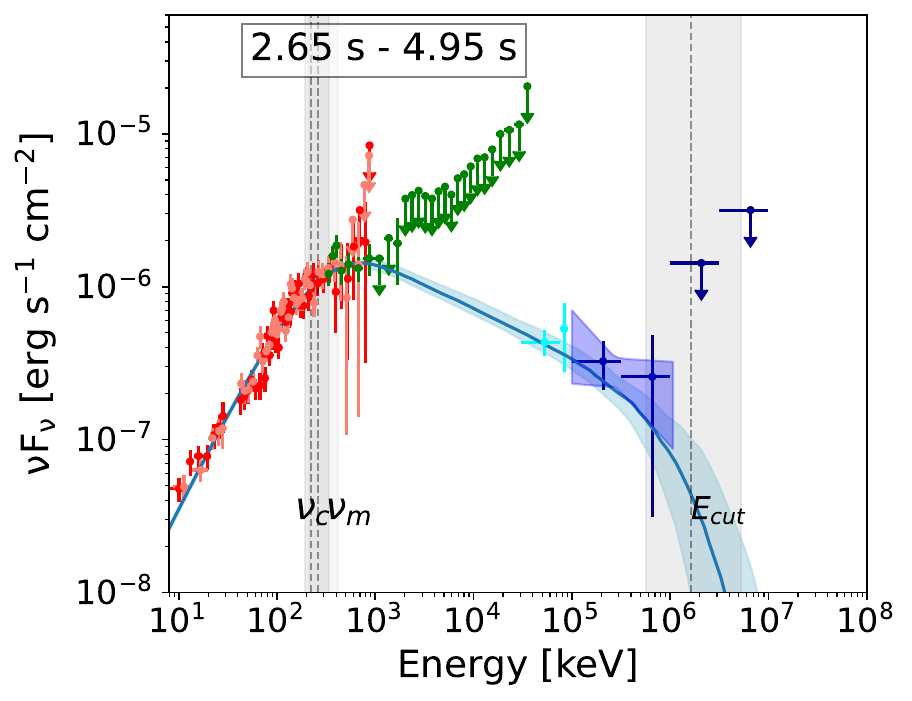}
    \includegraphics[width=0.33\linewidth]{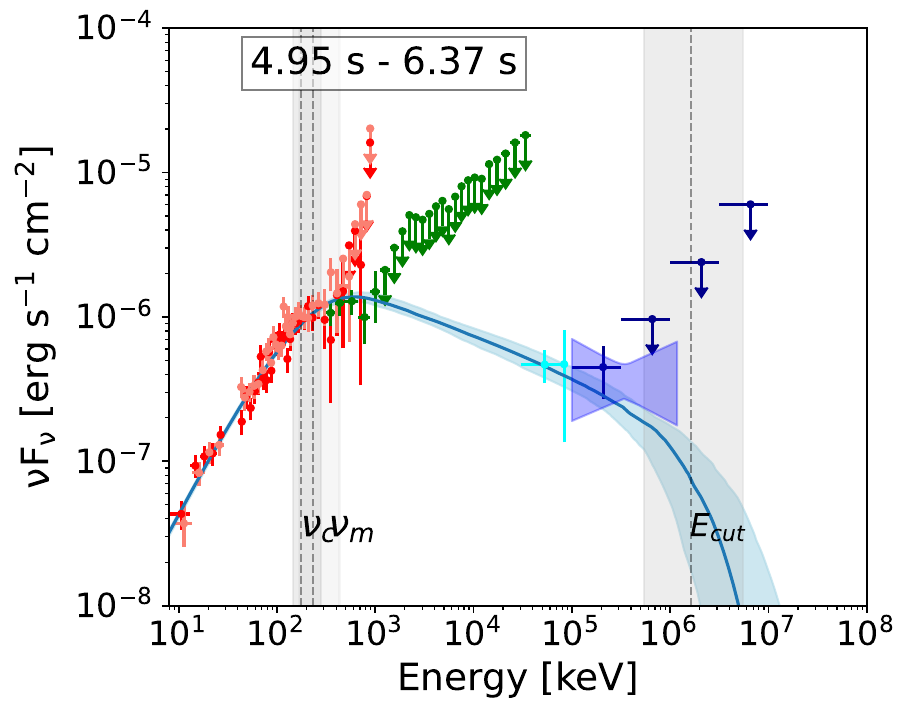}
    \includegraphics[width=0.33\linewidth]{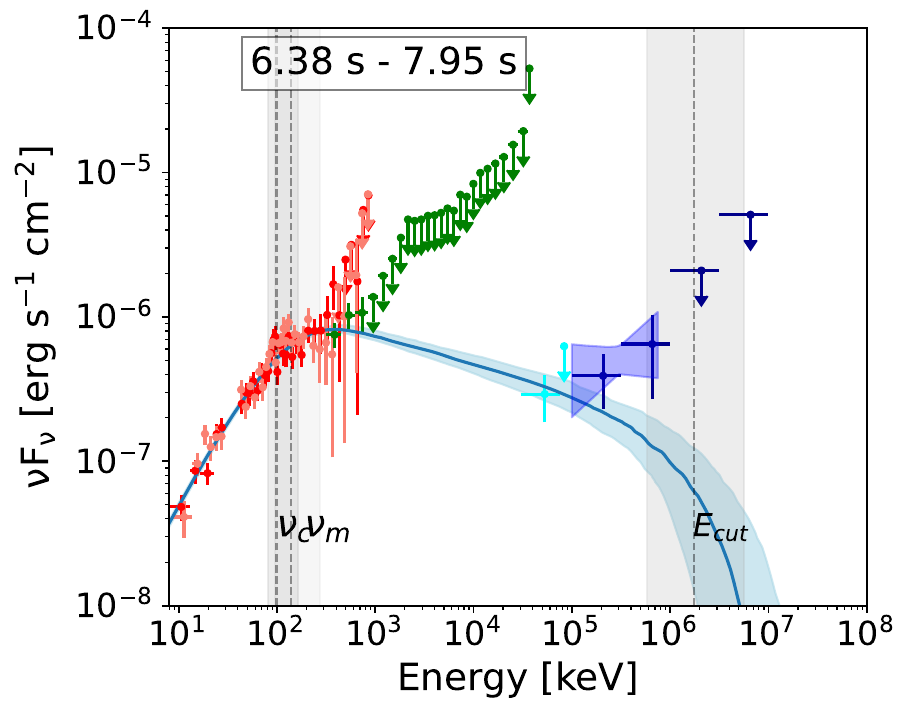}
    \includegraphics[width=0.33\linewidth]{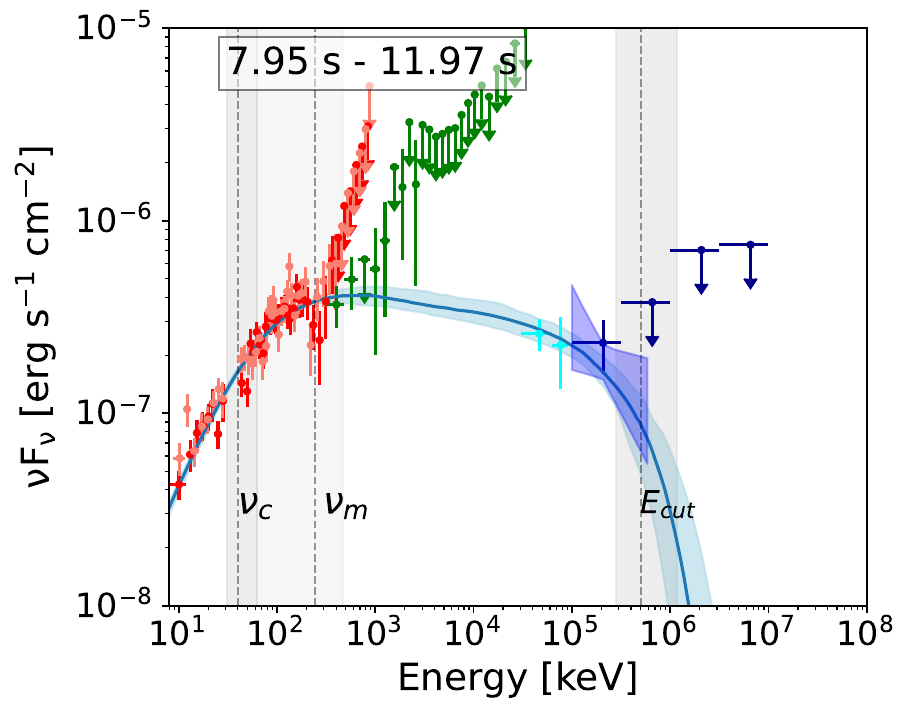}
\end{figure*}

\begin{figure*}[h]
    \centering \caption{Spectrum GRB 160509A.}
    \includegraphics[width=0.66\linewidth]{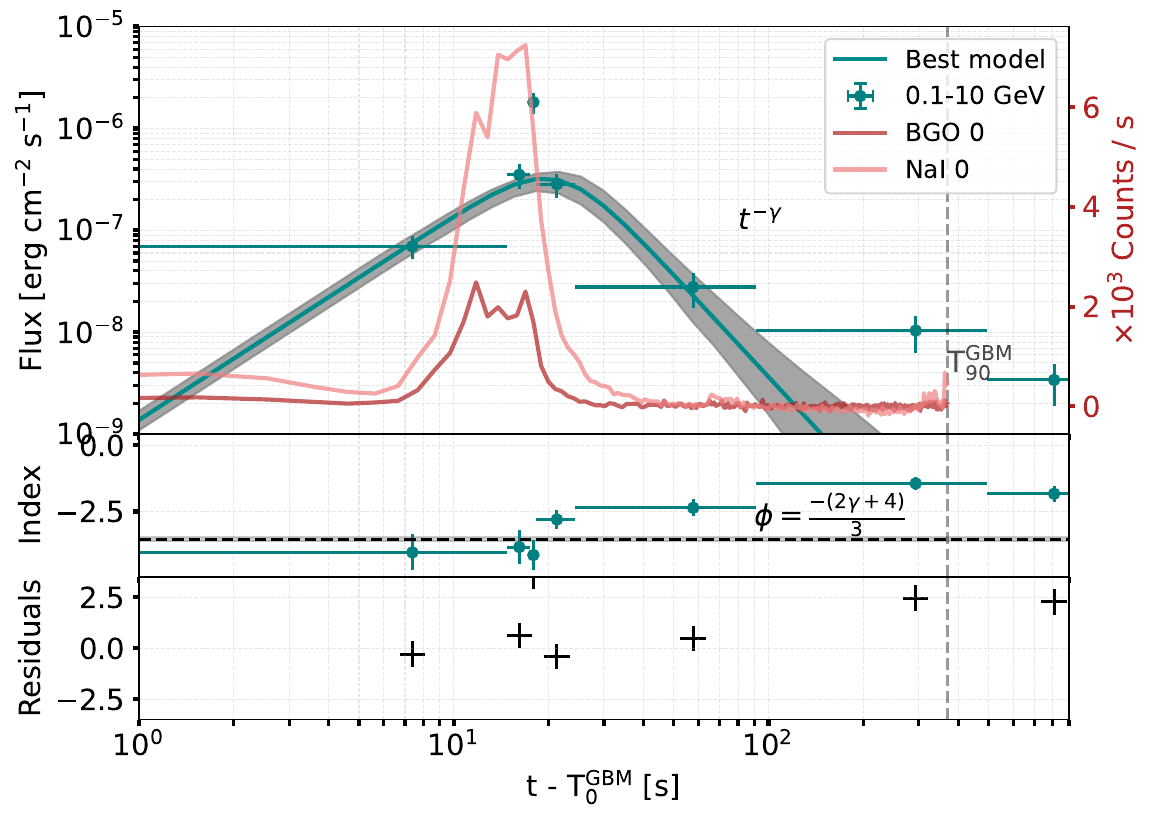} \newline
    \includegraphics[width=0.33\linewidth]{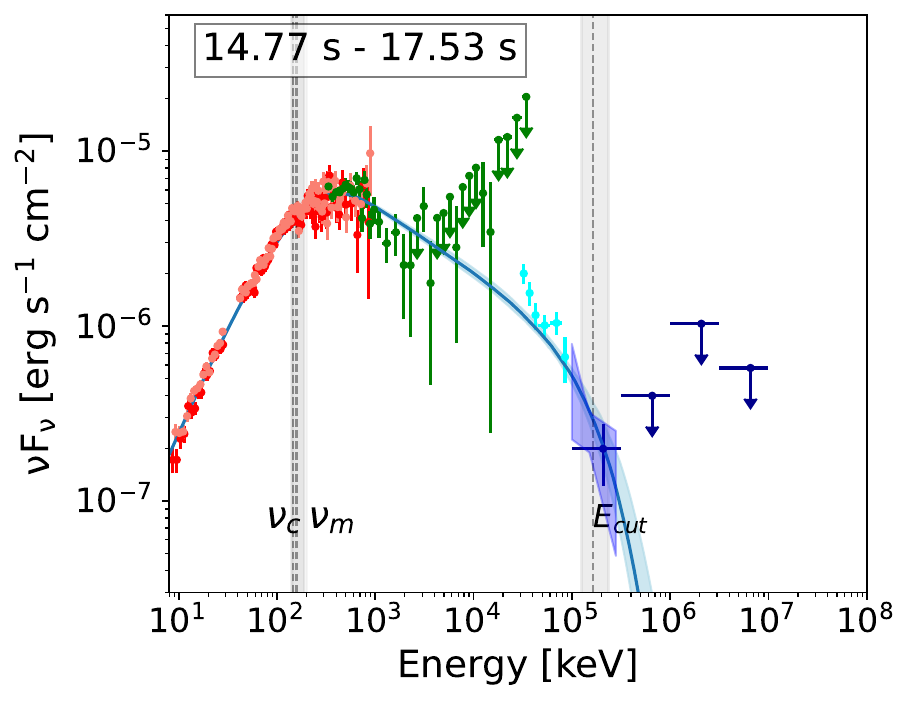}
    \includegraphics[width=0.33\linewidth]{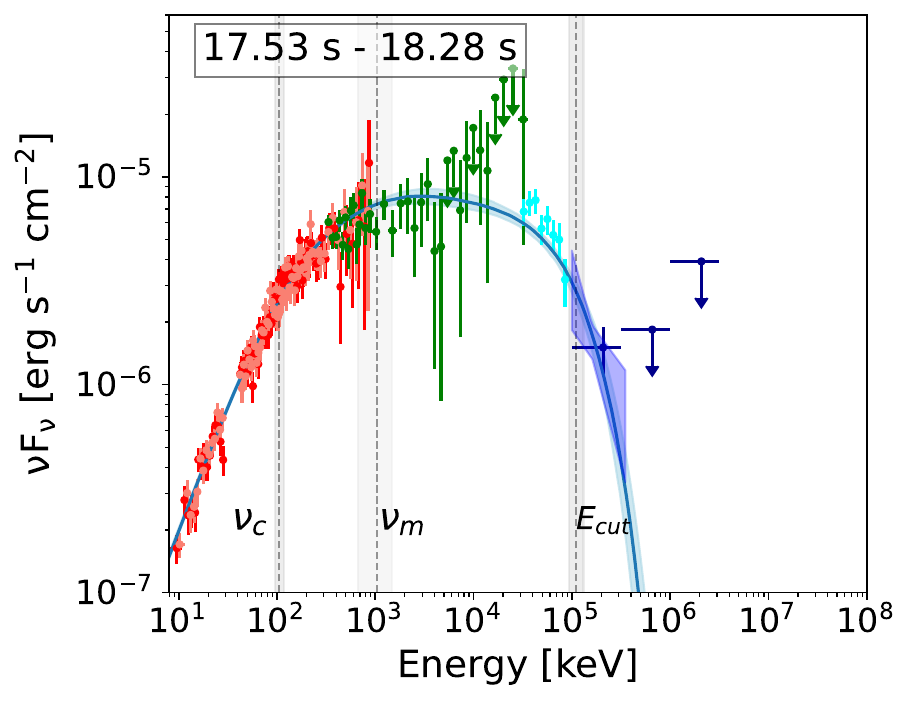}
    \includegraphics[width=0.33\linewidth]{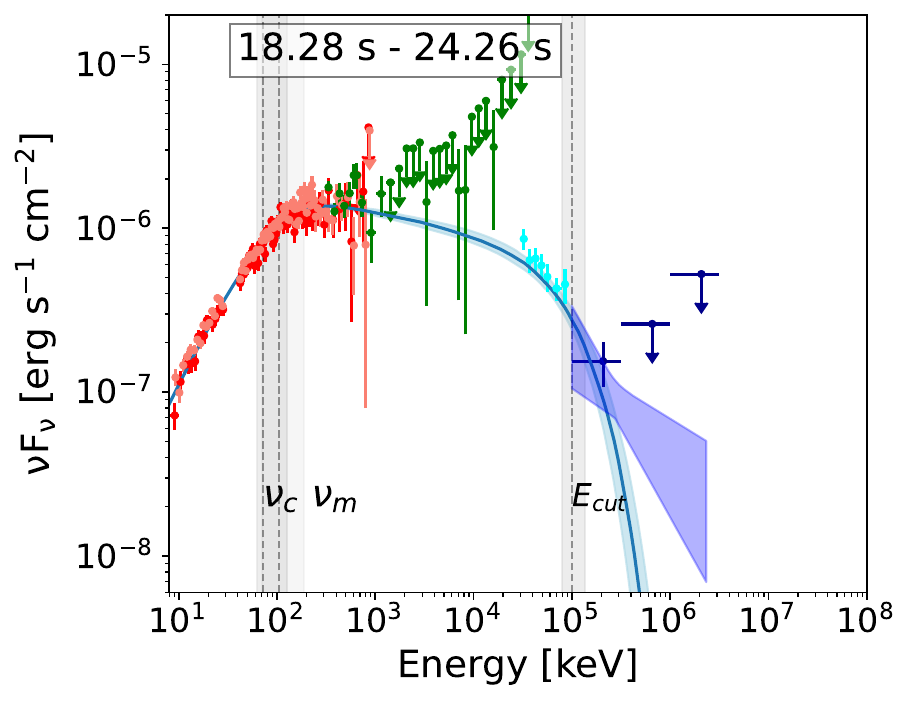}
\end{figure*}

\begin{figure*}[h]
    \centering \caption{GRB 160625B.}
    \label{fig:160625B}
    \includegraphics[width=0.66\linewidth]{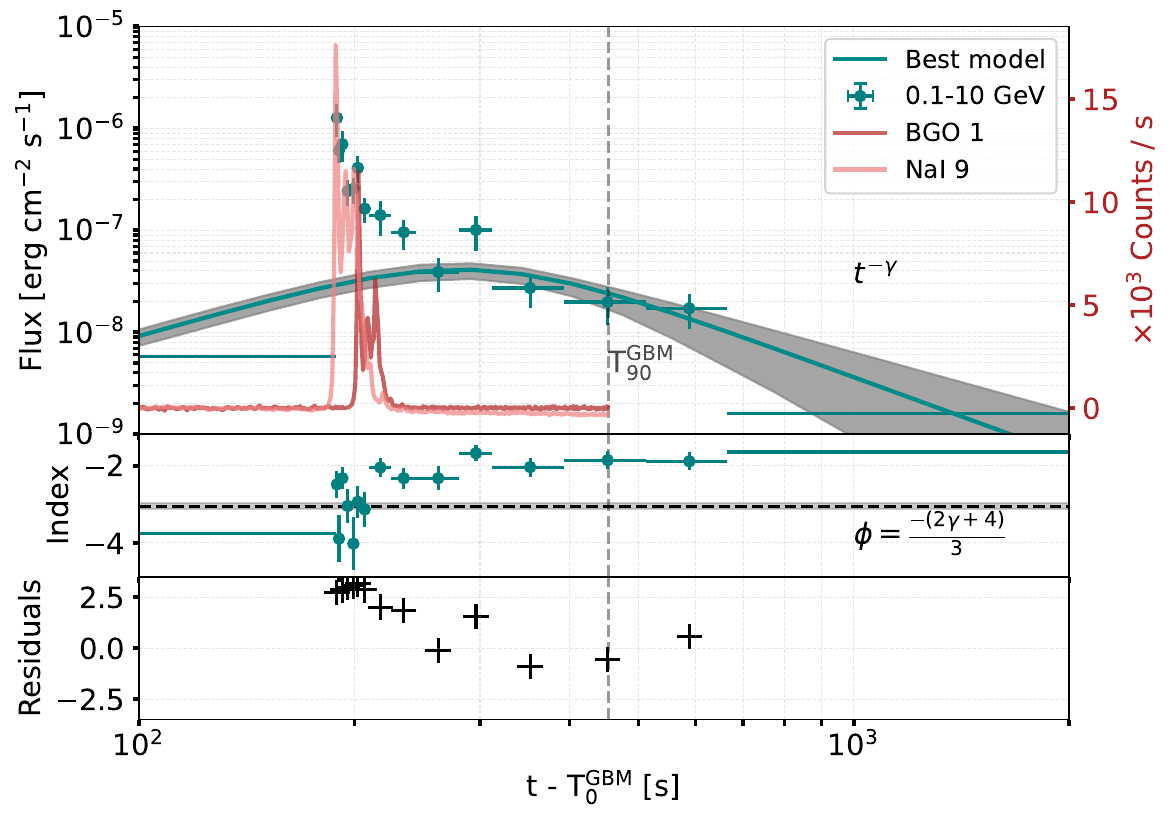} \newline
    \includegraphics[width=0.33\linewidth]{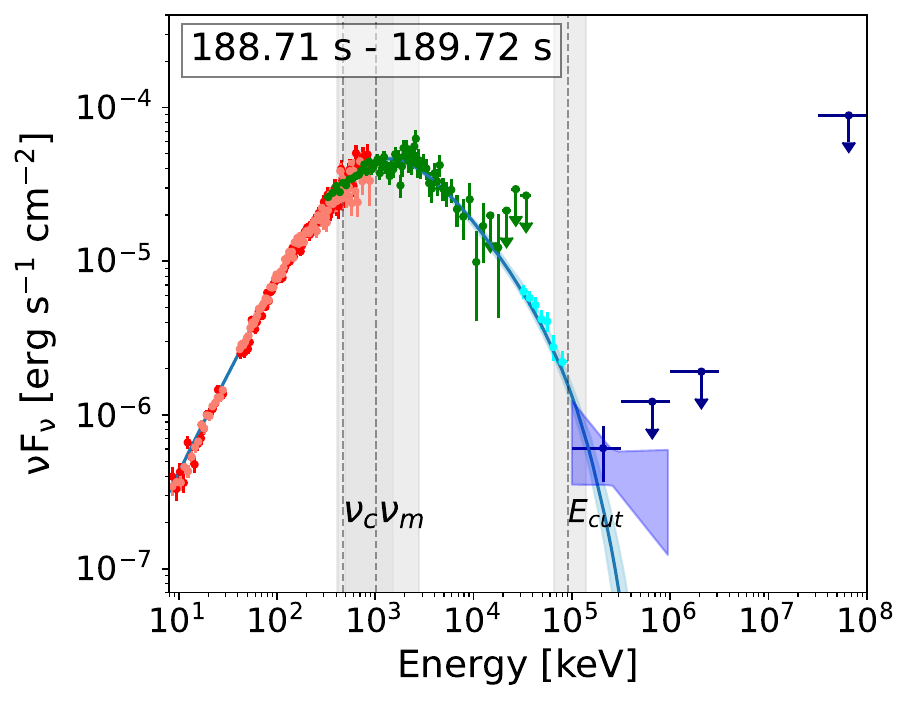}
    \includegraphics[width=0.33\linewidth]{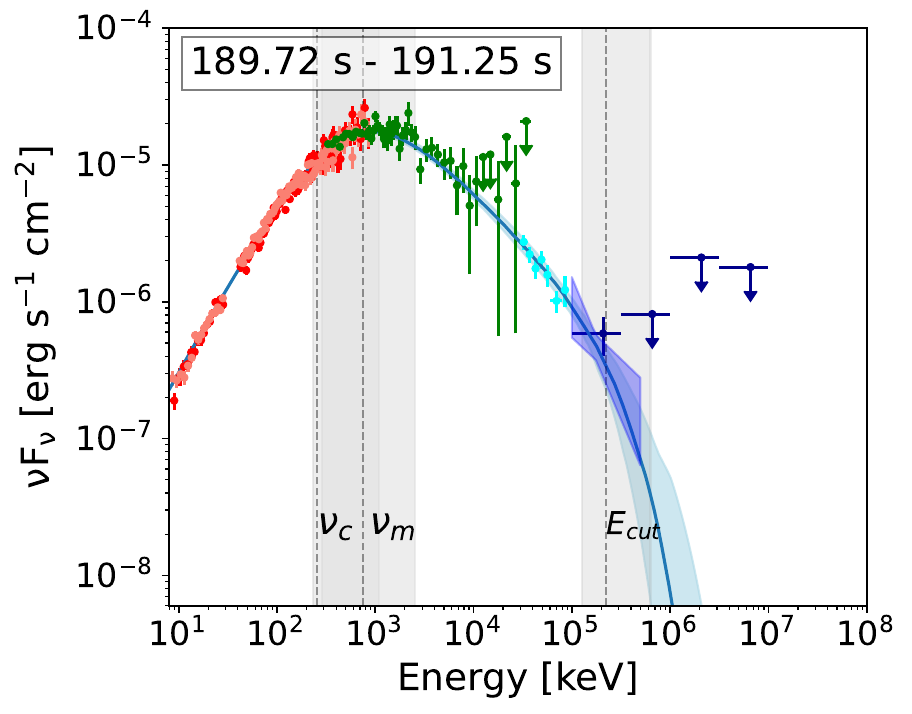}
    \includegraphics[width=0.33\linewidth]{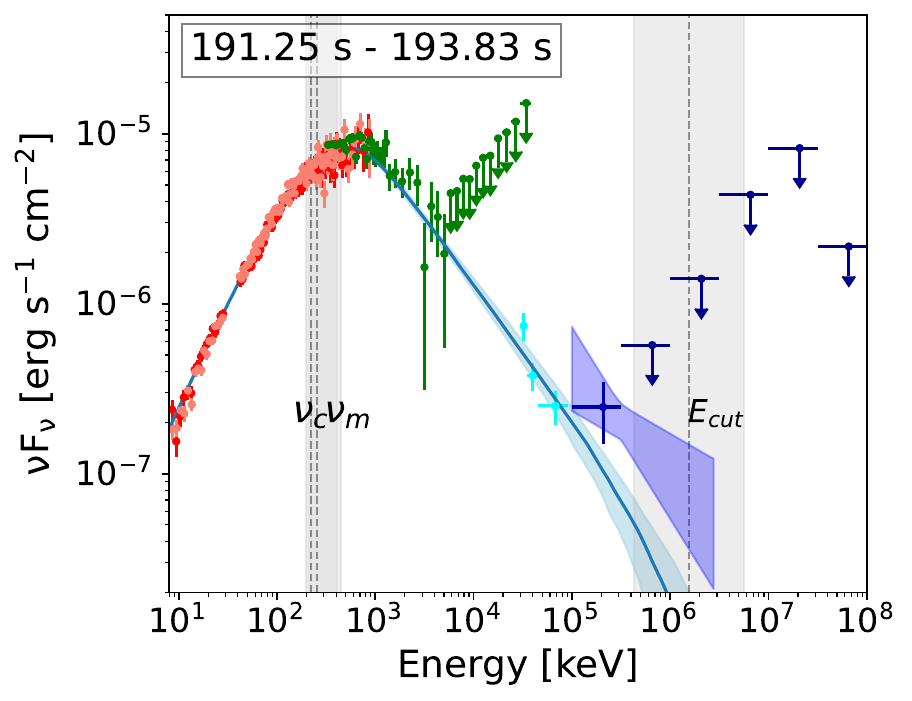}
    \includegraphics[width=0.33\linewidth]{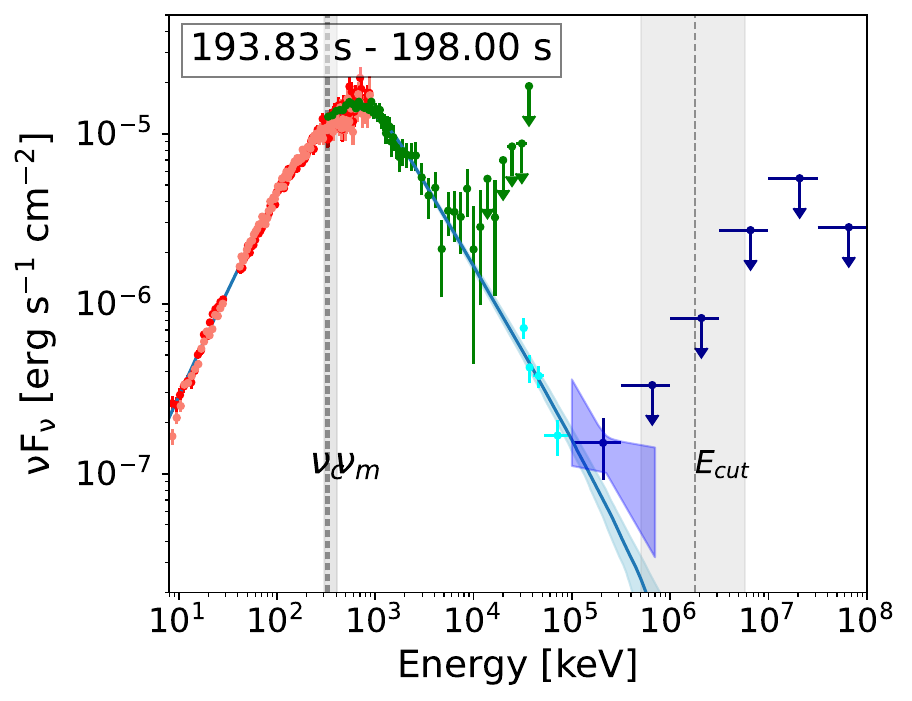}
    \includegraphics[width=0.33\linewidth]{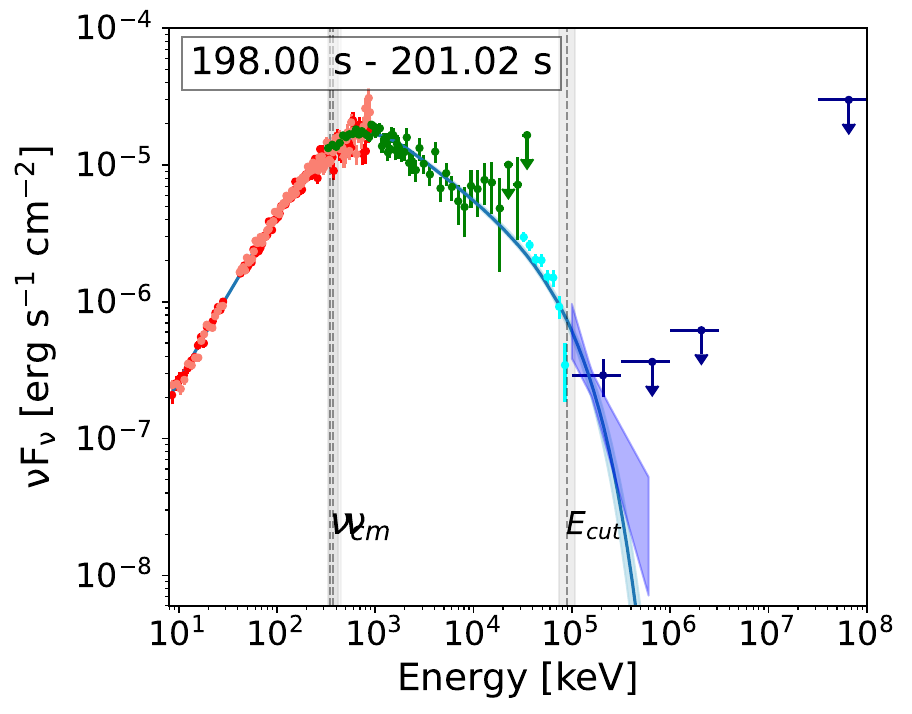}
    \includegraphics[width=0.33\linewidth]{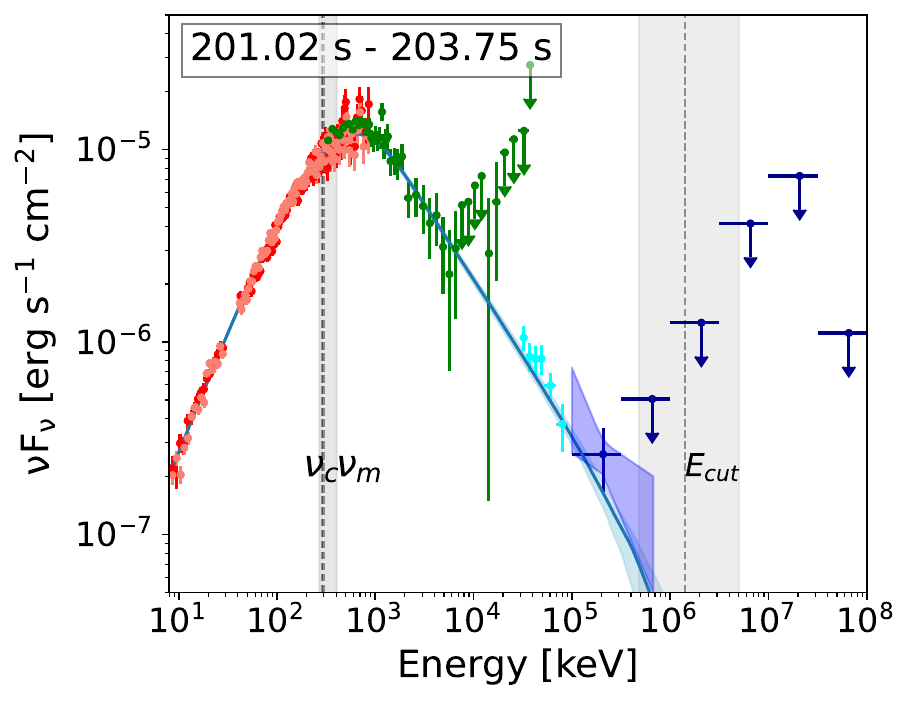}
\end{figure*}

\begin{figure*}[h]
    \centering \caption{GRB 170214A.}
    \label{fig:170214_1}
    \includegraphics[width=0.66\linewidth]{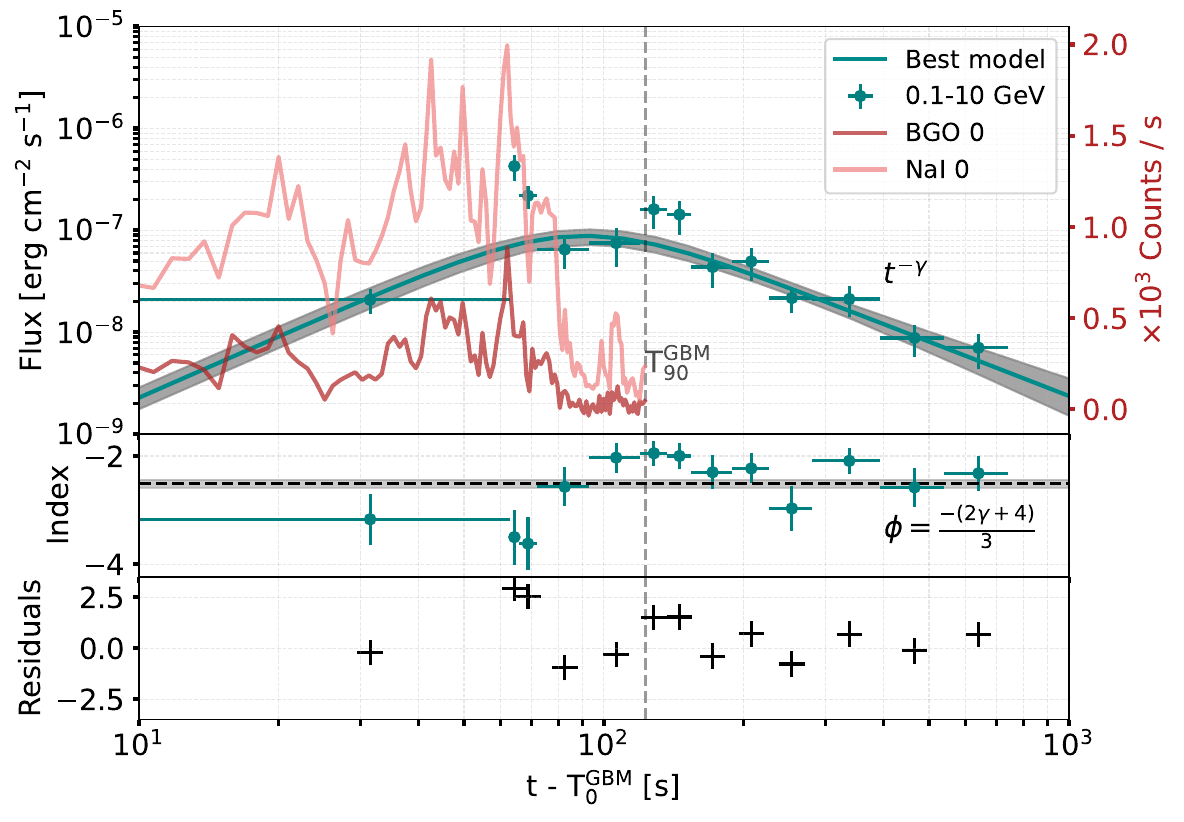} \newline
    \includegraphics[width=0.33\linewidth]{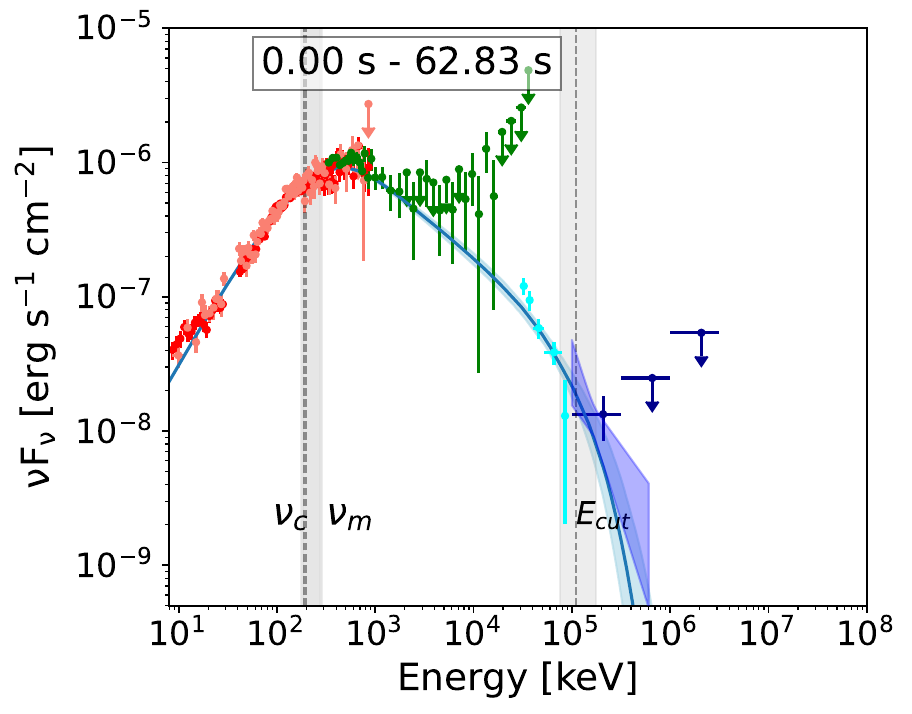}
    \includegraphics[width=0.33\linewidth]{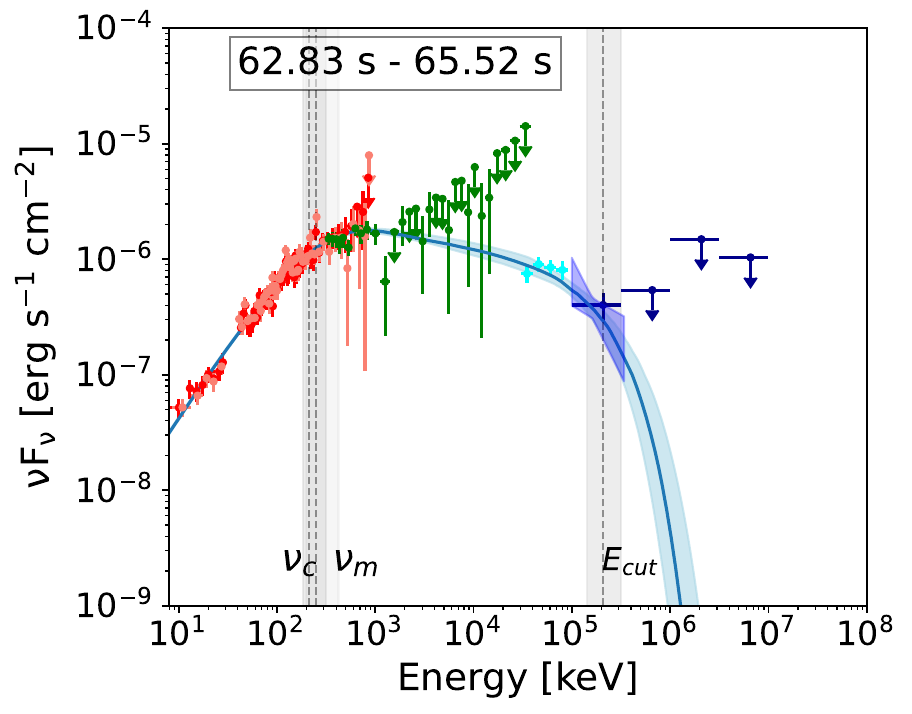}
    \includegraphics[width=0.33\linewidth]{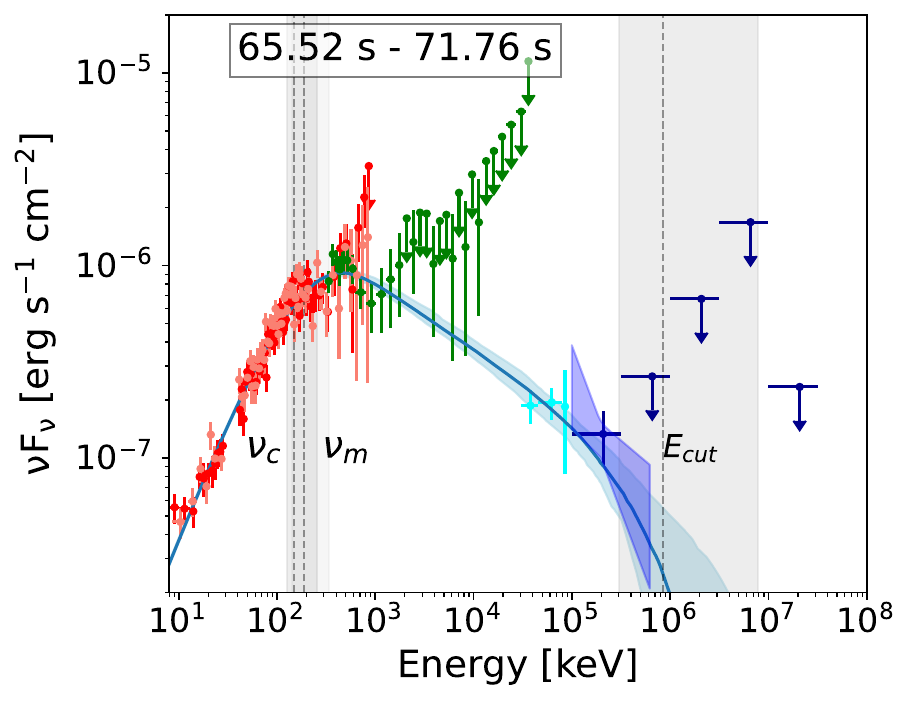} 
\end{figure*}

\clearpage
\section{GRBs in Sample-2}
In this section, we present the spectra for the single-bin GRB sample. The physical parameters derived from our fits are summarized in Table \ref{tab:Sample2}.

\begin{figure*}[p]\caption{Spectral fitting of the GRBs in Sample-2. See Table~\ref{tab:Sample2} for details.}
    \label{fig:label1}
    \centering
    \includegraphics[width=0.33\linewidth]{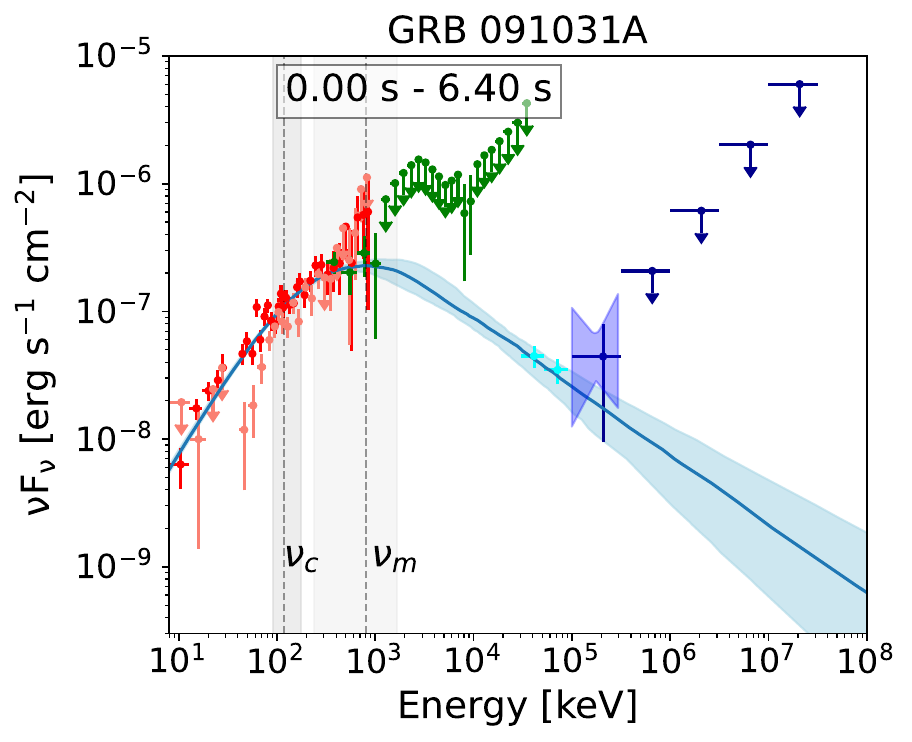}
    \includegraphics[width=0.33\linewidth]{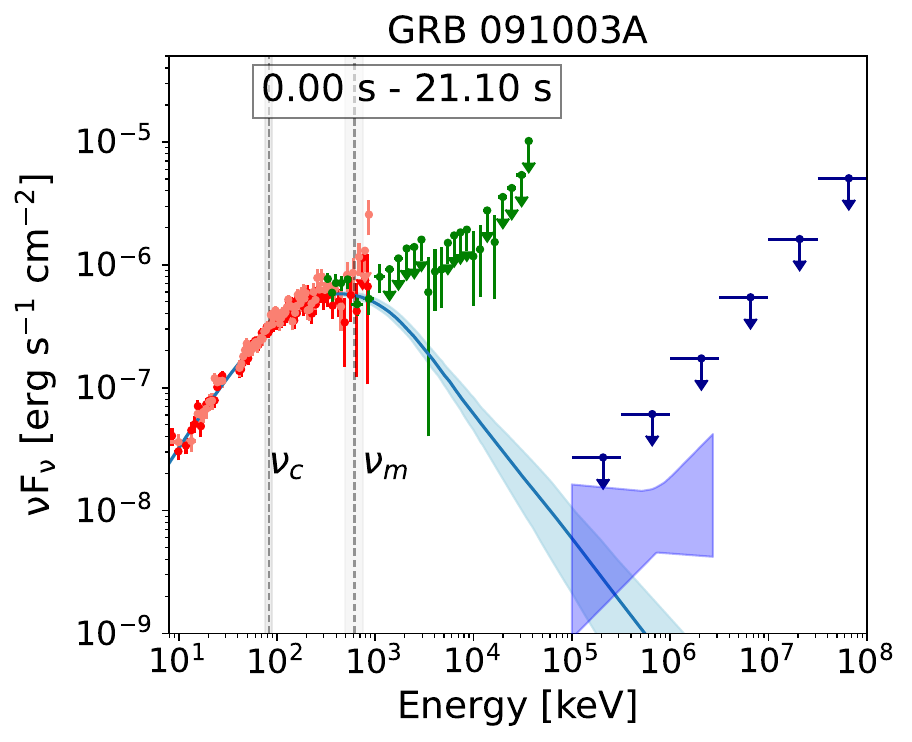}
    \includegraphics[width=0.33\linewidth]{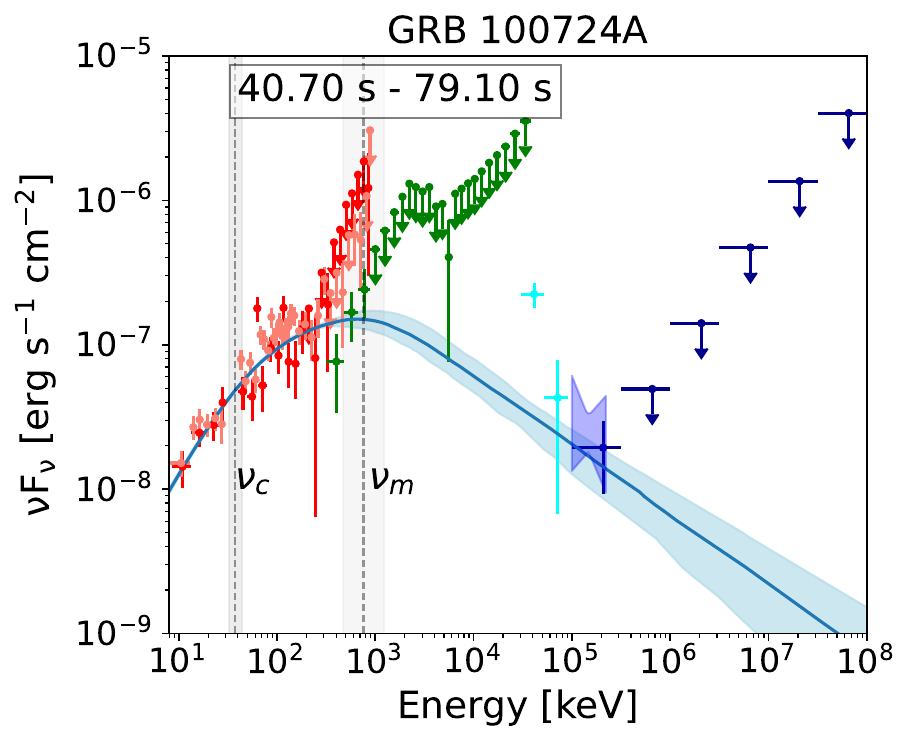}

    \includegraphics[width=0.33\linewidth]{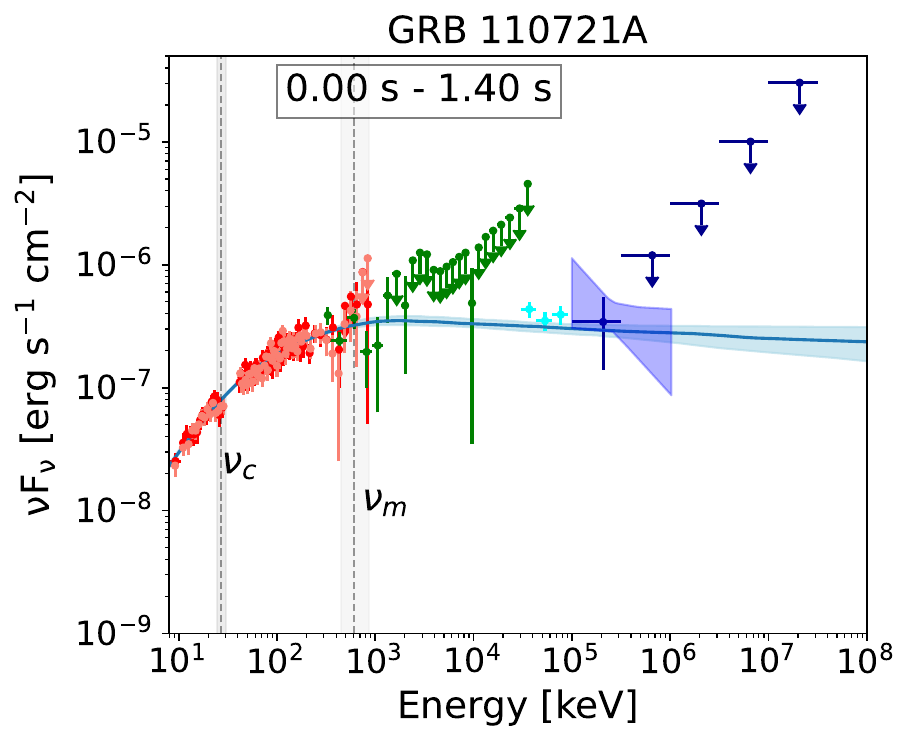}
\includegraphics[width=0.33\linewidth]{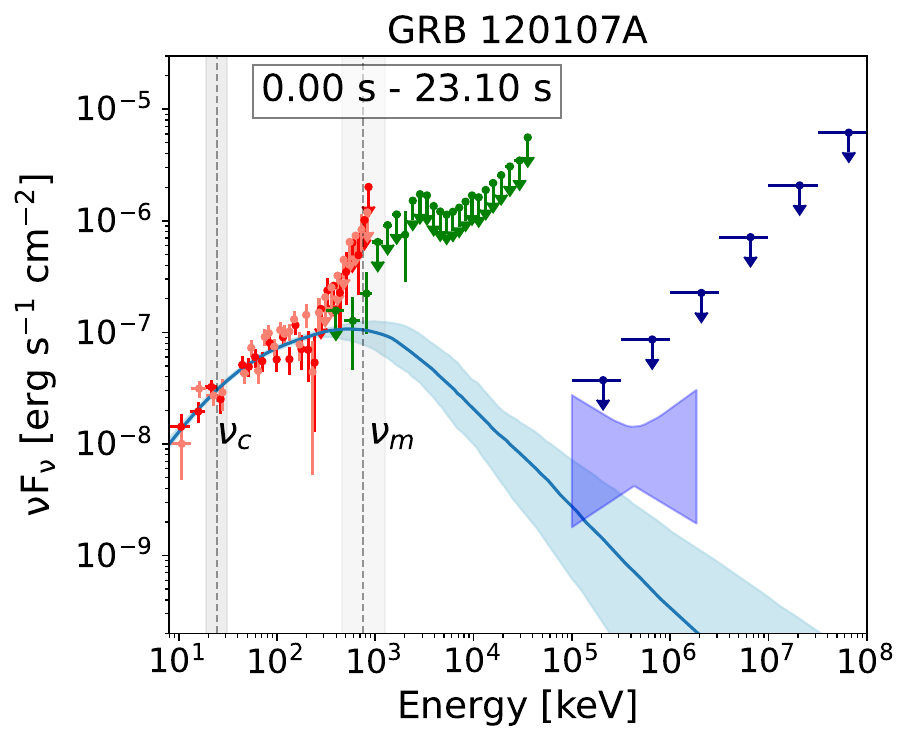}
\includegraphics[width=0.33\linewidth]{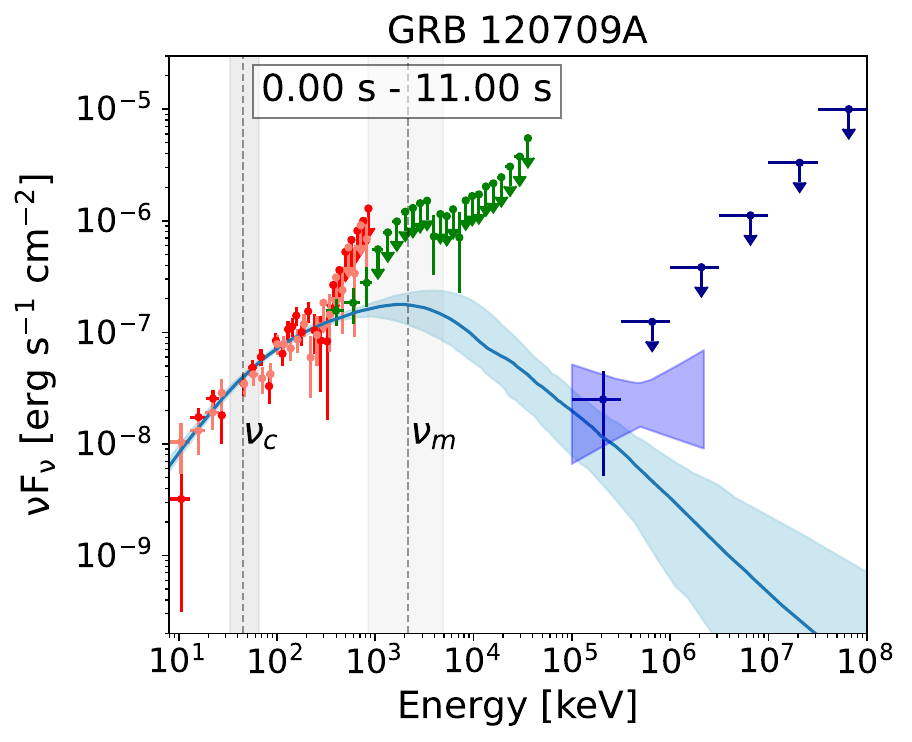}

\includegraphics[width=0.33\linewidth]{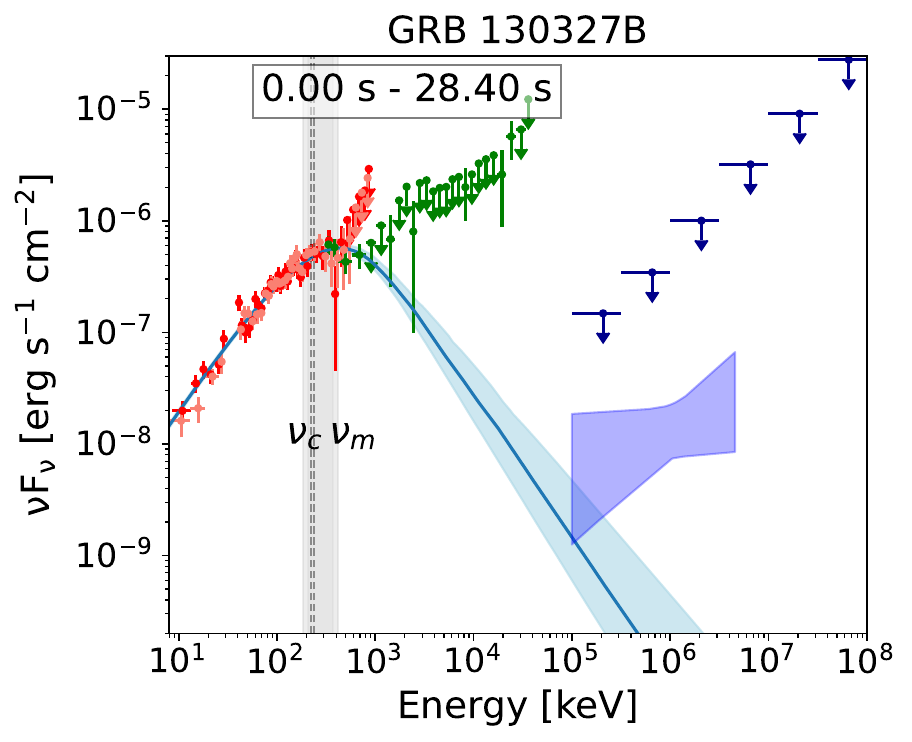}
\includegraphics[width=0.33\linewidth]{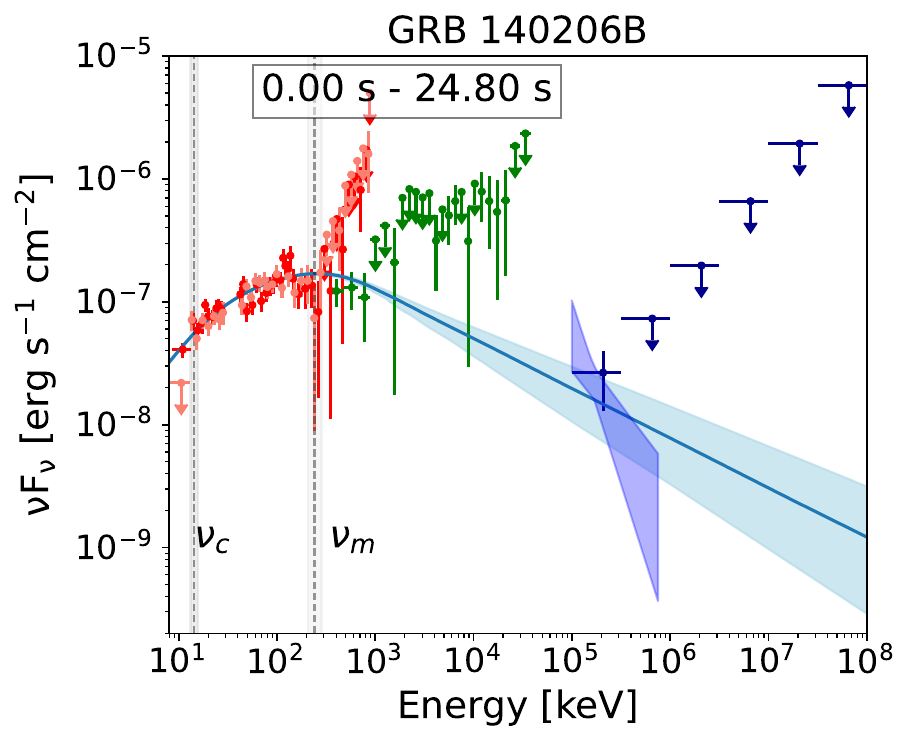}
\includegraphics[width=0.33\linewidth]{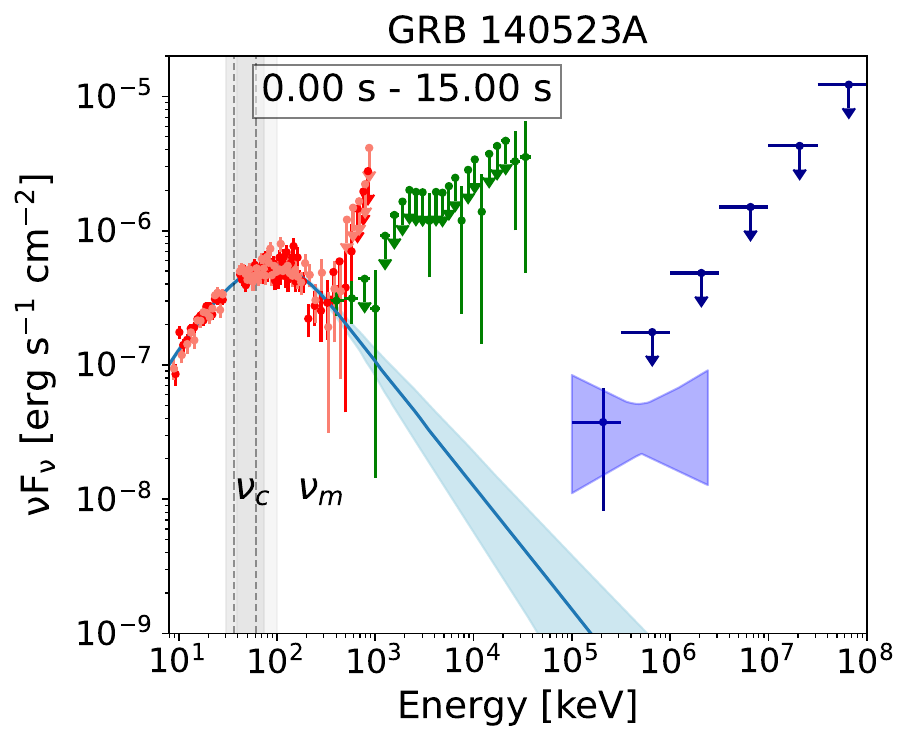}

\includegraphics[width=0.33\linewidth]{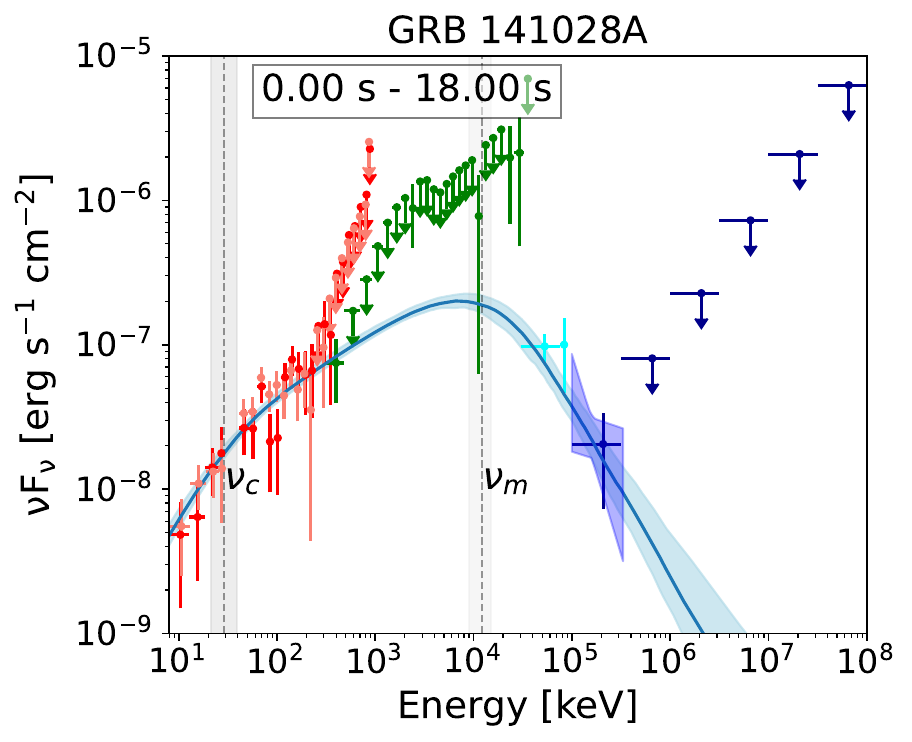}
\includegraphics[width=0.33\linewidth]{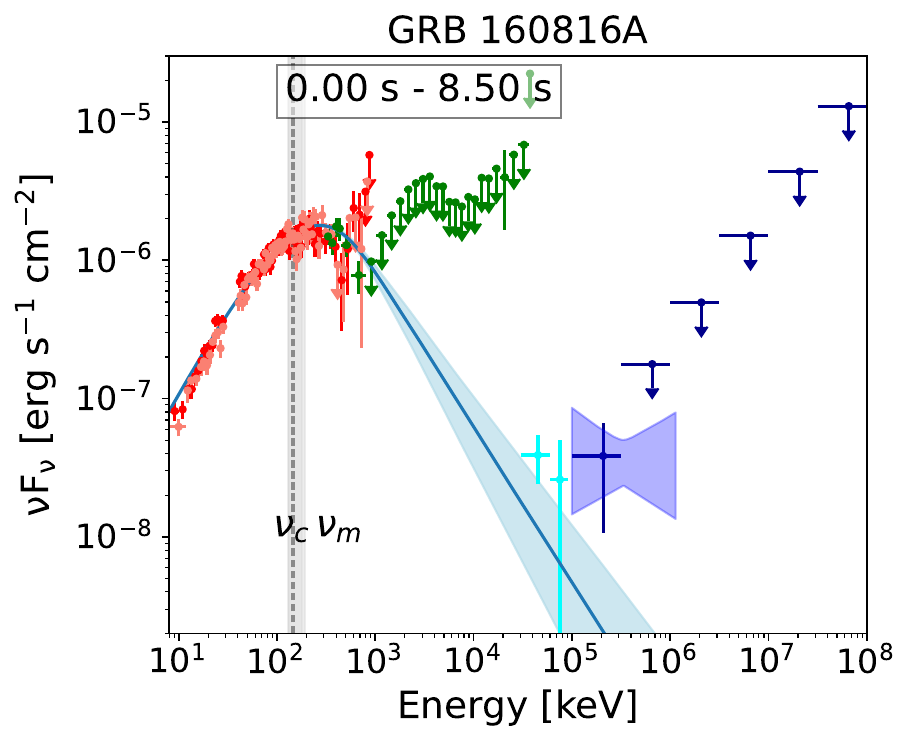}
\includegraphics[width=0.33\linewidth]{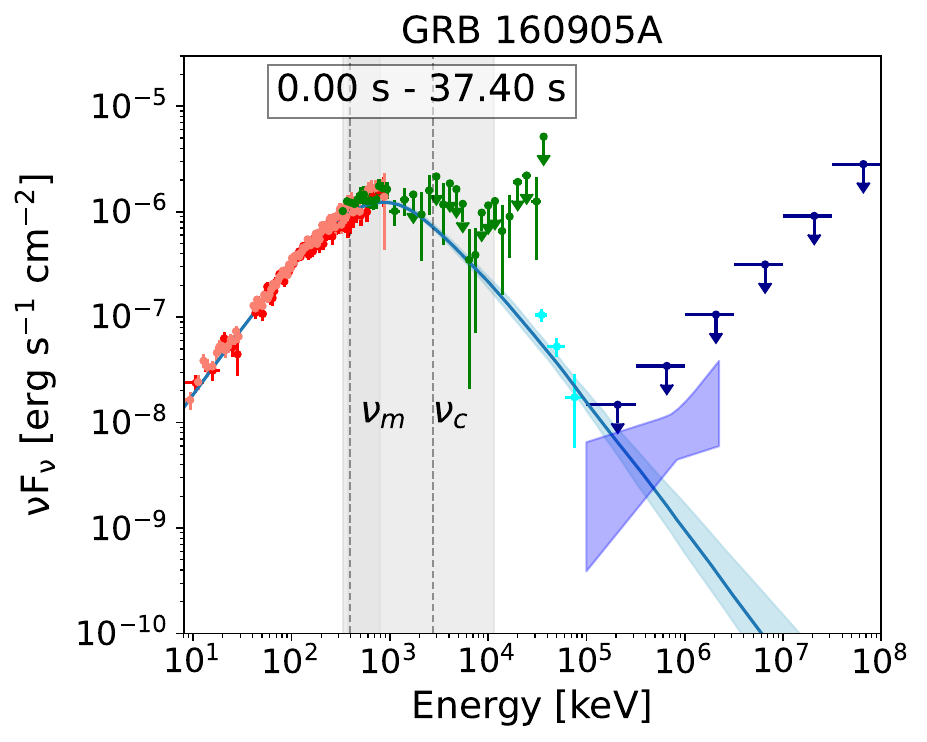}
    
\end{figure*}

\begin{figure*}[p]\ContinuedFloat
\centering
\includegraphics[width=0.33\linewidth]{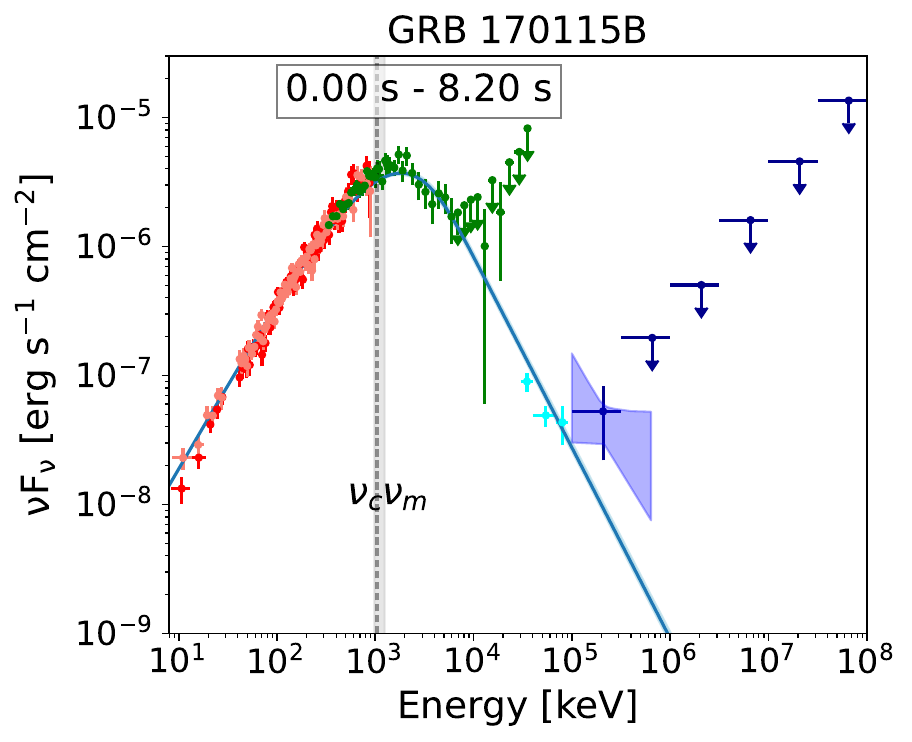}
\includegraphics[width=0.33\linewidth]{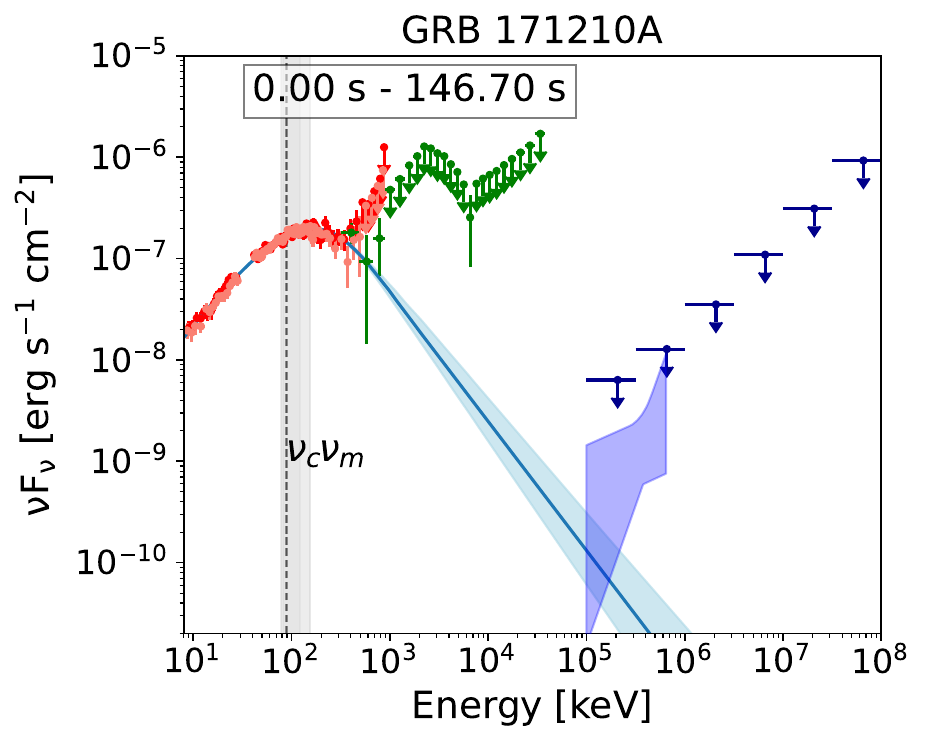}
\includegraphics[width=0.33\linewidth]{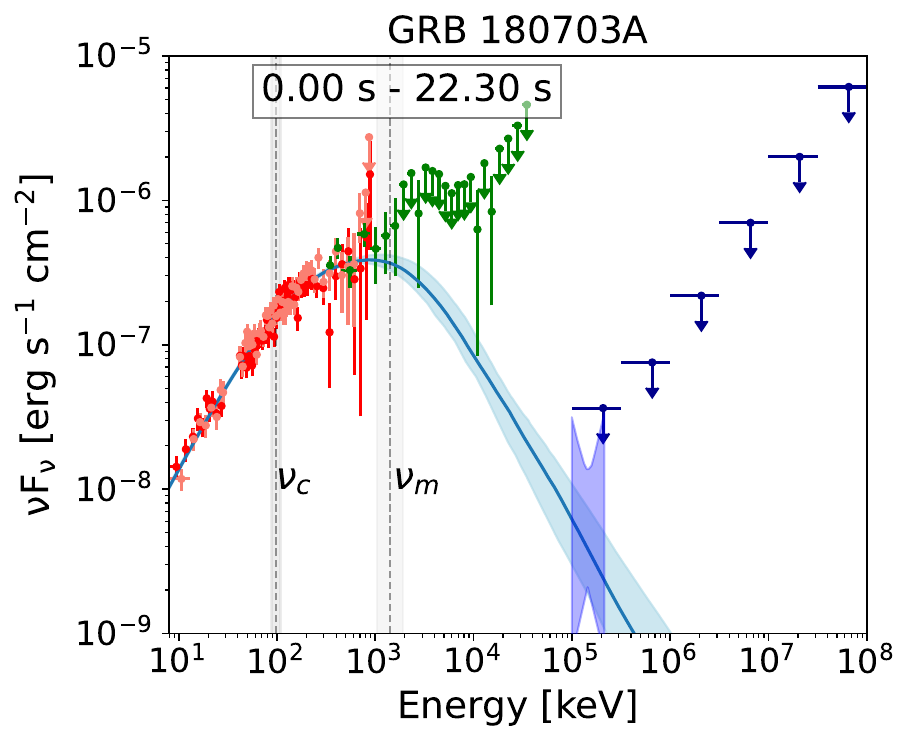}

\includegraphics[width=0.33\linewidth]{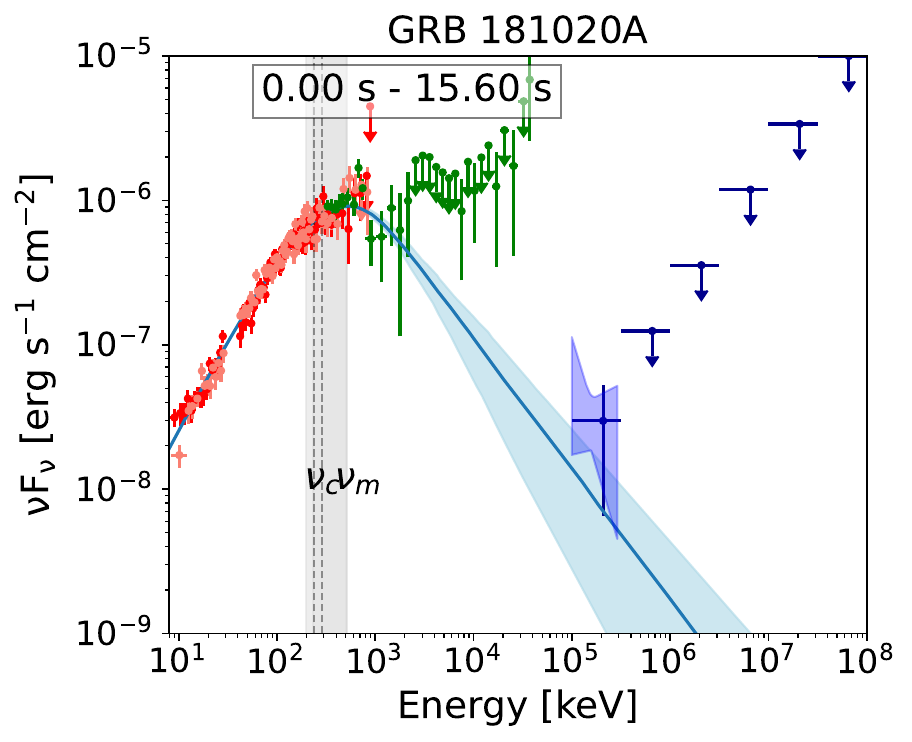}
\includegraphics[width=0.33\linewidth]{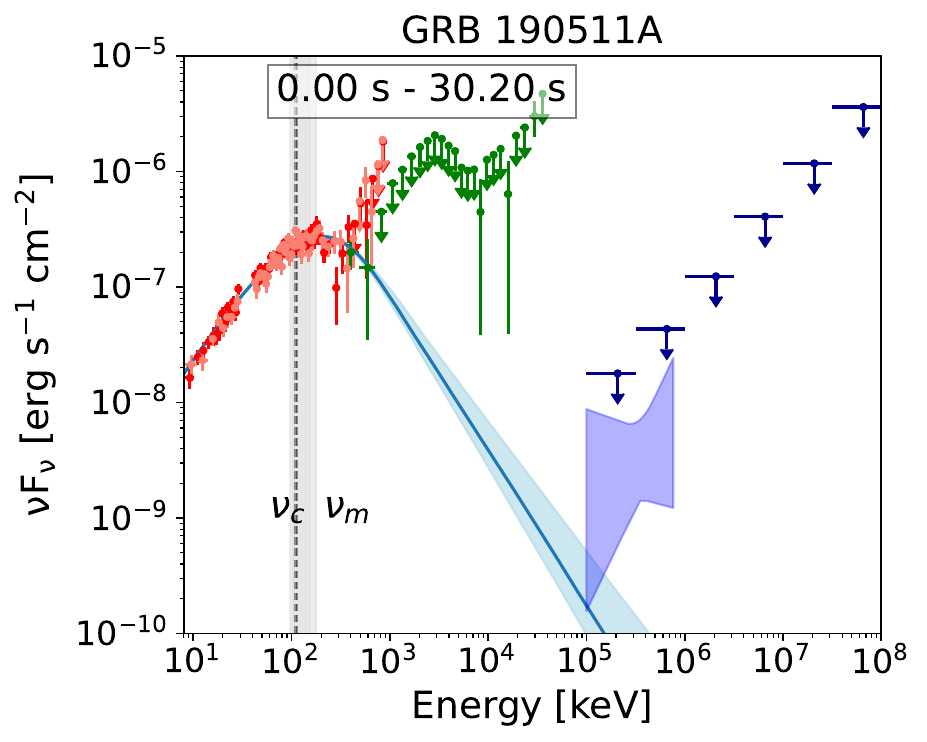}
\includegraphics[width=0.33\linewidth]{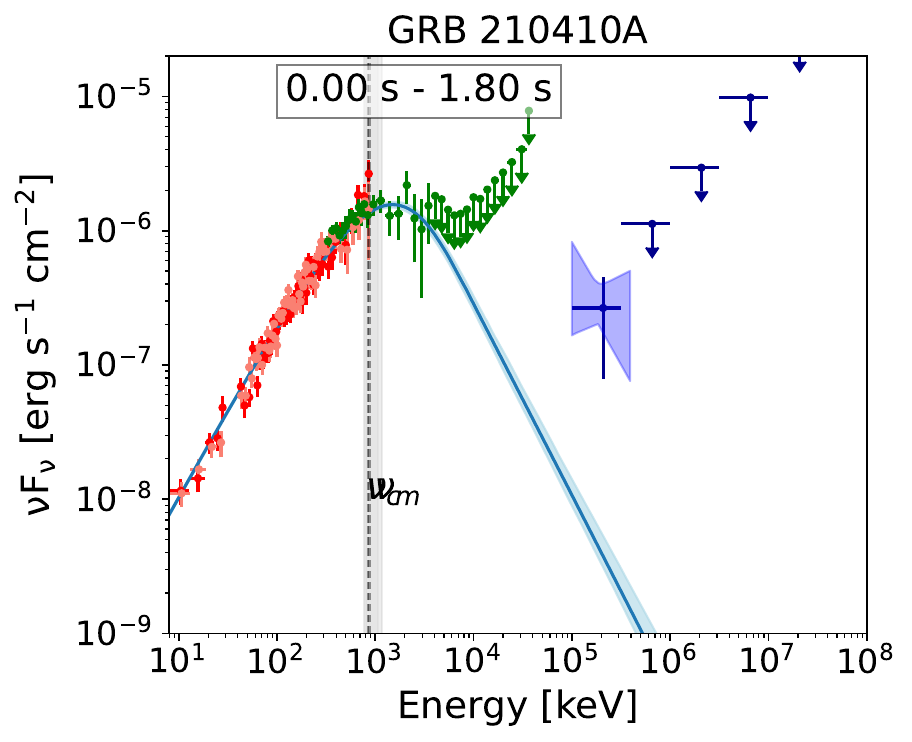}

\includegraphics[width=0.33\linewidth]{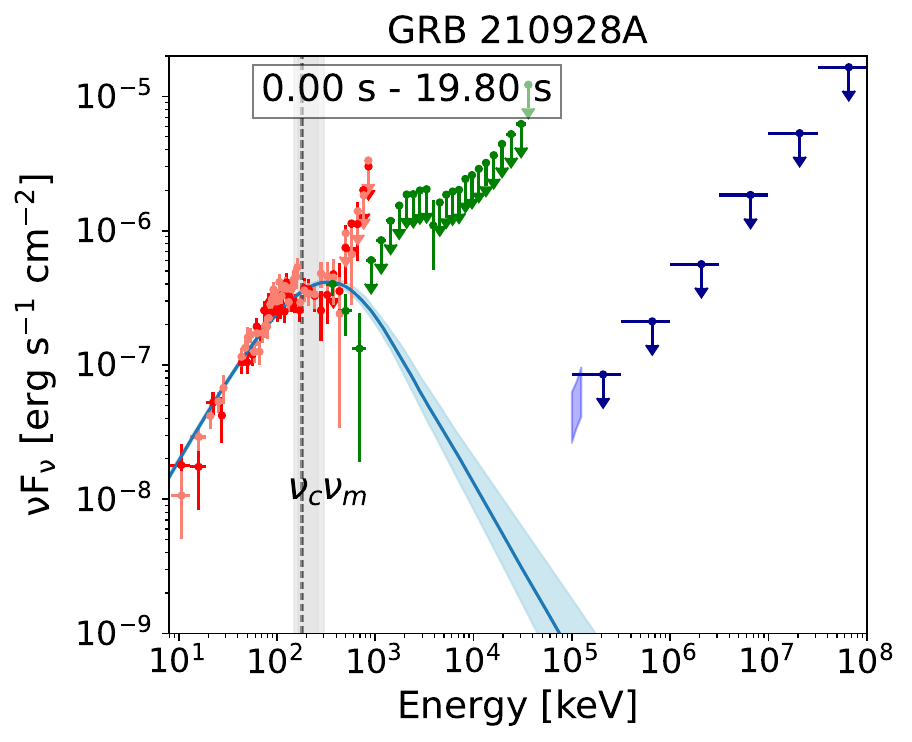}
\includegraphics[width=0.33\linewidth]{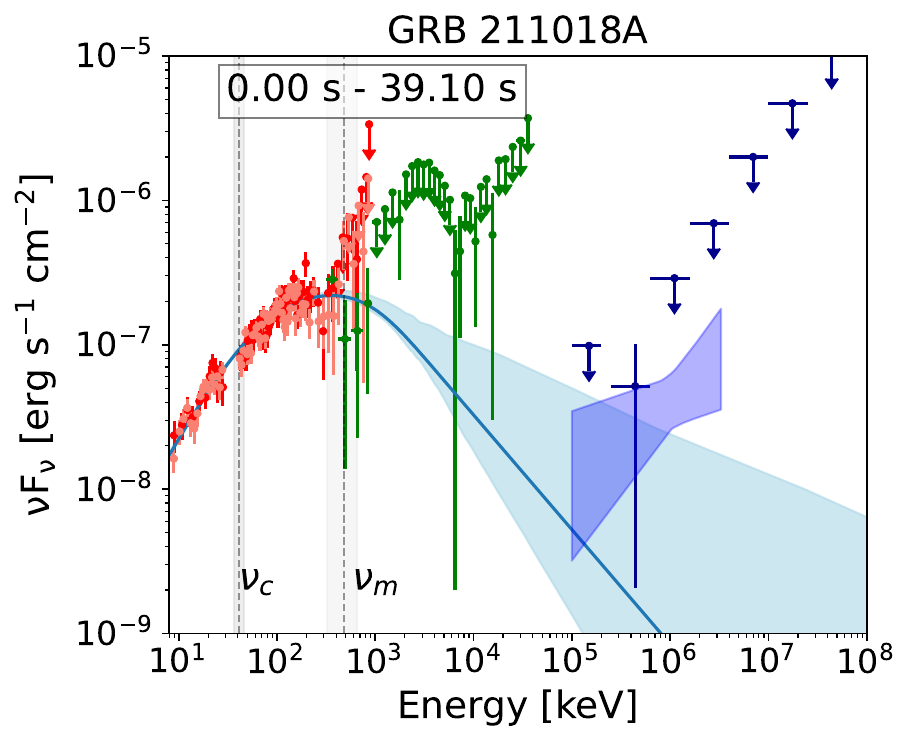}
\includegraphics[width=0.33\linewidth]{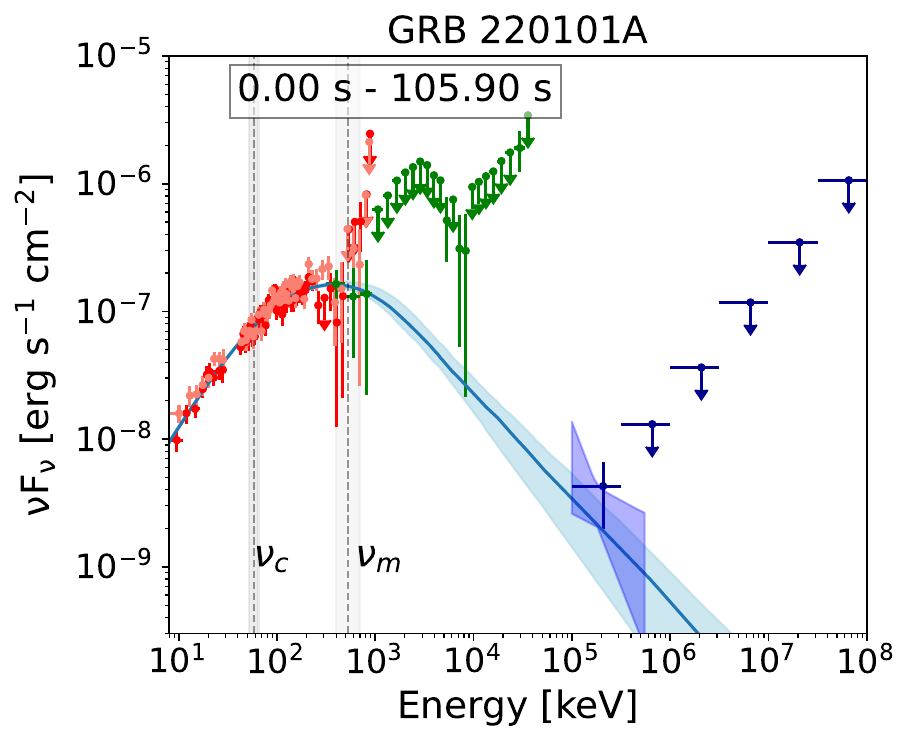}

\includegraphics[width=0.33\linewidth]{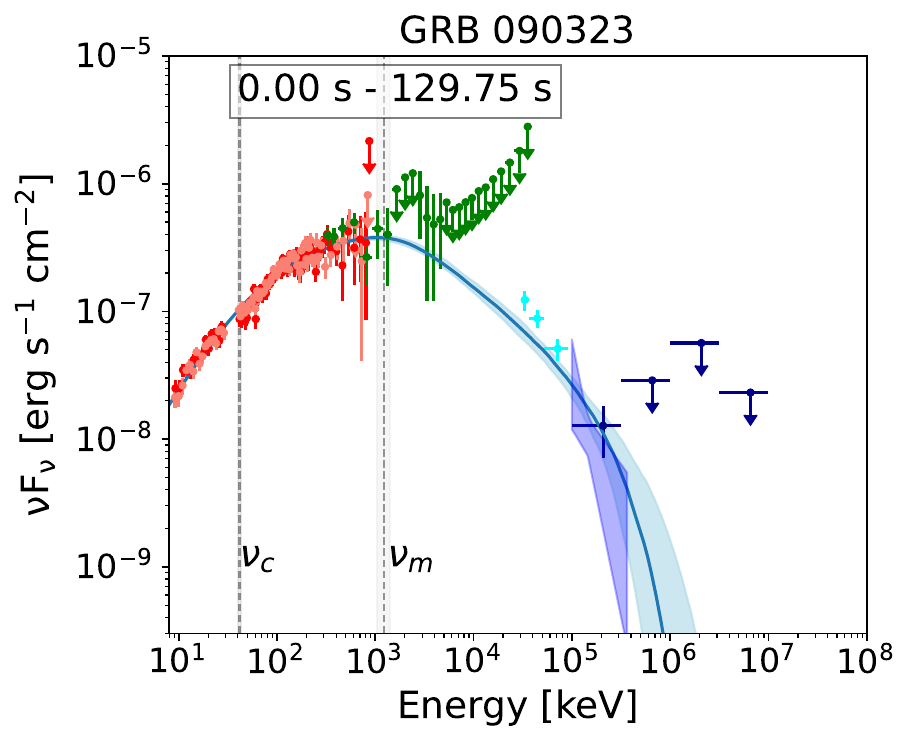}
\includegraphics[width=0.33\linewidth]{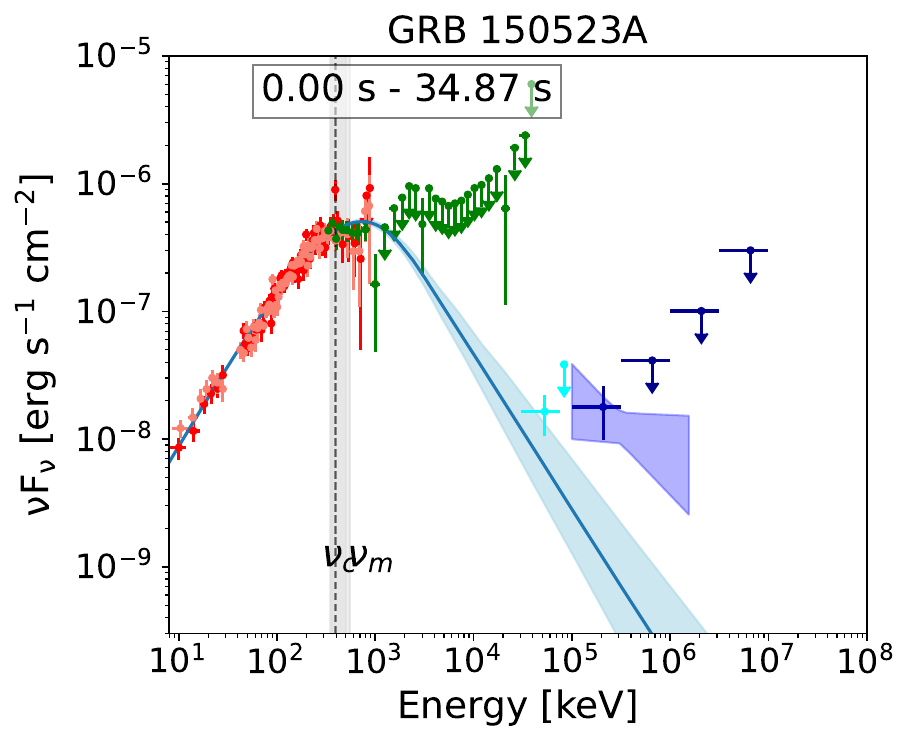}

    \caption{All GRBs in Sample-2, continued.}
    
\end{figure*}

\end{document}